\documentclass[aps,preprint,nofootinbib,prb]{revtex4-1}%
\usepackage{amssymb}
\usepackage{amsfonts}
\usepackage{amsmath}
\usepackage{graphicx}
\usepackage{natbib}%
\providecommand{\U}[1]{\protect\rule{.1in}{.1in}}
\newtheorem{theorem}{Theorem}
\newtheorem{acknowledgement}[theorem]{Acknowledgement}

\begin{document}
\preprint{ }
\title[Multi-Band Spin Magnetization Quantum Transport ]{Nonequilibrium Multi-Band Spin Quantum Transport Equations: Spin, Pseudo-Spin,
and Total Charge Coupling}
\author{Felix A. Buot}
\email{fbuot@gmu.edu}
\affiliation{Center for Computational Materials Science}
\affiliation{George Mason University, Fairfax, VA 22030, U.S.A}
\affiliation{}
\affiliation{C\&LB Research Institute, Carmen, Cebu, Philippines}

\begin{abstract}
Using the superfield nonequilibrium Green's function technique, we derive the
spatio-temporal spin magnetization quantum transport equations (SMQTEs) for a
two-band model of semiconductors. The relevant variables are the real
(Pauli-Dirac) spin, pseudo-spin, and the total charge. The results show that
the multi-band real SMQTEs are coupled to the pseudo-spin magnetization
transport equations by virtue of the presence of two additional discrete
quantum labels besides the up and down real-spin indices, namely, the
conduction and valence band quantum labels. The SMQTEs essentially consist of
three group of terms describing \ the rate of change, namely, (1) a group of
terms similar to the equation for particle quantum transport, i.e., with
spin-independent transport parameters, (2) a group of terms describing various
torques influencing the spin orientation and directional flow of spin
magnetization correlations or phase-space magnetization density, and (3) a
group of terms expressing the coupling of the real spin magnetization with the
pseudo-spin magnetization. Self-consistently, the pseudo-spin magnetization
equations incorporate the pseudo-spin/real spin coupling, as well as the
pseudo-spin coupling to the total charge.

\end{abstract}
\endpage{ }
\maketitle

\section{Introduction}

The need to analyze the ultrafast-switching-speed and power-dissipation
(speed-power product) performance of nanoelectronic devices has ushered the
extension of the classical Boltzmann transport equation to a fully
time-dependent and highly-nonlinear nonequilibrium quantum distribution
function (QDF) transport equations for charge carriers. This has been achieved
through the use of non-equilibrium Green's function, obtained either by the
time-contour quantum field formulation of Schwinger\cite{schwinger},
Keldysh\cite{keldysh}, and Kadanoff and Baym\cite{kadanoff}, or by the
real-time quantum superfield formulation of Buot, coupled with his lattice
Weyl (LW) transformation technique.\cite{LaRNC, buot1, buot2a, buot2} This
extension to QDF transport equation has proved to be highly crucial in
discovering autonomous THz current oscillations in resonant tunneling devices
through numerical simulations, and in resolving controversial issues
concerning the highly-nonlinear and bistable current-voltage characteristics
found in the experiments\cite{buotjensen, jbpaper}. Indeed, in the phase space
QDF kinetic approach has so far been the most successful technique in the
time-dependent analyses of open and active nanosystem and nanodevices, as
evidenced by the numerical work of Jensen and Buot on resonant tunneling
heterostructures\cite{jbpaper}.

To the authors' knowledge, the extension of the classical spin density
equation or the Bloch equation for spin transport\cite{DyakonovPerel}, which
is the analogue of the classical Boltzmann equation for charged-particle
transport, to fully space-time dependent and highly-nonlinear nonequilibrium
QDF transport equations for the magnetization, has only been recently reported
by the author for a single energy band in a co-authored paper\cite{arxiv} .
With the exploding surge of interest on spintronics and
nanomagnetics\cite{wolf}, there is an urgent need for this fully quantum
transport extension of the classical Bloch equation to guide the
time-dependent numerical simulation of the speed/power switching performance
and reliability analyses of realistic functional spin nanostructures and
transistors. In this paper, we report the multiband SMQTEs.

An immense activity with time-dependent spin magnetization equation, driven by
technological applications, occurs in the field of micromagnetics dealing
mostly with micro-domain wall dynamics. The so-called classical
Landau-Lifsitz-Gilbert equation, with its various modifications and
phenomenological damping terms, has become the major player in the analyses of
the dynamics of the magnetization field.\cite{gilbert, zhang, evans}

In magnetoelectronics and spintronics, the mutual dependence of electron
transport properties and magnetic properties have resulted in various
phenomena referred to as giant magnetoresistance\cite{baibich}, spin
modulator\cite{datta} based on spin precession with or without magnetic field
(based on spin-orbit coupling), and spin transfer torque\cite{ralph,
slonczewski} which have become springboards of several novel spin-based device concepts.

In analyzing the current-induced spin orientation of electrons in
semiconductors, with spin-orbit coupling, Dyakonov and
Perel\cite{DyakonovPerel} used the following transport equation for the spin
spatial-density vector, $\vec{S}$, as%
\begin{equation}
\frac{\partial}{\partial t}\vec{S}=\left[  -\vec{\nabla}\cdot\vec{Q}%
-\frac{\vec{S}}{\tau_{s}}\right]  +\vec{\Omega}\times\vec{S}, \label{dkeq}%
\end{equation}
where $\vec{Q}$ is the spin flux density,
\[
\vec{\Omega}=\frac{1}{\hbar}\mu_{B}g\vec{B},
\]
$\mu_{B}$ is the Bohr magneton, $g$ is the g-factor for electrons, and
$\vec{B}$ is the effective magnetic field. Earlier, in the 1950's
Torrey\cite{bediff} employed the classical Bloch equations for spin systems in
the diffusive regime as,%
\begin{equation}
\frac{\partial\vec{M}}{\partial t}=\left[  D\nabla^{2}\vec{M}-\vec{M}_{\tau
}\right]  +\gamma\vec{M}\times\vec{B},\text{ where }\vec{M}_{\tau}=\left\{
\frac{M_{x}}{\tau_{2}},\frac{M_{y}}{\tau_{2}},\frac{M_{z}}{\tau_{1}}\right\}
, \label{classBlochEq}%
\end{equation}
where $D$ is the diffusion coefficient, $\tau_{1}$ is the spin-magnetization
relaxation characteristic time, and $\tau_{2}$ is the spin-dephasing
characteristic time for the $M_{x}$ and $M_{y}$ components. Most of the works
that follows on spin transport also make use of classical spin transport
equations. These classical techniques are no longer valid for analyses of the
ultrafast switching speed and power-dissipation performance of the emerging
spintronic devices.

What these classical treatments have earlier shown is that spin transport
consist of terms similar to particle transport, i.e., with spin-independent
transport parameters, such as the diffusion coefficient, $D$, and relaxation
times in Eq. (\ref{classBlochEq}), and terms describing the torques in the
system.\cite{note1} However, the spin-charge coupling for a single band
transport\cite{johnson}, and spin/pseudo-spin/charge interaction\cite{gros} in
multiband transport are lacking in the equations or at best not fully treated
self-consistently within these classical and semiclassical
treatments.\cite{shipiro}

The purpose of this paper is to extend the above classical magnetization
transport equations to a fully time-dependent and highly-nonlinear
nonequilibrium spin magnetization QDF transport equations in phase space,
based on nonequilibrium Green's function technique,\cite{schwinger, keldysh,
kadanoff, LaRNC, buot1, buot2} on the same level as in the extension of the
classical Boltzmann kinetic equation to full QDF transport equation for
electrons. In the single-band magnetization QDF transport equations, the
spin-charge coupling is explicitly incorporated. In our present multi-band
generalization, we found that besides the highly-coupled transport equations
for the Pauli-Dirac or real-spin density, nonlinear coupling to the
pseudo-spin density is also present in all of the spin magnetization quantum
transport equations for coupled electrons and holes. Conversely, the
pseudo-spin magnetization quantum transport equations incorporate the
pseudo-spin/real-spin coupling, as well as the pseudo-spin coupling with the
total charge in the system. The SMQTEs are summarily cast in more physically
transparent forms in Sec. \ref{sumSMQTE}

\section{Nonequilibrium Quantum Superfield Theory}

Our starting point is the general quantum transport expressions for fermions
as obtained from the real-time quantum superfield theoretical formulation of
Buot\cite{buot1,buot2}:
\begin{align}
i\hbar\left(  \frac{\partial}{\partial t_{1}}+\frac{\partial}{\partial t_{2}%
}\right)  G^{\gtrless}  &  =\left[  vG^{\gtrless}-G^{\gtrless}v^{T}\right]
\nonumber\\
&  +\left[  \Sigma^{r}G^{\gtrless}-G^{\gtrless}\Sigma^{a}\right]  +\left[
\Sigma^{\gtrless}G^{a}-G^{r}\Sigma^{\gtrless}\right] \nonumber\\
&  +\left[  \Delta_{hh}^{r}g_{ee}^{\gtrless}-g_{hh}^{\gtrless}\Delta_{ee}%
^{a}\right] \nonumber\\
&  +\left[  \Delta_{hh}^{\gtrless}g_{ee}^{a}-g_{hh}^{r}\Delta_{ee}^{\gtrless
}\right]  . \label{grnlesseq}%
\end{align}
The last two brackets account for the Cooper pairings between Fermions of the
same specie. These do not concern us in this paper (their corresponding
transport equations\cite{buot1,buot2} are important in nonequilibrium
superconductivity). In what follows we will drop these last two brackets of
the RHS of Eq. (\ref{grnlesseq}).\cite{note2}

\section{Multi-Band Quantum Transport Equations}

In the absence of pairing between fermions of the same-specie Eq.
(\ref{grnlesseq}) becomes, by explicitly writing the quantum arguments, as%
\begin{align}
&  i\hbar\left(  \frac{\partial}{\partial t_{1}}+\frac{\partial}{\partial
t_{2}}\right)  G_{\alpha\beta}^{\lessgtr}\left(  12\right) \nonumber\\
&  =\left[  v_{\alpha}\delta_{\alpha\gamma}\delta_{1\bar{2}}G_{\gamma\beta
}^{\lessgtr}\left(  \bar{2}2\right)  -G_{\alpha\gamma}^{\lessgtr}\left(
1\bar{2}\right)  v_{\gamma}^{T}\delta_{\gamma\beta}\delta_{\bar{2}2}\right]
\nonumber\\
&  +\left[  \Sigma_{\alpha\gamma}^{r}\left(  1\bar{2}\right)  G_{\gamma\beta
}^{\lessgtr}\left(  \bar{2}2\right)  -G_{\alpha\gamma}^{\lessgtr}\left(
1\bar{2}\right)  \Sigma_{\gamma\beta}^{a}\left(  \bar{2}2\right)  \right]
\nonumber\\
&  +\left[  \Sigma_{\alpha\gamma}^{\lessgtr}\left(  1\bar{2}\right)
G_{\gamma\beta}^{a}\left(  \bar{2}2\right)  -G_{\alpha\gamma}^{r}\left(
1\bar{2}\right)  \Sigma_{\gamma\beta}^{\lessgtr}\left(  \bar{2}2\right)
\right]  , \label{multiband1}%
\end{align}
where the Greek subscript indices correspond to discrete band indices, and the
numeral indices correspond to the two-point space-time arguments. In what
follows we will treat the two-band model of a semiconductor and replace by $v$
and $c,$ the Greek indices for the valence and conduction band quantum labels, respectively.

\subsection{Electron-Hole Picture}

The transport equation for the conduction-band electrons in the electron-hole
or defect representation reduces to
\begin{align}
&  i\hbar\left(  \frac{\partial}{\partial t_{1}}+\frac{\partial}{\partial
t_{2}}\right)  G_{cc}^{e-h,<}\left(  12\right) \nonumber\\
&  =\left[  v_{c}\left(  1\xi\right)  G_{cc}^{<}\left(  \xi2\right)
-G_{cc}^{<}\left(  1\xi\right)  \ v_{c}^{T}\left(  \xi2\right)  \right]
\nonumber\\
&  +\left[  \Sigma_{cc}^{r}\ \left(  1\xi\right)  G_{cc}^{<}\left(
\xi2\right)  -G_{cc}^{<}\left(  1\xi\right)  \ \Sigma_{cc}^{a}\left(
\xi2\right)  \right] \nonumber\\
&  +\left[  \Sigma_{cc}^{<}\left(  1\xi\right)  \ G_{cc}^{a}\left(
\xi2\right)  -G_{cc}^{r}\left(  1\xi\right)  \ \Sigma_{cc}^{<}\left(
\xi2\right)  \right] \nonumber\\
&  +\left[  \Delta_{hh,cv}^{e-h,r}\left(  1\xi\right)  g_{ee,vc}%
^{e-h,<}\left(  \xi2\right)  -g_{hh,cv}^{e-h,<}\left(  1\xi\right)
\ \Delta_{ee,vc}^{e-h,a}\left(  \xi2\right)  \right] \nonumber\\
&  +\left[  \Delta_{hh,cv}^{e-h,<}\left(  1\xi\right)  \ g_{ee,vc}%
^{e-h,a}\left(  \xi2\right)  -g_{hh,cv}^{e-h,r}\left(  1\xi\right)
\ \Delta_{ee,vc}^{e-h,<}\left(  \xi2\right)  \right]  , \label{eq8}%
\end{align}
where the off-diagonal self-energies, $\Sigma_{\alpha\beta}$, and Green's
functions, $G_{\alpha\beta}$ are denoted by $\Delta^{e-h}$ and $g^{e-h}$,
respectively, in analogy to the $\Delta$-function and anomalous Green's
function, $\mathcal{F}$, i.e., $g^{e-h}\Rightarrow\mathcal{F}$ in the theory
of superconductivity.\cite{moreo,gubankova,liu} In Eq. (\ref{eq8}), there is
no Cooper pairing between electrons; moreover, the electron and electron-hole
pictures coincides for the conduction band.

To obtain the equation for the hole density from the general superfield
formulation for electrons, we make use of the electron-hole conversion table
given in Ref. \cite{buot2a, buot2}, where the relevant portion is reproduced below,

\begin{center}
$\overset{\text{Table 1. Mapping from Electron to Electron-Hole Picture}}{%
\begin{tabular}
[c]{|c|c|c|c|}\hline
$\mathbf{electron}\ picture$ & $\left\langle e-\ field\right\rangle $ &
$\left\langle e-h\ \ field\right\rangle $ & $\mathbf{e-h}\ \ picture$\\\hline
$-i\hslash G_{vv}^{<}\left(  12\right)  $ & $\left\langle \psi_{v}^{\dagger
}\left(  2\right)  \psi_{v}\left(  1\right)  \right\rangle $ & $\left\langle
\phi_{v}\left(  2\right)  \phi_{v}^{\dagger}\left(  1\right)  \right\rangle $
& $i\hslash G_{vv}^{h,>T}\left(  12\right)  $\\\hline
$-i\hslash G_{vc}^{<}\left(  12\right)  $ & $\left\langle \psi_{c}^{\dagger
}\left(  2\right)  \psi_{v}\left(  1\right)  \right\rangle $ & $\left\langle
\psi_{c}^{\dagger}\left(  2\right)  \phi_{v}^{\dagger}\left(  1\right)
\right\rangle $ & $-i\hslash g_{ee,vc}^{e-h,<}\left(  12\right)  $\\\hline
$-i\hslash G_{cv}^{<}\left(  12\right)  $ & $\left\langle \psi_{v}^{\dagger
}\left(  2\right)  \psi_{c}\left(  1\right)  \right\rangle $ & $\left\langle
\phi_{v}\left(  2\right)  \psi_{c}\left(  1\right)  \right\rangle $ &
$-i\hslash g_{hh,cv}^{e-h,<}\left(  12\right)  $\\\hline
$-i\hslash G_{cc}^{<}\left(  12\right)  $ & $\left\langle \psi_{c}^{\dagger
}\left(  2\right)  \psi_{c}\left(  1\right)  \right\rangle $ & $\left\langle
\psi_{c}^{\dagger}\left(  2\right)  \psi_{c}\left(  1\right)  \right\rangle $
& $-i\hslash G_{cc}^{e-h,<}\left(  12\right)  $\\\hline
$i\hslash G_{vv}^{>T}\left(  12\right)  $ & $\left\langle \psi_{v}\left(
2\right)  \psi_{v}^{\dagger}\left(  1\right)  \right\rangle $ & $\left\langle
\phi_{v}^{\dagger}\left(  2\right)  \phi_{v}\left(  1\right)  \right\rangle $
& $-i\hslash G_{vv}^{h,<}\left(  12\right)  $\\\hline
$i\hslash G_{vv}^{>}\left(  12\right)  $ & $\left\langle \psi_{v}\left(
1\right)  \psi_{v}^{\dagger}\left(  2\right)  \right\rangle $ & $\left\langle
\phi_{v}^{\dagger}\left(  1\right)  \phi_{v}\left(  2\right)  \right\rangle $
& $-i\hslash G_{vv}^{h,<T}\left(  12\right)  $\\\hline
\end{tabular}
}$
\end{center}

We also have the following Tables, which can also be similarly applied to the self-energies,

\begin{center}
$\overset{\text{Table 2.}}{%
\begin{tabular}
[c]{|c|c|}\hline
$\mathbf{electron}\ picture$ & $\mathbf{e-h}\ \ picture$\\\hline
$G_{vv}^{r}\left(  12\right)  $ & $-G_{vv}^{e-h,aT}\left(  12\right)
$\\\hline
$G_{vc}^{r}\left(  12\right)  $ & $g_{ee,vc}^{e-h,r}\left(  12\right)
$\\\hline
$G_{cv}^{r}\left(  12\right)  $ & $g_{hh,cv}^{e-h,r}\left(  12\right)
$\\\hline
$G_{cc}^{r}\left(  12\right)  $ & $G_{cc}^{e-h,r}\left(  12\right)  $\\\hline
\end{tabular}
}\overset{\text{Table 3.}}{\ \
\begin{tabular}
[c]{|c|c|}\hline
$\mathbf{electron}\ picture$ & $\mathbf{e-h}\ \ picture$\\\hline
$G_{vv}^{a}\left(  12\right)  $ & $-G_{vv}^{e-h,rT}\left(  12\right)
$\\\hline
$G_{vc}^{a}\left(  12\right)  $ & $g_{ee,vc}^{e-h,a}\left(  12\right)
$\\\hline
$G_{cv}^{a}\left(  12\right)  $ & $g_{hh,cv}^{e-h,a}\left(  12\right)
$\\\hline
$G_{cc}^{a}\left(  12\right)  $ & $G_{cc}^{e-h,a}\left(  12\right)  $\\\hline
\end{tabular}
}$
\end{center}

From the first table, we see that the hole density, $-i\hslash G_{vv}%
^{e-h,<}\left(  12\right)  $ correspond to $i\hslash G_{vv}^{>T}\left(
12\right)  $ in the electron energy-band representation. Thus, to obtain the
equation for the hole density given by $-i\hslash G_{vv}^{e-h,<}\left(
12\right)  ,$ one may look for the equation for $i\hslash G_{vv}^{>T}\left(
12\right)  $ in the general nonequilibrium formulation of electrons. The
equation for $i\hslash G_{vv}^{>T}\left(  12\right)  $ can be obtained from
the equation for $G_{vv}^{>},$by taking the complex conjugate of the above
equation and making use of the relation
\begin{equation}
G^{>\dagger}\left(  12\right)  =-G^{>T}\left(  12\right)  . \label{eq3}%
\end{equation}
We have the transport equation for $G^{>}\left(  12\right)  $ given by
\begin{align}
&  i\hbar\left(  \frac{\partial}{\partial t_{1}}+\frac{\partial}{\partial
t_{2}}\right)  G_{vv}^{>}\left(  12\right) \nonumber\\
&  =\left[  v_{vv}^{T}\left(  1\xi\right)  G_{vv}^{>}\left(  \xi2\right)
-G_{vv}^{>}\left(  1\xi\right)  v_{vv}\left(  \xi2\right)  \right] \nonumber\\
&  +\left[  \Sigma_{vv}^{r}\left(  1\xi\right)  G_{vv}^{>}\left(  \xi2\right)
-G_{vv}^{>}\left(  1\xi\right)  \Sigma_{vv}^{a}\left(  \xi2\right)  \right]
\nonumber\\
&  +\left[  \Sigma_{vc}^{r}\left(  1\xi\right)  G_{cv}^{>}\left(  \xi2\right)
-G_{vc}^{>}\left(  1\xi\right)  \Sigma_{cv}^{a}\left(  \xi2\right)  \right]
\nonumber\\
&  +\left[  \Sigma_{vv}^{>}\left(  1\xi\right)  G_{vv}^{a}\left(  \xi2\right)
-G_{vv}^{r}\left(  1\xi\right)  \Sigma_{vv}^{>}\left(  \xi2\right)  \right]
\nonumber\\
&  +\left[  \Sigma_{vc}^{>}\left(  1\xi\right)  G_{cv}^{a}\left(  \xi2\right)
-G_{vc}^{r}\left(  1\xi\right)  \Sigma_{cv}^{>}\left(  \xi2\right)  \right]
\label{eq4}%
\end{align}
Taking the complex conjugate of Eq. (\ref{eq4}), we obtain
\begin{align}
&  i\hbar\left(  \frac{\partial}{\partial t_{1}}+\frac{\partial}{\partial
t_{2}}\right)  G_{vv}^{>T}\left(  12\right) \nonumber\\
&  =\left[  -G_{vv}^{>}\left(  2\xi\right)  v_{vv}^{T}\left(  \xi1\right)
+v_{vv}^{T}\left(  \xi2\right)  G_{vv}^{>T}\left(  1\xi\right)  \right]
\nonumber\\
&  +\left[  -G_{vv}^{>}\left(  2\xi\right)  \Sigma_{vv}^{a}\left(
\xi1\right)  +\Sigma_{vv}^{r}\left(  2\xi\right)  G_{vv}^{>}\left(
\xi1\right)  \right] \nonumber\\
&  +\left[  -G_{cv}^{>}\left(  2\xi\right)  \Sigma_{vc}^{a}\left(
\xi1\right)  +\Sigma_{cv}^{r}\left(  2\xi\right)  G_{vc}^{>}\left(
\xi1\right)  \right] \nonumber\\
&  +\left[  -G_{vv}^{r}\left(  2\xi\right)  \Sigma_{vv}^{>}\left(
\xi1\right)  +\Sigma_{vv}^{>}\left(  2\xi\right)  G_{vv}^{a}\left(
\xi1\right)  \right] \nonumber\\
&  +\left[  -G_{cv}^{r}\left(  2\xi\right)  \Sigma_{vc}^{>}\left(
\xi1\right)  +\Sigma_{cv}^{>}\left(  2\xi\right)  G_{vc}^{a}\left(
\xi1\right)  \right]  . \label{eq5}%
\end{align}
Applying the relation
\[
-G_{vv}^{>T}\left(  12\right)  =-G_{vv}^{>}\left(  21\right)  =G_{vv}%
^{e-h,<}\left(  12\right)  ,
\]
and going entirely to the defect representation for the rest of the terms, we
obtain for the time evolution of the hole density without Cooper pairing
between holes.
\begin{align}
&  -i\hbar\left(  \frac{\partial}{\partial t_{1}}+\frac{\partial}{\partial
t_{2}}\right)  G_{vv}^{e-h,<}\left(  12\right) \nonumber\\
&  =-\left[  -v_{vv}\left(  1\xi\right)  G_{vv}^{e-h,<}\left(  \xi2\right)
+G_{vv}^{e-h,<}\left(  1\xi\right)  v_{vv}^{T}\left(  \xi2\right)  \right]
\nonumber\\
&  -\left[  -\Sigma_{vv}^{e-h,r}\left(  1\xi\right)  G_{vv}^{e-h,<}\left(
\xi2\right)  +G_{vv}^{e-h,<}\left(  1\xi\right)  \Sigma_{vv}^{e-h,a}\left(
\xi2\right)  \right] \nonumber\\
&  -\left[  -\Sigma_{vc}^{e-h,r}\left(  1\xi\right)  G_{cv}^{e-h,<}\left(
\xi2\right)  +G_{vc}^{e-h,<}\left(  1\xi\right)  \Sigma_{cv}^{e-h,a}\left(
\xi2\right)  \right] \nonumber\\
&  -\left[  -\Sigma_{vv}^{e-h,<}\left(  1\xi\right)  G_{vv}^{e-h,a}\left(
\xi2\right)  +G_{vv}^{e-h,r}\left(  1\xi\right)  \Sigma_{vv}^{e-h,<}\left(
\xi2\right)  \right] \nonumber\\
&  -\left[  -\Sigma_{vc}^{e-h,<}\left(  1\xi\right)  G_{cv}^{e-h,a}\left(
\xi2\right)  +G_{vc}^{e-h,r}\left(  1\xi\right)  \Sigma_{cv}^{e-h,<}\left(
\xi2\right)  \right]  . \label{eq6-2}%
\end{align}
Replacing the off-diagonal self-energies, $\Sigma_{\alpha\beta}$, and Green's
functions, $G_{\alpha\beta}$ by $\Delta^{e-h}$ and $g^{e-h}$, respectively, as
was done in Eq. (\ref{eq8}), allow us to rewrite the equation as%
\begin{align}
&  i\hbar\left(  \frac{\partial}{\partial t_{1}}+\frac{\partial}{\partial
t_{2}}\right)  G_{vv}^{e-h,<}\left(  12\right) \nonumber\\
&  =-\left[  v_{vv}\left(  1\xi\right)  G_{vv}^{e-h,<}\left(  \xi2\right)
-G_{vv}^{e-h,<}\left(  1\xi\right)  v_{vv}^{T}\left(  \xi2\right)  \right]
\nonumber\\
&  -\left[  \Sigma_{vv}^{e-h,r}\left(  1\xi\right)  G_{vv}^{e-h,<}\left(
\xi2\right)  -G_{vv}^{e-h,<}\left(  1\xi\right)  \Sigma_{vv}^{e-h,a}\left(
\xi2\right)  \right] \nonumber\\
&  -\left[  \Delta_{ee,vc}^{e-h,r}\left(  1\xi\right)  g_{hh,cv}%
^{e-h,<}\left(  \xi2\right)  -g_{ee,vc}^{e-h,<}\left(  1\xi\right)  \Delta
e_{hh,cv}^{e-h,a}\left(  \xi2\right)  \right] \nonumber\\
&  -\left[  \Sigma_{vv}^{e-h,<}\left(  1\xi\right)  G_{vv}^{e-h,a}\left(
\xi2\right)  -G_{vv}^{e-h,r}\left(  1\xi\right)  \Sigma_{vv}^{e-h,<}\left(
\xi2\right)  \right] \nonumber\\
&  -\left[  \Delta_{ee,vc}^{e-h,<}\left(  1\xi\right)  g_{hh,cv}%
^{e-h,a}\left(  \xi2\right)  -g_{ee,vc}^{e-h,r}\left(  1\xi\right)
\Delta_{cv}^{e-h,<}\left(  \xi2\right)  \right]  . \label{vbHoleEq}%
\end{align}
Compared with Eq. (\ref{eq8}) for the conduction band, one can readily see
that for flat bands, or atomic limit, the residual terms in these two
equations are equal,%
\begin{align}
i\hbar\left(  \frac{\partial}{\partial t_{1}}+\frac{\partial}{\partial t_{2}%
}\right)  G_{vv}^{e-h,<}\left(  12\right)   &  \Rightarrow-\left[
\Delta_{ee,vc}^{e-h,r}g_{hh,cv}^{e-h,<}-g_{ee,vc}^{e-h,<}\Delta e_{hh,cv}%
^{e-h,a}\right] \nonumber\\
&  =\left[  \Delta_{hh,cv}^{e-h,r}g_{ee,vc}^{e-h,<}-g_{hh,cv}^{e-h,<}%
\ \Delta_{ee,vc}^{e-h,a}\right]  ,\nonumber\\
i\hbar\left(  \frac{\partial}{\partial t_{1}}+\frac{\partial}{\partial t_{2}%
}\right)  G_{cc}^{e-h,<}\left(  12\right)   &  \Rightarrow\left[
\Delta_{hh,cv}^{e-h,r}g_{ee,vc}^{e-h,<}-g_{hh,cv}^{e-h,<}\ \Delta
_{ee,vc}^{e-h,a}\right]  , \label{equality}%
\end{align}
since $\Delta_{ee,vc}^{e-h,r}=\Delta_{ee,vc}^{e-h,a}$, and $\Delta
e_{hh,cv}^{e-h,a}=\Delta_{hh,cv}^{e-h,r}$. This equality simply states that
the rate of change of the holes in the valence band is the same as the rate of
change of the electrons in the conduction band, or the rate of electron
creation in conduction band equals the rate of hole creation in the valence
band, and \textit{vice versa}.

If we use the electron picture to write the transport equation for the valence
band, the resulting equation would be similar to Eq. (\ref{vbHoleEq}) without
the superscript $^{e-h}$ and with the right hand side of that equation
multiplied by $-1$. Then one can also easily see that for flat-band case, the
rate of change of electrons in the conduction band and that of the valence
band would have opposite sign, i.e., the rate of creation of electrons in the
conduction band occurs at the expense of equal rate of loss of electrons in
the valence band. Later in this paper, we will use the electron picture to
investigate the pseudo-spin transport equations. There, the total charge
represents the total net charge to be used in the Poisson equation. Of course
the total net charge at each point of the phase-space is the charge
represented by the total number of electrons in the conduction and valence
band minus the total positive background charge of the crystal lattice.

Note that the single particle Hamiltonian $v_{vv}$ is not really a two-point
function, unlike the correlation functions. The above equation reduces to
\begin{align}
&  i\hbar\left(  \frac{\partial}{\partial t_{1}}+\frac{\partial}{\partial
t_{2}}\right)  G_{vv}^{e-h,<}\left(  12\right) \nonumber\\
&  =-\left[  v_{vv}\left(  1\xi\right)  G_{vv}^{e-h,<}\left(  \xi2\right)
-G_{vv}^{e-h,<}\left(  1\xi\right)  v_{vv}^{T}\left(  \xi2\right)  \right]
\nonumber\\
&  +\left[  \Sigma_{vv}^{e-h,r}\left(  1\xi\right)  G_{vv}^{e-h,<}\left(
\xi2\right)  -G_{vv}^{e-h,<}\left(  1\xi\right)  \Sigma_{vv}^{e-h,a}\left(
\xi2\right)  \right] \nonumber\\
&  +\left[  \Sigma_{vv}^{e-h,<}\left(  1\xi\right)  G_{vv}^{e-h,a}\left(
\xi2\right)  -G_{vv}^{e-h,r}\left(  1\xi\right)  \Sigma_{vv}^{e-h,<}\left(
\xi2\right)  \right] \nonumber\\
&  +\left[  g_{ee,vc}^{e-h,>}\left(  2\xi\right)  \Delta_{hh,cv}%
^{e-h,a}\left(  \xi1\right)  -\Delta_{ee,vc}^{e-h,r}\left(  2\xi\right)
g_{hh,cv}^{e-h,>}\left(  \xi1\right)  \right] \nonumber\\
&  +\left[  g_{ee,vc}^{e-h,r}\left(  2\xi\right)  \Delta_{hh,cv}%
^{e-h,>}\left(  \xi1\right)  -\Delta_{ee,vc}^{e-h,>}\left(  2\xi\right)
g_{hh,cv}^{e-h,a}\left(  \xi1\right)  \right]  . \label{eq6}%
\end{align}

The equations for the interband or polarization terms, $g_{hh}^{e-h,<},$
$g_{ee}^{e-h,<},$ etc. can be obtained from the equations for $G_{\alpha\beta
}^{<}\left(  12\right)  $, where $\alpha\neq\beta$. In the electron-hole or
defect representation, the equation for $g_{hh,cv}^{e-h,<}$ is determined from
the general superfield formalism given by the equation for $G_{cv}^{<}\left(
12\right)  $. Upon transforming to the electron-hole picture, the change of
polarization due to the destruction of electron-hole pairs brought about by
Auger recombination, optical de-excitation, and other recombination processes
is given in the defect representation as,%

\begin{align}
&  i\hbar\left(  \frac{\partial}{\partial t_{1}}+\frac{\partial}{\partial
t_{2}}\right)  g_{hh,cv}^{e-h,<}\left(  12\right) \nonumber\\
&  =\left[  v_{cc}\left(  \xi2\right)  g_{hh,cv}^{e-h,<}\left(  1\xi\right)
-g_{hh,cv}^{e-h,<}\left(  \xi2\right)  v_{vv}^{T}\left(  1\xi\right)  \right]
\nonumber\\
&  +\left[  \Sigma_{cc}^{e-h,r}\left(  1\xi\right)  g_{hh,cv}^{e-h,<}\left(
\xi2\right)  -G_{cc}^{e-h,<}\left(  1\xi\right)  \Delta_{hh,cv}^{e-h,a}\left(
\xi2\right)  \right] \nonumber\\
&  +\left[  \Sigma_{cc}^{e-h,<}\left(  1\xi\right)  g_{hh,cv}^{e-h,a}\left(
\xi2\right)  -G_{cc}^{e-h,r}\left(  1\xi\right)  \Delta_{hh,cv}^{e-h,<}\left(
\xi2\right)  \right] \nonumber\\
&  +\left[  -\Delta_{hh,cv}^{e-h,r}\left(  1\xi\right)  G_{vv}^{e-h,>T}\left(
\xi2\right)  +g_{hh,cv}^{e-h,<}\left(  1\xi\right)  \Sigma_{vv}^{e-h,rT}%
\left(  \xi2\right)  \right] \nonumber\\
&  +\left[  -\Delta_{hh,cv}^{e-h,<}\left(  1\xi\right)  G_{vv}^{e-h,rT}\left(
\xi2\right)  +g_{hh,cv}^{e-h,r}\left(  1\xi\right)  \Sigma_{vv}^{e-h,>T}%
\left(  \xi2\right)  \right]  . \label{eq10}%
\end{align}

Similarly, upon going over to the defect representation of the reverse
process, i.e., the change of polarization due to the creation of electron-hole
pairs by impact ionization, optical excitation, Zener tunneling, and other
excitation processes is therefore given by
\begin{align}
&  i\hbar\left(  \frac{\partial}{\partial t_{1}}+\frac{\partial}{\partial
t_{2}}\right)  g_{ee,vc}^{e-h,<}\left(  12\right) \nonumber\\
&  =\left[  v_{vv}\left(  1\xi\right)  g_{ee,vc}^{e-h,<}\left(  \xi2\right)
-g_{ee,vc}^{e-h,<}\left(  1\xi\right)  v_{c}^{T}\left(  \xi2\right)  \right]
\nonumber\\
&  +\left[  -\Sigma_{vv}^{e-h,aT}\left(  1\xi\right)  g_{ee,vc}^{e-h,<}\left(
\xi2\right)  +G_{vv}^{e-h,>T}\left(  1\xi\right)  \Delta_{ee,vc}%
^{e-h,a}\left(  \xi2\right)  \right] \nonumber\\
&  +\left[  -\Sigma_{vv}^{e-h,>T}\left(  1\xi\right)  g_{ee,vc}^{e-h,a}\left(
\xi2\right)  +G_{vv}^{e-h,aT}\left(  1\xi\right)  \Delta_{ee,vc}%
^{e-h,<}\left(  \xi2\right)  \right] \nonumber\\
&  +\left[  \Delta_{ee,vc}^{e-h,r}\left(  1\xi\right)  G_{cc}^{e-h,<}\left(
\xi2\right)  -g_{ee,vc}^{e-h,<}\left(  1\xi\right)  \Sigma_{cc}^{e-h,a}\left(
\xi2\right)  \right] \nonumber\\
&  +\left[  \Delta_{ee,vc}^{e-h,<}\left(  1\xi\right)  G_{cc}^{e-h,a}\left(
\xi2\right)  -g_{ee,vc}^{e-h,r}\left(  1\xi\right)  \Sigma_{cc}^{e-h,<}\left(
\xi2\right)  \right]  . \label{eq9}%
\end{align}
Note the presence of the 'transposed' terms, namely, $G_{vv}^{e-h,>T}\left(
\xi2\right)  $, $\Sigma_{vv}^{e-h,rT}\left(  \xi2\right)  $, $G_{vv}%
^{e-h,rT}\left(  \xi2\right)  $, and $\Sigma_{vv}^{e-h,>T}\left(  \xi2\right)
$ in Eq. (\ref{eq10}) and $\Sigma_{vv}^{e-h,aT}\left(  1\xi\right)  $,
$G_{vv}^{e-h,aT}\left(  1\xi\right)  $, $G_{vv}^{e-h,>T}\left(  1\xi\right)
$, and $\Sigma_{vv}^{e-h,>T}\left(  1\xi\right)  $ in Eq. (\ref{eq9}) of the
interband 'pairing' correlation functions.

In the calculation of the \textit{pseudo-spin} transport equations the
transposed quantities in Eqs. (\ref{eq10}) and (\ref{eq9}) do not enter since
we use the electron picture to provide the two states for the electrons,
namely the two quantum labels $c$ and $v$ in order to derive the pseudo-spin
quantum transport equations. Since there is a one-to-one mapping between the
\textit{electron picture} and \textit{electron-hole picture} except for the
valence band as shown in the tables above, all one does to go from
electron-hole picture to the electron picture is to change the sign of the
right-hand side of Eq. (\ref{eq6}) and substitute for the transposed
quantities their equivalent expressions in the electron picture.

\section{Equations for Bloch Electrons with Spin}

In the presence of Pauli-Dirac spin degree of freedom, Eqs. (\ref{eq8}),
(\ref{eq6}), (\ref{eq10}), and (\ref{eq9}) become matrix equations. These are
given in the Appendix.

\subsection{Spin Canonical $2\times2$ Matrix Forms}

In spintronics, we are interested in the time-dependent evolution of the
multi-band polarization and magnetization densities, $S_{\alpha\beta,z}%
=i\hbar\left(  G_{\alpha\beta,\uparrow\uparrow}^{<}-G_{\alpha\beta
,\downarrow\downarrow}^{<}\right)  ,$ as these are transported across the
device. This leads us to transform all $2\times2$ matrices into their spin
canonical forms, defined here as expansion in terms of the Pauli spin matrices
and identity.

First let us partition the \textit{total} nonequilibrium Green's functions
(TNEGF) into $2\times2$ submatrix components, by virtue of the spin indices,
as%
\begin{equation}
\left(
\begin{array}
[c]{cccc}%
G_{cc,\uparrow\uparrow}^{e-h,<}\left(  12\right)  & G_{cc,\uparrow\downarrow
}^{e-h,<}\left(  12\right)  & g_{hh,cv,\uparrow\uparrow}^{e-h,<}\left(
12\right)  & g_{hh,cv,\uparrow\downarrow}^{e-h,<}\left(  12\right) \\
G_{cc,\downarrow\uparrow}^{e-h,<}\left(  12\right)  & G_{cc,\downarrow
\downarrow}^{e-h,<}\left(  12\right)  & g_{hh,cv,\downarrow\uparrow}%
^{e-h,<}\left(  12\right)  & g_{hh,cv,\downarrow\downarrow}^{e-h,<}\left(
12\right) \\
g_{ee,vc,\downarrow\downarrow}^{e-h,<}\left(  12\right)  & g_{ee,vc,\downarrow
\uparrow}^{e-h,<}\left(  12\right)  & G_{vv,\uparrow\uparrow}^{e-h,<}\left(
12\right)  & G_{vv,\uparrow\downarrow}^{e-h,<}\left(  12\right) \\
g_{ee,vc,\uparrow\downarrow}^{e-h,<}\left(  12\right)  & g_{ee,vc,\uparrow
\uparrow}^{e-h,<}\left(  12\right)  & G_{vv,\downarrow\uparrow}^{e-h,<}\left(
12\right)  & G_{vv,\downarrow\downarrow}^{e-h,<}\left(  12\right)
\end{array}
\right)  =\left(
\begin{array}
[c]{cc}%
\left(  CC\right)  & \left(  CV\right) \\
\left(  VC\right)  & \left(  VV\right)
\end{array}
\right)  , \label{eq15}%
\end{equation}
where $\left(  CC\right)  $, $\left(  CV\right)  $, $\left(  VC\right)  $, and
$\left(  VV\right)  $ are $2\times2$ submatrices. We can form four coupled
spin magnetization quantum transport equations for each submatrix, giving us
$16$ coupled magnetization quantum transport equations.

Thus, we first transform the $2\times2$ matrix variables in Eq. (\ref{eq15})
into their spin-canonical forms using the Pauli-matrices,%
\begin{align}
\left(  CC\right)   &  =\frac{1}{2}\left(  S_{cc,o}I+\vec{S}_{cc}\cdot
\vec{\sigma}\right)  ,\label{sep1}\\
\left(  CV\right)   &  =\frac{1}{2}\left(  \ S_{cv,o}I+\vec{S}_{cv}\cdot
\vec{\sigma}\right)  ,\label{sep2}\\
\left(  VC\right)   &  =\frac{1}{2}\left(  S_{vc,o}I+\vec{S}_{vc}\cdot
\vec{\sigma}\right)  ,\label{sep3}\\
\left(  VV\right)   &  =\frac{1}{2}\left(  S_{vv,o}I+\vec{S}_{vv}\cdot
\vec{\sigma}\right)  , \label{sep4}%
\end{align}
where we dropped the superscript $^{<}$ in the above spin-canonical form, this
is to be understood in what follows unless otherwise specifically specified.
The spin canonical expansion in terms of the Pauli matrices basically
separates the spin-independent terms from the spin-dependent terms.

In the case of the conduction, Eq. (\ref{sep1}), and valence bands,
Eq.(\ref{sep4}), the coefficients of the identity matrix, namely, $S_{cc,o}$
and $S_{vv,o}$ represent the charge density of electrons in the conduction
band and charge density of holes in the valence band, respectively. Similarly,
$S_{cv,o}$ and $S_{vc,o}$ represent the probability density for the
annihilation and creation of electron-hole pairs, respectively. The
coefficients of the vector, $\vec{\sigma}$, represent the corresponding spin
magnetization vectors. Thus the canonical form basically performs the
essential function of separating particle charge and spin.

We will later show that the coefficients of the identity matrix given above
can also be expressed in terms of the pseudo-spin vector and the total charge
in our multi-band system.

\subsection{Intraband Nonequilibrium Spin Correlation Functions}

To establish the notations used in this paper, we give the spin canonical
forms for the single particle Hamiltonian, various correlation functions, and self-energies.

For the intraband canonical terms, we have for the intraband
spin-\textit{subtraces} scalar correlation functions,%
\begin{align}
S_{cc,o}^{r,a,<}  &  =\left(  G_{cc,\uparrow\uparrow}^{e-h,r,a,<}%
+G_{cc,\downarrow\downarrow}^{e-h,r,a,<}\right)  ,\label{scc}\\
S_{vv,o}^{r,a,<}  &  =\left(  G_{vv,\uparrow\uparrow}^{e-h,r,a,<}%
+G_{vv,\downarrow\downarrow}^{e-h,r,a,<}\right)  . \label{svv}%
\end{align}
The intraband spin-correlation vector components for electrons are,%
\begin{align*}
S_{cc,x}^{r,a,<}  &  =\left(  G_{cc,\downarrow\uparrow}^{e-h,r,a,<}%
+G_{cc,\uparrow\downarrow}^{e-h,r,a,<}\right)  ,\\
iS_{cc,y}^{r,a,<}  &  =\left(  G_{cc,\downarrow\uparrow}^{e-h,r,a,<}%
-G_{cc,\uparrow\downarrow}^{e-h,r,a,<}\right)  ,\\
S_{cc,z}^{r,a,<}  &  =\left(  G_{cc,\uparrow\uparrow}^{e-h,r,a,<}%
-G_{cc,\downarrow\downarrow}^{e-h,r,a,<}\right)  ,
\end{align*}
and for holes,%
\begin{align*}
S_{vv,x}^{r,a,<}  &  =\left(  G_{vv,\downarrow\uparrow}^{e-h,r,a,<}%
+G_{vv,\uparrow\downarrow}^{e-h,r,a,<}\right)  ,\\
iS_{vv,y}^{r,a,<}  &  =\left(  G_{vv,\downarrow\uparrow}^{e-h,r,a,<}%
-G_{vv,\uparrow\downarrow}^{e-h,r,a,<}\right)  ,\\
S_{vv,z}^{r,a,<}  &  =\left(  G_{vv,\uparrow\uparrow}^{e-h,r,a,<}%
-G_{vv,\downarrow\downarrow}^{e-h,r,a,<}\right)  .
\end{align*}

\subsection{Interband Nonequilibrium Spin Correlation Functions}

For the interband canonical terms, we have interband spin-\textit{subtraces}
scalar correlation functions,%
\begin{align}
\ S_{cv,o}^{r,a,<}  &  =\left(  g_{hh,cv,\uparrow\uparrow}^{e-h,r,a,<}%
+g_{hh,cv,\downarrow\downarrow}^{e-h,r,a,<}\right)  ,\label{scv}\\
S_{vc,o}^{r,a,<}  &  =\left(  g_{ee,vc,\downarrow\downarrow}^{e-h,r,a,<}%
+g_{ee,vc,\uparrow\uparrow}^{e-h,r,a,<}\right)  , \label{svc}%
\end{align}
and for the interband spin-vector correlation components,%
\begin{align*}
S_{cv,x}^{r,a,<}  &  =\left(  g_{hh,cv,\downarrow\uparrow}^{e-h,r,a,<}%
+g_{hh,cv,\uparrow\downarrow}^{e-h,r,a,<}\right)  ,\\
iS_{cv,y}^{r,a,<}  &  =\left(  g_{hh,cv,\downarrow\uparrow}^{e-h,r,a,<}%
-g_{hh,cv,\uparrow\downarrow}^{e-h,r,a,<}\right)  ,\\
S_{cv,z}^{r,a,<}  &  =\left(  g_{hh,cv,\uparrow\uparrow}^{e-h,r,a,<}%
-g_{hh,cv,\downarrow\downarrow}^{e-h,r,a,<}\right)  .
\end{align*}
The corresponding interband reverse process 'conjugate' correlations are,
\begin{align*}
S_{vc,x}^{r,a,<}  &  =\left(  g_{ee,vc,\uparrow\downarrow}^{e-h,r,a,<}%
+g_{ee,vc,\downarrow\uparrow}^{e-h,r,a,<}\right)  ,\\
iS_{vc,y}^{r,a,<}  &  =\left(  g_{ee,vc,\uparrow\downarrow}^{e-h,r,a,<}%
-g_{ee,vc,\downarrow\uparrow}^{e-h,r,a,<}\right)  ,\\
S_{vc,z}^{r,a,<}  &  =\left(  g_{ee,vc,\downarrow\downarrow}^{e-h,r,a,<}%
-g_{ee,vc,\uparrow\uparrow}^{e-h,r,a,<}\right)  .
\end{align*}

\subsection{Single-Particle Hamiltonian Spin Canonical Forms}

We have for the single-particle Hamiltonian for the electrons expressed in
canonical forms, assuming the presence of external magnetic field and/or
spin-orbit coupling,
\[
v_{c,\sigma\sigma^{\prime}}=\frac{1}{2}\left(  \bar{H}_{c}\hat{I}%
+\mathcal{\vec{B}}_{c}\cdot\vec{\sigma}\right)  ,
\]
where the upper bar in $\bar{H}_{c}$ indicates the subtrace of the $2\times2 $
spin matrix for the conduction band. Similarly, for the holes we have,%
\[
v_{v,\sigma\sigma^{\prime}}=\frac{1}{2}\left(  \bar{H}_{v}\hat{I}%
+\mathcal{\vec{B}}_{v}\cdot\vec{\sigma}\right)  .
\]
These lead to the matrix expression for electrons,%
\[
v_{c,\sigma\sigma^{\prime}}=\frac{1}{2}\left(
\begin{array}
[c]{cc}%
\bar{H}_{c}+\mathcal{B}_{c,z} & \mathcal{B}_{c,x}-i\mathcal{B}_{c,y}\\
\mathcal{B}_{c,x}+i\mathcal{B}_{c,y} & \bar{H}_{c}-\mathcal{B}_{c,z}%
\end{array}
\right)  ,
\]
and for holes,%
\[
v_{v,\sigma\sigma^{\prime}}=\frac{1}{2}\left(
\begin{array}
[c]{cc}%
\bar{H}_{v}+\mathcal{B}_{v,z} & \mathcal{B}_{v,x}-i\mathcal{B}_{v,y}\\
\mathcal{B}_{v,x}+i\mathcal{B}_{v,y} & \bar{H}_{v}-\mathcal{B}_{v,z}%
\end{array}
\right)  .
\]

The components of $\mathcal{\vec{B}}_{c}$ and $\mathcal{\vec{B}}_{v}$ are real
valued, since the single-particle Hamiltonian, $v$, is Hermitian. These are
the effective magnetic field components, multiplied by $\mu_{B}g$, which
accounts for the external magnetic fields as well as the effects of spin-orbit coupling.

\subsection{Canonical Forms for the Spin-Dependent Electrons Self-Energy}

Similarly, we have for the spin-canonical forms for the electron
self-energies,%
\[
\Sigma_{cc,\sigma\sigma^{\prime}}^{r}=\frac{1}{2}\left(
\begin{array}
[c]{cc}%
\bar{\Sigma}_{cc}^{r}+\Xi_{cc,z}^{r} & \Xi_{cc,x}^{r}-i\Xi_{cc,y}^{r}\\
\Xi_{cc,x}^{r}+i\Xi_{cc,y}^{r} & \bar{\Sigma}_{cc}^{r}-\Xi_{cc,z}^{r}%
\end{array}
\right)  ,
\]%
\[
\Sigma_{cc,\sigma\sigma^{\prime}}^{a}=\frac{1}{2}\left(
\begin{array}
[c]{cc}%
\bar{\Sigma}_{cc}^{a}+\Xi_{cc,z}^{a} & \Xi_{cc,x}^{a}-i\Xi_{cc,y}^{a}\\
\Xi_{cc,x}^{a}+i\Xi_{cc,y}^{a} & \bar{\Sigma}_{cc}^{a}-\Xi_{cc,z}^{a}%
\end{array}
\right)  ,
\]%
\[
\Sigma_{cc,\sigma\sigma^{\prime}}^{<}=\frac{1}{2}\left(
\begin{array}
[c]{cc}%
\bar{\Sigma}_{cc}^{<}+\Xi_{cc,z}^{<} & \Xi_{cc,x}^{<}-i\Xi_{cc,y}^{<}\\
\Xi_{cc,x}^{<}+i\Xi_{cc,y}^{<} & \bar{\Sigma}_{cc}^{<}-\Xi_{cc,z}^{<}%
\end{array}
\right)  .
\]

\subsection{Spin-Canonical Forms for the Holes Self-Energy}%

\[
\Sigma_{\ vv,\sigma\sigma^{\prime}}^{r}=\frac{1}{2}\left(
\begin{array}
[c]{cc}%
\bar{\Sigma}_{\ vv}^{r}+\Xi_{\ vv,z}^{r} & \Xi_{\ vv,x}^{r}-i\Xi_{\ vv,y}%
^{r}\\
\Xi_{\ vv,x}^{r}+i\Xi_{\ vv,y}^{r} & \bar{\Sigma}_{\ vv}^{r}-\Xi_{\ vv,z}^{r}%
\end{array}
\right)  ,
\]

\[
\Sigma_{\ vv,\sigma\sigma^{\prime}}^{a}=\frac{1}{2}\left(
\begin{array}
[c]{cc}%
\bar{\Sigma}_{\ vv}^{a}+\Xi_{\ vv,z}^{a} & \Xi_{\ vv,x}^{a}-i\Xi_{\ vv,y}%
^{a}\\
\Xi_{\ vv,x}^{a}+i\Xi_{\ vv,y}^{a} & \bar{\Sigma}_{\ vv}^{a}-\Xi_{\ vv,z}^{a}%
\end{array}
\right)  ,
\]

\[
\Sigma_{\ vv,\sigma\sigma^{\prime}}^{<}=\frac{1}{2}\left(
\begin{array}
[c]{cc}%
\bar{\Sigma}_{\ vv}^{<}+\Xi_{\ vv,z}^{<} & \Xi_{\ vv,x}^{<}-i\Xi_{\ vv,y}%
^{<}\\
\Xi_{\ vv,x}^{<}+i\Xi_{\ vv,y}^{<} & \bar{\Sigma}_{\ vv}^{<}-\Xi_{\ vv,z}^{<}%
\end{array}
\right)  .
\]

\subsection{Canonical Forms of Electron-Hole Pairing Self-Energy}%

\[
\Delta_{hh,cv,\sigma\sigma^{\prime}}^{e-h,r}=\frac{1}{2}\left(  \bar{\Delta
}_{hh,cv}^{r}+\vec{\delta}_{hh,cv}^{r}\cdot\vec{\sigma}\right)  ,
\]

\[
\Delta_{hh,cv,\sigma\sigma^{\prime}}^{e-h,<}=\frac{1}{2}\left(  \bar{\Delta
}_{hh,cv}^{\,<}+\vec{\delta}_{hh,cv}^{<}\cdot\vec{\sigma}\right)  ,
\]

\[
\Delta_{ee,vc,\sigma\sigma^{\prime}}^{e-h,a}=\frac{1}{2}\left(  \bar{\Delta
}_{ee,vc}^{a}+\vec{\delta}_{ee,vc}^{a}\cdot\vec{\sigma}\right)  ,
\]

\[
\Delta_{ee,vc,\sigma\uparrow}^{e-h,<}=\frac{1}{2}\left(  \bar{\Delta}%
_{ee,vc}^{<}+\vec{\delta}_{ee,vc}^{<}\cdot\vec{\sigma}\right)  ,
\]

\[
\Delta_{hh,cv,\sigma\sigma^{\prime}}^{e-h,r}=\frac{1}{2}\left(
\begin{array}
[c]{cc}%
\bar{\Delta}_{hh,cv}^{r}+\delta_{hh,cv,z}^{r} & \delta_{hh,cv,x}^{r}%
-i\delta_{hh,cv,y}^{r}\\
\delta_{hh,cv,x}^{r}+i\delta_{hh,cv,y}^{r} & \bar{\Delta}_{hh,cv}^{r}%
-\delta_{hh,cv,z}^{r}%
\end{array}
\right)  ,
\]

\[
\Delta_{hh,cv,\sigma\sigma^{\prime}}^{e-h,<}=\frac{1}{2}\left(
\begin{array}
[c]{cc}%
\bar{\Delta}_{hh,cv}^{<}+\delta_{hh,cv,z}^{<} & \delta_{hh,cv,x}^{<}%
-i\delta_{hh,cv,y}^{<}\\
\delta_{hh,cv,x}^{<}+i\delta_{hh,cv,y}^{<} & \bar{\Delta}_{hh,cv}^{<}%
-\delta_{hh,cv,z}^{<}%
\end{array}
\right)  ,
\]

\[
\Delta_{ee,vc,\sigma\sigma^{\prime}}^{e-h,a}=\frac{1}{2}\left(
\begin{array}
[c]{cc}%
\bar{\Delta}_{ee,vc}^{a}+\delta_{ee,vc,z}^{a} & \delta_{ee,vc,x}^{a}%
-i\delta_{ee,vc,y}^{a}\\
\delta_{ee,vc,x}^{a}+i\delta_{ee,vc,y}^{a} & \bar{\Delta}_{ee,vc}^{a}%
-\delta_{ee,vc,z}^{a}%
\end{array}
\right)  ,
\]

\[
\Delta_{ee,vc,\sigma\sigma^{\prime}}^{e-h,<}=\frac{1}{2}\left(
\begin{array}
[c]{cc}%
\bar{\Delta}_{ee,vc}^{<}+\delta_{ee,vc,z}^{<} & \delta_{ee,vc,x}^{<}%
-i\delta_{ee,vc,y}^{<}\\
\delta_{ee,vc,x}^{<}+i\delta_{ee,vc,y}^{<} & \bar{\Delta}_{ee,vc}^{<}%
-\delta_{ee,vc,z}^{<}%
\end{array}
\right)  .
\]

\section{Nonequilibrium Pauli-Dirac Spin Equations}

The twelve transport equations for the components of the multi-band spin
magnetization transport equations can be written is more compact form as
spin-vector equations, a generalization of Eq. (\ref{dkeq}) to multi-band
spin-magnetization quantum transport equations.

The results can be summarized as four-coupled \textit{vector} equations for
the spin magnetization-distribution functions,%
\begin{align}
&  i\hbar\left(  \frac{\partial}{\partial t_{1}}+\frac{\partial}{\partial
t_{2}}\right)  \vec{S}_{cc}^{<}\nonumber\\
&  =\frac{1}{2}\left\{  \left[  \bar{H}_{c},\vec{S}_{cc}^{<}\right]  +\left[
\mathcal{\vec{B}}_{cc},S_{cc,o}^{<}\right]  +i\left[  \mathcal{\vec{B}}%
_{cc}\times\vec{S}_{cc}^{<}-\vec{S}_{cc}^{<}\times\mathcal{\vec{B}}%
_{cc}\right]  \right\} \nonumber\\
&  +\frac{1}{2}\left[
\begin{array}
[c]{c}%
+\left[  \bar{\Sigma}_{cc}^{r}\vec{S}_{cc}^{<}-\ \vec{S}_{cc}^{<}\bar{\Sigma
}_{cc}^{a}\right]  +\left[  \vec{\Xi}_{cc}^{r}\ S_{cc,o}^{<}-S_{cc,o}^{<}%
\vec{\Xi}_{cc}^{a}\right]  \ \ \\
+i\left[  \vec{\Xi}_{cc}^{r}\ \times\vec{S}_{cc}^{<}-\ \vec{S}_{cc}^{<}%
\times\vec{\Xi}_{cc}^{a}\right]
\end{array}
\right] \nonumber\\
&  +\frac{1}{2}\left[
\begin{array}
[c]{c}%
\left[  \bar{\Sigma}_{cc}^{<}\vec{S}_{cc}^{a}-\ \vec{S}_{cc}^{r}\bar{\Sigma
}_{cc}^{<}\right]  +\left[  \vec{\Xi}_{cc}^{<}S_{cc,o}^{a}-S_{cc,o}^{r}%
\vec{\Xi}_{cc}^{<}\right]  \ \\
+i\left[  \vec{\Xi}_{cc}^{<}\times\vec{S}_{cc}^{a}-\ \vec{S}_{cc}^{r}%
\times\vec{\Xi}_{cc}^{<}\right]  \
\end{array}
\right] \nonumber\\
&  +\frac{1}{2}\left[
\begin{array}
[c]{c}%
\frac{1}{i\hbar}\left[  \bar{\Delta}_{hh,cv}^{r}\vec{S}_{vc}^{<}-\vec{S}%
_{cv}^{<}\bar{\Delta}_{ee,vc}^{a}\right]  +\frac{1}{i\hbar}\left[  \vec
{\delta}_{hh,cv}^{r}S_{vc,o}^{<}-S_{cv,o}^{<}\vec{\delta}_{ee,vc}^{a}\right]
\\
+\frac{1}{\hbar}\left[  \vec{\delta}_{hh,cv}^{r}\times\vec{S}_{vc}^{<}-\vec
{S}_{cv}^{<}\times\vec{\delta}_{ee,vc}^{a}\right]
\end{array}
\right] \nonumber\\
&  +\frac{1}{2}\left[
\begin{array}
[c]{c}%
+\left[  \bar{\Delta}_{hh,cv}^{<}\ \vec{S}_{vc}^{a}-\vec{S}_{cv}^{r}%
\bar{\Delta}_{ee,vc}^{<}\right]  +\left[  \vec{\delta}_{hh,cv}^{<}%
\ S_{vc,o}^{a}-S_{cv,o}^{r}\vec{\delta}_{ee,vc}^{<}\right] \\
+i\left[  \vec{\delta}_{hh,cv}^{<}\ \times\vec{S}_{vc}^{a}-\vec{S}_{cv}%
^{r}\times\vec{\delta}_{ee,vc}^{<}\right]
\end{array}
\right]  ,\nonumber\\
&  \label{genBloch1}%
\end{align}

\begin{align}
&  i\hbar\left(  \frac{\partial}{\partial t_{1}}+\frac{\partial}{\partial
t_{2}}\right)  \vec{S}_{cv}^{<}\nonumber\\
&  =\frac{1}{2}\left[
\begin{array}
[c]{c}%
\left[  \bar{H}_{c}\vec{S}_{cv}^{<}-\vec{S}_{cv}^{<}\bar{H}_{v}\right]
+\left[  \mathcal{\vec{B}}_{cc}S_{cv,o}^{<}-S_{cv,o}^{<}\mathcal{\vec{B}}%
_{vv}\right] \\
+i\left[  \mathcal{\vec{B}}_{cc}\times\vec{S}_{cv}^{<}-\vec{S}_{cv}^{<}%
\times\mathcal{\vec{B}}_{vv}\right]
\end{array}
\right] \nonumber\\
&  \frac{1}{2}\left[
\begin{array}
[c]{c}%
\left[  \bar{\Sigma}_{cc}^{r}\vec{S}_{cv}^{<}-\vec{S}_{cc}^{<}\bar{\Delta
}_{hh,cv}^{\ a}\right]  +\left[  \vec{\Xi}_{cc}^{r}S_{cv,o}^{<}-S_{cc,o}%
^{<}\vec{\delta}_{hh,cv}^{\ a}\right] \\
+i\left[  \vec{\Xi}_{cc}^{r}\times\vec{S}_{cv}^{<}\ -\vec{S}_{cc}^{<}%
\times\vec{\delta}_{hh,cv}^{\ a}\right]
\end{array}
\right] \nonumber\\
\frac{1}{2}  &  \left[
\begin{array}
[c]{c}%
\left[  \bar{\Sigma}_{cc}^{<}\vec{S}_{cv}^{a}-\vec{S}_{cc}^{r}\bar{\Delta
}_{hh,cv}^{<}\right]  +\left[  \vec{\Xi}_{cc}^{<}S_{cv,o}^{a}-S_{cc,o}%
^{r}\ \vec{\delta}_{hh,cv}^{<}\right] \\
+i\left[  \vec{\Xi}_{cc}^{<}\times\vec{S}_{cv}^{a}\ \ -\vec{S}_{cc}^{r}%
\times\vec{\delta}_{hh,cv}^{<}\right]  \
\end{array}
\right] \nonumber\\
&  \frac{1}{2}\left[
\begin{array}
[c]{c}%
-\left[  \bar{\Delta}_{hh,cv}^{r}\vec{S}_{vv}^{>T}-\vec{S}_{cv}^{<}\bar
{\Sigma}_{\ vv}^{rT}\right]  -\left[  \vec{\delta}_{hh,cv}^{r}S_{vv,o}%
^{>T}-S_{cv,o}^{<}\vec{\Xi}_{\ vv}^{rT}\right] \\
+i\left[  \vec{S}_{cv}^{<}\times\vec{\Xi}_{\ vv}^{rT}-\vec{\delta}_{hh,cv}%
^{r}\times\vec{S}_{vv}^{,>T}\right]  \ \ \
\end{array}
\right] \nonumber\\
\frac{1}{2}  &  \left[
\begin{array}
[c]{c}%
-\left[  \bar{\Delta}_{hh,cv}^{<}\vec{S}_{vv}^{rT}-\vec{S}_{cv}^{r}\bar
{\Sigma}_{\ vv}^{>T}\right]  -\left[  \vec{\delta}_{hh,cv}^{<}S_{vv,o}%
^{rT}-S_{cv,o}^{r}\vec{\Xi}_{\ vv}^{>T}\right] \\
+i\left[  \vec{S}_{cv}^{r}\times\vec{\Xi}_{\ vv}^{>T}-\vec{\delta}_{hh,cv}%
^{<}\times\vec{S}_{vv}^{rT}\right]  \
\end{array}
\right]  , \label{genBloch2}%
\end{align}%
\begin{align}
&  i\hbar\left(  \frac{\partial}{\partial t_{1}}+\frac{\partial}{\partial
t_{2}}\right)  \vec{S}_{vc}^{<}\nonumber\\
&  =\frac{1}{2}\left[
\begin{array}
[c]{c}%
\left[  \bar{H}_{v}\vec{S}_{vc}^{<}-\vec{S}_{vc}^{<}\bar{H}_{c}\right]
+\left[  \mathcal{\vec{B}}_{vv}S_{vc,o}^{<}-S_{vc,o}^{<}\mathcal{\vec{B}}%
_{cc}\right] \\
+i\left[  \mathcal{\vec{B}}_{vv}\times\vec{S}_{vc}^{<}-\vec{S}_{vc}^{<}%
\times\mathcal{\vec{B}}_{cc}\right]
\end{array}
\right] \nonumber\\
&  +\frac{1}{2}\left[
\begin{array}
[c]{c}%
-\left[  \bar{\Sigma}_{\ vv}^{a,T}\vec{S}_{vc}^{<}-\vec{S}_{vv}^{>T}%
\bar{\Delta}_{ee,vc}^{a}\right]  -\left[  \vec{\Xi}_{\ vv}^{a,T}S_{vc,o}%
^{<}-S_{vv,o}^{>T}\vec{\delta}_{ee,vc}^{a}\right] \\
+i\left[  \vec{S}_{vv}^{>T}\times\vec{\delta}_{ee,vc}^{a}-\vec{\Xi}%
_{\ vv}^{a,T}\times\vec{S}_{vc}^{<}\right]  \ \
\end{array}
\right] \nonumber\\
&  +\frac{1}{2}\left[
\begin{array}
[c]{c}%
-\left[  \bar{\Sigma}_{\ vv}^{>T}\ \vec{S}_{vc}^{a}-\vec{S}_{vv}^{aT}%
\bar{\Delta}_{ee,vc}^{<}\right]  -\left[  \vec{\Xi}_{\ vv}^{>T}S_{vc,o}%
^{a}-S_{vv,o}^{aT}\vec{\delta}_{ee,vc}^{<}\right] \\
+i\left[  \vec{S}_{vv}^{aT}\times\vec{\delta}_{ee,vc}^{<}-\vec{\Xi}%
_{\ vv}^{>T}\times\vec{S}_{vc}^{a}\right]  \ \
\end{array}
\right] \nonumber\\
&  +\frac{1}{2}\left[
\begin{array}
[c]{c}%
\left[  \bar{\Delta}_{ee,vc}^{r}\vec{S}_{cc}^{<}-\vec{S}_{vc}^{<}\bar{\Sigma
}_{cc}^{a}\right]  +\left[  \vec{\delta}_{ee,vc}^{r}S_{cc,o}^{<}-S_{vc,o}%
^{<}\vec{\Xi}_{cc}^{a}\right] \\
+i\left[  \vec{\delta}_{ee,vc}^{r}\times\vec{S}_{cc}^{<}-\vec{S}_{vc}%
^{<}\times\vec{\Xi}_{cc}^{a}\right]  \ \
\end{array}
\right] \nonumber\\
&  +\frac{1}{2}\left[
\begin{array}
[c]{c}%
\left[  \bar{\Delta}_{ee,vc}^{<}\vec{S}_{cc}^{a}-\vec{S}_{vc}^{r}\bar{\Sigma
}_{cc}^{<}\right]  +\left[  \vec{\delta}_{ee,vc}^{<}\ S_{cc,o}^{a}%
-S_{vc,o}^{r}\vec{\Xi}_{cc}^{<}\right] \\
+i\left[  \vec{\delta}_{ee,vc}^{<}\ \times\vec{S}_{cc}^{a}-\vec{S}_{vc}%
^{r}\times\vec{\Xi}_{cc}^{<}\right]  \ \
\end{array}
\right]  , \label{genBloch3}%
\end{align}

\begin{align}
&  i\hbar\left(  \frac{\partial}{\partial t_{1}}+\frac{\partial}{\partial
t_{2}}\right)  \vec{S}_{vv}^{<}\nonumber\\
&  =-\frac{1}{2}\left\{  \left[  \bar{H}_{v},\vec{S}_{vv}^{<}\right]  +\left[
\mathcal{\vec{B}}_{vv},S_{vv,o}^{<}\right]  +i\left[  \mathcal{\vec{B}}%
_{vv}\times\vec{S}_{vv}^{<}-\vec{S}_{vv}^{<}\times\mathcal{\vec{B}}%
_{vv}\right]  \right\} \nonumber\\
&  -\frac{1}{2}\left[
\begin{array}
[c]{c}%
\left[  \bar{\Sigma}_{\ vv}^{r}\vec{S}_{vv}^{<}-\vec{S}_{vv}^{<}\bar{\Sigma
}_{\ vv}^{a}\right]  +\left[  \vec{\Xi}_{\ vv}^{r}S_{vv,o}^{<}-S_{vv,o}%
^{<}\vec{\Xi}_{\ vv}^{a}\right] \\
+i\left[  \vec{\Xi}_{\ vv}^{r}\times\vec{S}_{vv}^{<}\ -\vec{S}_{vv}^{<}%
\times\vec{\Xi}_{\ vv}^{a}\right]  \
\end{array}
\right] \nonumber\\
&  -\frac{1}{2}\left[
\begin{array}
[c]{c}%
\left[  \bar{\Sigma}_{\ vv}^{<}\vec{S}_{vv}^{a}-\vec{S}_{vv}^{r}\bar{\Sigma
}_{\ vv}^{<}\right]  +\left[  \vec{\Xi}_{\ vv}^{<}S_{vv,o}^{a}-S_{vv,o}%
^{r}\vec{\Xi}_{\ vv}^{<}\right] \\
+i\left[  \vec{\Xi}_{\ vv}^{<}\times\vec{S}_{vv}^{a}-\vec{S}_{vv}^{r}%
\times\vec{\Xi}_{\ vv}^{<}\right]  \
\end{array}
\right] \nonumber\\
&  -\frac{1}{2}\left[
\begin{array}
[c]{c}%
\frac{1}{i\hbar}\left[  \bar{\Delta}_{ee,vc}^{r}\vec{S}_{cv}^{<}-\vec{S}%
_{vc}^{<}\bar{\Delta}_{hh,cv}^{\ a}\right]  +\frac{1}{i\hbar}\left[
\vec{\delta}_{ee,vc}^{r}S_{cv,o}^{<}-S_{vc,o}^{<}\vec{\delta}_{hh,cv}%
^{\ a}\right] \\
+\frac{1}{\hbar}\left[  \vec{\delta}_{ee,vc}^{r}\times\vec{S}_{cv}^{<}%
\ -\vec{S}_{vc}^{<}\times\vec{\delta}_{hh,cv}^{\ a}\right]  \
\end{array}
\right] \nonumber\\
&  -\frac{1}{2}\left[
\begin{array}
[c]{c}%
\left[  \bar{\Delta}_{ee,vc}^{<}\vec{S}_{cv}^{a}-\vec{S}_{vc}^{r}\bar{\Delta
}_{hh,cv}^{<}\right]  +\left[  \vec{\delta}_{ee,vc}^{<}S_{cv,o}^{a}%
-S_{vc,o}^{r}\vec{\delta}_{hh,cv}^{<}\right] \\
+i\left[  \vec{\delta}_{ee,vc}^{<}\times\vec{S}_{cv}^{a}\ -\vec{S}_{vc}%
^{r}\times\vec{\delta}_{hh,cv}^{<}\right]  \ \
\end{array}
\right]  . \label{genBloch4}%
\end{align}

\subsection{Single Conduction Band Limit}

In the single conduction band limit for the electrons, Eq. (\ref{genBloch1})
reduces to%
\begin{align}
&  i\hbar\left(  \frac{\partial}{\partial t_{1}}+\frac{\partial}{\partial
t_{2}}\right)  \vec{S}_{cc}^{<}\nonumber\\
&  =\frac{1}{2}\left[  \bar{H}_{c},\vec{S}_{cc}^{<}\right]  +\frac{1}%
{2}\left[  \mathcal{\vec{B}}_{cc},S_{cc,o}^{<}\right]  +\frac{i}{2}\left[
\mathcal{\vec{B}}_{cc}\times\vec{S}_{cc}^{<}-\vec{S}_{cc}^{<}\times
\mathcal{\vec{B}}_{cc}\right] \nonumber\\
&  +\frac{1}{2}\left[
\begin{array}
[c]{c}%
+\left[  \bar{\Sigma}_{cc}^{r}\vec{S}_{cc}^{<}-\ \vec{S}_{cc}^{<}\bar{\Sigma
}_{cc}^{a}\right]  +\left[  \vec{\Xi}_{cc}^{r}S_{cc,o}^{<}-S_{cc,o}^{<}%
\vec{\Xi}_{cc}^{a}\right]  \ \ \\
+i\left[  \vec{\Xi}_{cc}^{r}\ \times\vec{S}_{cc}^{<}-\ \vec{S}_{cc}^{<}%
\times\vec{\Xi}_{cc}^{a}\right]
\end{array}
\right] \nonumber\\
&  +\frac{1}{2}\left[
\begin{array}
[c]{c}%
\left[  \bar{\Sigma}_{cc}^{<}\vec{S}_{cc}^{a}-\ \vec{S}_{cc}^{r}\bar{\Sigma
}_{cc}^{<}\right]  +\left[  \vec{\Xi}_{cc}^{<}S_{cc,o}^{a}-S_{cc,o}^{r}%
\vec{\Xi}_{cc}^{<}\right]  \ \\
+i\left[  \vec{\Xi}_{cc}^{<}\times\vec{S}_{cc}^{a}-\ \vec{S}_{cc}^{r}%
\times\vec{\Xi}_{cc}^{<}\right]  \
\end{array}
\right]  . \label{singleBand}%
\end{align}
Upon ignoring the effect of $S_{cc,o}^{<}$,\cite{note3} the above further
reduces to,%
\begin{align*}
&  i\hbar\left(  \frac{\partial}{\partial t_{1}}+\frac{\partial}{\partial
t_{2}}\right)  \vec{S}_{cc}^{<}\\
&  =\frac{1}{2}\left[  \bar{H}_{c}+\operatorname{Re}\bar{\Sigma}_{cc}^{r}%
,\vec{S}_{cc}^{<}\right]  +\frac{i}{2}\left\{  \operatorname{Im}\bar{\Sigma
}_{cc}^{r},\vec{S}_{cc}^{<}\right\}  -\frac{i}{2}\left\{  \bar{\Sigma}%
_{cc}^{<},\operatorname{Im}\vec{S}_{cc}^{r}\right\}  +\left[  \bar{\Sigma
}_{cc}^{<},\operatorname{Re}\vec{S}_{cc}^{r}\right] \\
&  +\frac{i}{2}\left[  \mathcal{\vec{B}}_{cc}\times\vec{S}_{cc}^{<}-\vec
{S}_{cc}^{<}\times\mathcal{\vec{B}}_{cc}\right]  +\frac{i}{2}\left[
\operatorname{Re}\vec{\Xi}_{cc}^{r}\ \times\vec{S}_{cc}^{<}-\ \vec{S}_{cc}%
^{<}\times\operatorname{Re}\vec{\Xi}_{cc}^{r}\right] \\
&  -\frac{1}{2}\left\{  \operatorname{Im}\vec{\Xi}_{cc}^{r}\ \times\vec
{S}_{cc}^{<}+\ \vec{S}_{cc}^{<}\times\operatorname{Im}\vec{\Xi}_{cc}%
^{r}\right\} \\
&  +\frac{i}{2}\left[  \vec{\Xi}_{cc}^{<}\times\operatorname{Re}\vec{S}%
_{cc}^{r}-\ \operatorname{Re}\vec{S}_{cc}^{r}\times\vec{\Xi}_{cc}^{<}\right]
\ +\frac{1}{2}\left\{  \vec{\Xi}_{cc}^{<}\times\operatorname{Im}\vec{S}%
_{cc}^{r}+\operatorname{Im}\ \vec{S}_{cc}^{r}\times\vec{\Xi}_{cc}^{<}\right\}
.
\end{align*}
Further upon neglecting the effects of $\operatorname{Re}\vec{S}_{cc}^{r}%
$,\cite{note4} as was done in Ref. \cite{arxiv}, we finally obtain%
\begin{align*}
&  i\hbar\left(  \frac{\partial}{\partial t_{1}}+\frac{\partial}{\partial
t_{2}}\right)  \vec{S}_{cc}^{<}\\
&  =\frac{1}{2}\left[  \bar{H}_{c}+\operatorname{Re}\bar{\Sigma}_{cc}^{r}%
,\vec{S}_{cc}^{<}\right]  +\frac{i}{2}\left\{  \operatorname{Im}\bar{\Sigma
}_{cc}^{r},\vec{S}_{cc}^{<}\right\}  -\frac{i}{2}\left\{  \bar{\Sigma}%
_{cc}^{<},\operatorname{Im}\vec{S}_{cc}^{r}\right\} \\
&  +\frac{i}{2}\left[  \left(  \mathcal{\vec{B}}_{cc}+\operatorname{Re}%
\vec{\Xi}_{cc}^{r}\right)  \times\vec{S}_{cc}^{<}-\vec{S}_{cc}^{<}%
\times\left(  \mathcal{\vec{B}}_{cc}+\operatorname{Re}\vec{\Xi}_{cc}%
^{r}\right)  \right] \\
&  -\frac{1}{2}\left\{  \operatorname{Im}\vec{\Xi}_{cc}^{r}\ \times\vec
{S}_{cc}^{<}+\ \vec{S}_{cc}^{<}\times\operatorname{Im}\vec{\Xi}_{cc}%
^{r}\right\}  \ +\frac{1}{2}\left\{  \vec{\Xi}_{cc}^{<}\times\operatorname{Im}%
\vec{S}_{cc}^{r}+\operatorname{Im}\ \vec{S}_{cc}^{r}\times\vec{\Xi}_{cc}%
^{<}\right\}  .
\end{align*}
Using the following relations,%

\begin{align}
-\operatorname{Im}\vec{\Xi}_{cc}^{r}  &  =\vec{\gamma},\label{rel1}\\
-\operatorname{Im}\bar{\Sigma}_{cc}^{r}  &  =\bar{\Gamma},\label{rel2}\\
-\operatorname{Im}\vec{S}_{cc}^{r}  &  =\mathcal{\vec{A},}\label{rel3}\\
\mathcal{\vec{B}}_{cc}  &  \Rightarrow-\mathcal{\vec{B}}_{cc}=-g_{s}%
\frac{e\hbar}{2m^{\ast}c}\vec{B}_{eff}\text{ \ for electrons.} \label{rel4}%
\end{align}
Then, we obtain%
\begin{align}
&  i\hbar\left(  \frac{\partial}{\partial t_{1}}+\frac{\partial}{\partial
t_{2}}\right)  \vec{S}_{cc}^{<}\nonumber\\
&  =\frac{1}{2}\left[  \bar{H}_{c}+\operatorname{Re}\bar{\Sigma}_{cc}^{r}%
,\vec{S}_{cc}^{<}\right]  -\frac{i}{2}\left\{  \bar{\Gamma},\vec{S}_{cc}%
^{<}\right\}  +\frac{i}{2}\left\{  \bar{\Sigma}_{cc}^{<},\mathcal{\vec{A}%
}\right\} \nonumber\\
&  +\frac{i}{2}\left[  \vec{S}_{cc}^{<}\times\left(  \mathcal{\vec{B}}%
_{cc}-\operatorname{Re}\vec{\Xi}_{cc}^{r}\right)  -\left(  \mathcal{\vec{B}%
}_{cc}-\operatorname{Re}\vec{\Xi}_{cc}^{r}\right)  \times\vec{S}_{cc}%
^{<}\right] \nonumber\\
&  +\frac{1}{2}\left\{  \vec{\gamma}\ \times\vec{S}_{cc}^{<}+\ \vec{S}%
_{cc}^{<}\times\vec{\gamma}\right\}  \ -\frac{1}{2}\left\{  \vec{\Xi}_{cc}%
^{<}\times\mathcal{\vec{A}}+\mathcal{\vec{A}}\times\vec{\Xi}_{cc}^{<}\right\}
, \label{singleBlim}%
\end{align}
which agrees with the single-band spin-vector magnetization transport equation
given in Ref.\cite{arxiv}, within the approximation used and using Eqs.
(\ref{rel1})-(\ref{rel4}) except for the presence of the extra term in Eq.
(\ref{singleBlim}) given by%
\[
\frac{i}{2}\left[  \operatorname{Re}\vec{\Xi}_{cc}^{r}\ \times\vec{S}_{cc}%
^{<}-\ \vec{S}_{cc}^{<}\times\operatorname{Re}\vec{\Xi}_{cc}^{r}\right]  .
\]
This terms was overlooked in casting the $\operatorname{Re}\Sigma^{r}$
single-particle Hamiltonian $\mathcal{\tilde{H}}$ in Ref.\cite{arxiv} into its
$2\times2$ spin canonical form. Equation (\ref{singleBand}) is the exact
expression for the single-band spin magnetization vector transport equation.

\section{Pseudo-Spin Correlation Functions, $S_{\alpha\beta,o}^{<}$}

In addition to the twelve multi-band magnetization transport equations of the
Pauli-Dirac spin components, we still have four more equations for the
$S_{\alpha\beta,o}^{<}$ making a total of $16$ equations. By the same token as
was done for the $2\times2$ real-spin matrices, we can separate the $2\times2$
matrix containing $S_{\alpha\beta,o}^{<}$ into a term independent of the band
indices, the total charge in the system, plus the pseudo-spin magnetization
terms. To do this, we need to revert to the electron picture as discussed
before. Then, the additional magnetization transport equations has to do with
pseudo-spin magnetization by virtue of our two-band model, i.e., the presence
of two-band discrete quantum labels $v $ and $c$.

From Eqs. (\ref{scc}), (\ref{svv}), (\ref{scv}), and (\ref{svc}), we have the
following four transport equations for the Pauli-Dirac spin scalar
$S_{\alpha\beta,o}^{<}$, where the subscripts $\alpha$ and $\beta$ have the
range on the set $\left\{  v,c\right\}  $,%

\begin{align}
&  i\hbar\left(  \frac{\partial}{\partial t_{1}}+\frac{\partial}{\partial
t_{2}}\right)  S_{cc,o}^{<}\nonumber\\
&  =\frac{1}{2}\left[  \bar{H}_{c},S_{cc,o}^{<}\right]  +\frac{1}{2}\left[
\mathcal{\vec{B}}_{cc}\cdot\vec{S}_{cc}^{<}-\vec{S}_{cc}^{<}\cdot
\mathcal{\vec{B}}_{cc}\right] \nonumber\\
&  +\frac{1}{2}\left[  \bar{\Sigma}_{cc}^{r}S_{cc,o}^{<}-\ S_{cc,o}^{<}%
\bar{\Sigma}_{cc}^{a}\right]  +\frac{1}{2}\left[  \vec{\Xi}_{cc}^{r}\cdot
\vec{S}_{cc}^{<}-\vec{S}_{cc}^{<}\cdot\vec{\Xi}_{cc}^{a}\right]
\ \ \nonumber\\
&  +\frac{1}{2}\left[  \bar{\Sigma}_{cc}^{<}S_{cc,o}^{a}-\ S_{cc,o}^{r}%
\bar{\Sigma}_{cc}^{<}\right]  +\frac{1}{2}\left[  \vec{\Xi}_{cc}^{<}\cdot
\vec{S}_{cc}^{a}-\vec{S}_{cc}^{r}\cdot\vec{\Xi}_{cc}^{<}\right]
\ \ \nonumber\\
&  +\frac{1}{2}\left[  \bar{\Delta}_{hh,cv}^{r}S_{vc,o}^{<}-\ S_{cv,o}^{<}%
\bar{\Delta}_{ee,vc}^{a}\right]  +\frac{1}{2}\left[  \vec{\delta}_{hh,cv}%
^{r}\cdot\vec{S}_{vc}^{<}-\vec{S}_{cv}^{<}\cdot\vec{\delta}_{ee,vc}^{a}\right]
\nonumber\\
&  +\frac{1}{2}\left[  \bar{\Delta}_{hh,cv}^{<}S_{vc,o}^{a}-S_{cv,o}^{r}%
\bar{\Delta}_{ee,vc}^{<}\right]  +\frac{1}{2}\left[  \vec{\delta}_{hh,cv}%
^{<}\cdot\vec{S}_{vc}^{a}-\vec{S}_{cv}^{r}\cdot\vec{\delta}_{ee,vc}%
^{<}\right]  , \label{scalarEqCC}%
\end{align}

\begin{align}
&  i\hbar\left(  \frac{\partial}{\partial t_{1}}+\frac{\partial}{\partial
t_{2}}\right)  S_{cv,o}^{<}\nonumber\\
&  =\frac{1}{2}\left[  \mathcal{\vec{B}}_{cc}\cdot\vec{S}_{cv}^{<}-\vec
{S}_{cv}^{<}\cdot\mathcal{\vec{B}}_{vv}\right]  +\frac{1}{2}\left[  \bar
{H}_{c}S_{cv,o}^{<}-S_{cv,o}^{<}\bar{H}_{v}\right] \nonumber\\
&  +\frac{1}{2}\left[  \vec{\Xi}_{cc}^{r}\cdot\ \vec{S}_{cv}^{<}-\vec{S}%
_{cc}^{<}\cdot\vec{\delta}_{hh,cv}^{\ a}\right]  +\frac{1}{2}\left[
\bar{\Sigma}_{cc}^{r}S_{cv,o}^{<}-S_{cc,o}^{<}\bar{\Delta}_{hh,cv}%
^{\ a}\right]  \ \ \ \nonumber\\
&  +\frac{1}{2}\left[  \vec{\Xi}_{cc}^{<}\cdot\vec{S}_{cv}^{a}-\vec{S}%
_{cc}^{r}\cdot\vec{\delta}_{hh,cv}^{<}\right]  \ +\frac{1}{2}\left[
\bar{\Sigma}_{cc}^{<}S_{cv,o}^{a}-S_{cc,o}^{r}\bar{\Delta}_{hh,cv}^{<}\right]
\ \ \ \nonumber\\
&  -\frac{1}{2}\left[  \bar{\Delta}_{hh,cv}^{r}\ S_{vv,o}^{>T}-S_{cv,o}%
^{<}\bar{\Sigma}_{\ vv}^{rT}\right]  -\frac{1}{2}\left[  \vec{\delta}%
_{hh,cv}^{r}\ \cdot\vec{S}_{vv}^{>T}-\vec{S}_{cv}^{<}\cdot\vec{\Xi}%
_{\ vv}^{rT}\right]  \ \nonumber\\
&  -\frac{1}{2}\left[  \bar{\Delta}_{hh,cv}^{<}S_{vv,o}^{rT}-S_{cv,o}^{r}%
\bar{\Sigma}_{\ vv}^{>T}\right]  -\frac{1}{2}\left[  \vec{\delta}%
_{hh,cv,z}^{<}\cdot\vec{S}_{z,vv}^{rT}-\vec{S}_{z,cv}^{r}\cdot\vec{\Xi
}_{\ vv,z}^{>T}\right]  ,\ \ \label{scalarEqCV}%
\end{align}%
\begin{align}
&  i\hbar\left(  \frac{\partial}{\partial t_{1}}+\frac{\partial}{\partial
t_{2}}\right)  S_{vc,o}^{<}\nonumber\\
&  =\frac{1}{2}\left[  \left[  \bar{H}_{v}S_{vc,o}^{<}-S_{vc,o}^{<}\bar{H}%
_{c}\right]  +\frac{1}{2}\left[  \mathcal{\vec{B}}_{vv}\cdot\vec{S}_{vc}%
^{<}-\vec{S}_{vc}^{<}\cdot\mathcal{\vec{B}}_{cc}\right]  \right] \nonumber\\
&  -\frac{1}{2}\left[  \bar{\Sigma}_{\ vv}^{aT}\ S_{vc,o}^{<}-S_{vv,o}%
^{>T}\bar{\Delta}_{ee,vc}^{a}\right]  -\frac{1}{2}\left[  \vec{\Xi}%
_{\ vv}^{aT}\cdot\vec{S}_{vc}^{<}-\vec{S}_{vv}^{>T}\cdot\vec{\delta}%
_{ee,vc}^{a}\right]  \ \nonumber\\
&  -\frac{1}{2}\left[  \bar{\Sigma}_{\ vv}^{>T}S_{vc,o}^{a}-S_{vv,o}^{aT}%
\bar{\Delta}_{ee,vc}^{<}\right]  -\frac{1}{2}\left[  \vec{\Xi}_{\ vv}%
^{>T}\cdot\vec{S}_{vc}^{a}-\vec{S}_{vv}^{aT}\cdot\vec{\delta}_{ee,vc}%
^{<}\right]  \ \ \ \nonumber\\
&  +\frac{1}{2}\left[  \bar{\Delta}_{ee,vc}^{r}\ S_{cc,o}^{<}-\ S_{vc,o}%
^{<}\bar{\Sigma}_{cc}^{a}\right]  +\frac{1}{2}\left[  \vec{\delta}_{ee,vc}%
^{r}\cdot\vec{S}_{cc}^{<}-\vec{S}_{vc}^{<}\cdot\vec{\Xi}_{cc}^{a}\right]
\ \ \nonumber\\
&  +\frac{1}{2}\left[  \bar{\Delta}_{ee,vc}^{<}S_{cc,o}^{a}-\ S_{vc,o}^{r}%
\bar{\Sigma}_{cc}^{<}\right]  +\frac{1}{2}\left[  \vec{\delta}_{ee,vc,z}%
^{<}\cdot\vec{S}_{z,cc}^{a}-\vec{S}_{z,vc}^{r}\cdot\vec{\Xi}_{cc,z}%
^{<}\right]  ,\ \ \label{scalarEqVC}%
\end{align}%
\begin{align}
&  i\hbar\left(  \frac{\partial}{\partial t_{1}}+\frac{\partial}{\partial
t_{2}}\right)  S_{vv,o}^{<}\nonumber\\
&  =\frac{1}{2}\left[  \left[  \bar{H}_{v}S_{vv,o}^{<}-S_{vv,o}^{<}\bar{H}%
_{v}\right]  +\frac{1}{2}\left[  \mathcal{\vec{B}}_{vv}\cdot\vec{S}_{vv}%
^{<}-\vec{S}_{vv}^{<}\cdot\mathcal{\vec{B}}_{vv}\right]  \ \right] \nonumber\\
&  +\frac{1}{2}\left[  \bar{\Sigma}_{\ vv}^{r}S_{vv,o}^{<}-S_{vv,o}^{<}%
\bar{\Sigma}_{\ vv}^{a}\right]  +\frac{1}{2}\left[  \vec{\Xi}_{\ vv}^{r}%
\cdot\vec{S}_{vv}^{<}-\vec{S}_{vv}^{<}\cdot\vec{\Xi}_{\ vv}^{a}\right]
\ \nonumber\\
&  +\frac{1}{2}\left[  \bar{\Sigma}_{\ vv}^{<}S_{vv,o}^{a}-S_{vv,o}^{r}%
\bar{\Sigma}_{\ vv}^{<}\right]  +\frac{1}{2}\left[  \vec{\Xi}_{\ vv}^{<}%
\cdot\vec{S}_{vv}^{a}-\vec{S}_{vv}^{r}\cdot\vec{\Xi}_{\ vv}^{<}\right]
\ \nonumber\\
&  +\frac{1}{2}\left[  \bar{\Delta}_{ee,vc}^{r}S_{cv,o}^{<}-S_{vc,o}^{<}%
\bar{\Delta}_{hh,cv}^{\ a}\right]  +\frac{1}{2}\left[  \vec{\delta}%
_{ee,vc}^{r}\cdot\vec{S}_{cv}^{<}-\vec{S}_{vc}^{<}\cdot\vec{\delta}%
_{hh,cv}^{\ a}\right]  \ \ \nonumber\\
&  +\frac{1}{2}\left[  \bar{\Delta}_{ee,vc}^{<}S_{cv,o}^{a}-S_{vc,o}^{r}%
\bar{\Delta}_{hh,cv}^{<}\right]  +\frac{1}{2}\left[  \vec{\delta}_{ee,vc}%
^{<}\cdot\vec{S}_{cv}^{a}-\vec{S}_{vc}^{r}\cdot\vec{\delta}_{hh,cv}%
^{<}\right]  .\ \ \label{scalarEqVV}%
\end{align}

Note that Eqs. (\ref{genBloch1})-(\ref{genBloch4}) and Eqs. (\ref{scalarEqCC}%
)-(\ref{scalarEqVV}) yield $16$ coupled transport equations. By separating
$S_{\alpha\beta,o}^{<}$ into the total charge in the system plus the
pseudo-spin, the multi-band spin magnetization quantum transport equations
nonlinearly incorporates the effects of pseudo-spin magnetization. We have
already perform the sign change of Eq. (\ref{vbHoleEq}) to obtain the electron
picture of Eq. (\ref{scalarEqVV}) above. What remains to be done for Eqs.
(\ref{scalarEqCV}) and (\ref{scalarEqVC}) in order to revert to electron
picture is to replaced all the transposed quantities, $F_{vv}^{\alpha T}$ with
their equivalent expressions given in the Table 1-3.

\subsection{\label{limcase}Spinless Two-Level Atom and Flat-Band Limits}

We observe that in Eqs. (\ref{scalarEqCC})-(\ref{scalarEqVV}) each equation
contains spin-indpendent terms and corresponding terms which involved
spin-vector dot products expressing the summation (i.e., the process of
'integrating out') of the real-spin degree of freedom. To gain some
understanding into these equations, we examine Eqs. (\ref{scalarEqCC}%
)-(\ref{scalarEqVV}) by first retaining only the spin-independent terms, i.e.,
for the moment ignoring the vector dot product portions.

We have the resulting transport equations for the scalar $S_{\alpha\beta
,o}^{<}$ involving only the spin-independent terms of Eqs. (\ref{scalarEqCC}%
)-(\ref{scalarEqVV}),%
\begin{align}
&  i\hbar\left(  \frac{\partial}{\partial t_{1}}+\frac{\partial}{\partial
t_{2}}\right)  S_{cc,o}^{<}=\frac{1}{2}\left[  \bar{H}_{c},S_{cc,o}^{<}\right]
\nonumber\\
&  \ \ +\frac{1}{2}\left[  \bar{\Delta}_{hh,cv}^{r}S_{vc,o}^{<}-\ S_{cv,o}%
^{<}\bar{\Delta}_{ee,vc}^{a}\right]  +\frac{1}{2}\left[  \bar{\Delta}%
_{hh,cv}^{<}S_{vc,o}^{a}-S_{cv,o}^{r}\bar{\Delta}_{ee,vc}^{<}\right]
\ \nonumber\\
&  +\frac{1}{2}\left[  \bar{\Sigma}_{cc}^{r}S_{cc,o}^{<}-\ S_{cc,o}^{<}%
\bar{\Sigma}_{cc}^{a}\right]  \ \ +\frac{1}{2}\left[  \bar{\Sigma}_{cc}%
^{<}S_{cc,o}^{a}-\ S_{cc,o}^{r}\bar{\Sigma}_{cc}^{<}\right]  , \label{pseudoA}%
\end{align}%
\begin{align}
i\hbar &  \left(  \frac{\partial}{\partial t_{1}}+\frac{\partial}{\partial
t_{2}}\right)  S_{cv,o}^{<}=\frac{1}{2}\left[  \bar{H}_{c}S_{cv,o}%
^{<}-S_{cv,o}^{<}\bar{H}_{v}\right] \nonumber\\
&  \ \ \ -\frac{1}{2}\left[  \bar{\Delta}_{hh,cv}^{r}\ S_{vv,o}^{>T}%
+S_{cc,o}^{<}\bar{\Delta}_{hh,cv}^{\ a}\right]  \ -\frac{1}{2}\left[
\bar{\Delta}_{hh,cv}^{<}S_{vv,o}^{rT}+S_{cc,o}^{r}\bar{\Delta}_{hh,cv}%
^{<}\right]  \ \ \nonumber\\
&  +\frac{1}{2}\left[  \bar{\Sigma}_{cc}^{r}S_{cv,o}^{<}+S_{cv,o}^{<}%
\bar{\Sigma}_{\ vv}^{rT}\right]  \ \ \ +\frac{1}{2}\left[  \bar{\Sigma}%
_{cc}^{<}S_{cv,o}^{a}+S_{cv,o}^{r}\bar{\Sigma}_{\ vv}^{>T}\right]  ,
\label{pseudo2_2}%
\end{align}%
\begin{align}
i\hbar &  \left(  \frac{\partial}{\partial t_{1}}+\frac{\partial}{\partial
t_{2}}\right)  S_{vc,o}^{<}=\frac{1}{2}\left[  \bar{H}_{v}S_{vc,o}%
^{<}-S_{vc,o}^{<}\bar{H}_{c}\right] \nonumber\\
&  +\frac{1}{2}\left[  \bar{\Delta}_{ee,vc}^{r}\ S_{cc,o}^{<}+S_{vv,o}%
^{>T}\bar{\Delta}_{ee,vc}^{a}\right]  \ \ +\frac{1}{2}\left[  \bar{\Delta
}_{ee,vc}^{<}S_{cc,o}^{a}+S_{vv,o}^{aT}\bar{\Delta}_{ee,vc}^{<}\right]
\nonumber\\
&  -\frac{1}{2}\left[  \bar{\Sigma}_{\ vv}^{aT}\ S_{vc,o}^{<}+S_{vc,o}^{<}%
\bar{\Sigma}_{cc}^{a}\right]  \ -\frac{1}{2}\left[  \bar{\Sigma}_{\ vv}%
^{>T}S_{vc,o}^{a}+S_{vc,o}^{r}\bar{\Sigma}_{cc}^{<}\right]  \ ,
\label{pseudoC}%
\end{align}%
\begin{align}
&  i\hbar\left(  \frac{\partial}{\partial t_{1}}+\frac{\partial}{\partial
t_{2}}\right)  S_{vv,o}^{<}=\frac{1}{2}\left[  \bar{H}_{v}S_{vv,o}%
^{<}-S_{vv,o}^{<}\bar{H}_{v}\right]  \ \nonumber\\
&  \ +\frac{1}{2}\left[  \bar{\Delta}_{ee,vc}^{r}S_{cv,o}^{<}-S_{vc,o}^{<}%
\bar{\Delta}_{hh,cv}^{\ a}\right]  \ \ +\frac{1}{2}\left[  \bar{\Delta
}_{ee,vc}^{<}S_{cv,o}^{a}-S_{vc,o}^{r}\bar{\Delta}_{hh,cv}^{<}\right]
\nonumber\\
&  +\frac{1}{2}\left[  \bar{\Sigma}_{\ vv}^{r}S_{vv,o}^{<}-S_{vv,o}^{<}%
\bar{\Sigma}_{\ vv}^{a}\right]  \ +\frac{1}{2}\left[  \bar{\Sigma}_{\ vv}%
^{<}S_{vv,o}^{a}-S_{vv,o}^{r}\bar{\Sigma}_{\ vv}^{<}\right]
\ .\ \label{pseudo4}%
\end{align}

To gain insights in these equations, as representing the pseudo-spin part of
the spin magnetization transport equations, let us reduce these equations to a
two-level flat energy bands or a two-level atomic system. Let us take the zero
of energy in the middle of the band or energy gap, and let $\bar{H}_{c}%
=-\bar{H}_{v}=\frac{\hbar\omega_{o}}{2}$. We will also ignore terms arising
from the particle self-energies, $\Sigma$, as well as terms involving
$\bar{\Delta}^{\lessgtr}$. We identify the interband matrix elements as
$\bar{\Delta}_{ee,vc}^{r}=\left\langle v\right\vert \mathcal{H}_{I}\left\vert
c\right\rangle $, and $\bar{\Delta}_{hh,cv}^{\ a}=\left\langle c\right\vert
\mathcal{H}_{I}\left\vert v\right\rangle $, so that $\bar{\Delta}%
_{hh,vc}^{\ a}=\bar{\Delta}_{ee,vc}^{r}$, and by virtue of the locality of
space-time dependence. The above four equations reduce to the following
expressions,
\begin{equation}
i\hbar\frac{\partial}{\partial t}S_{cc,o}^{<}=\frac{1}{2}\left[  \bar{\Delta
}_{hh,cv}^{r}S_{vc,o}^{<}-\ S_{cv,o}^{<}\bar{\Delta}_{ee,vc}^{r}\right]  ,
\label{atom1}%
\end{equation}%
\begin{equation}
i\hbar\frac{\partial}{\partial t}S_{cv,o}^{<}=\frac{1}{2}\left[  \hbar
\omega_{o}S_{cv,o}^{<}+\bar{\Delta}_{hh,cv}^{r}\ \left(  S_{vv,o}%
^{<}\ \ \ -S_{cc,o}^{<}\right)  \right]  , \label{atom2}%
\end{equation}%
\begin{equation}
i\hbar\frac{\partial}{\partial t}S_{vc,o}^{<}=\frac{1}{2}\left[  -\hbar
\omega_{o}S_{vc,o}^{<}+\bar{\Delta}_{ee,vc}^{r}\ \left(  S_{cc,o}%
^{<}\ \ -S_{vv,o}^{<}\right)  \right]  , \label{atom3}%
\end{equation}%
\begin{equation}
i\hbar\frac{\partial}{\partial t}S_{vv,o}^{<}=\frac{1}{2}\left[  \ \bar
{\Delta}_{ee,vc}^{r}S_{cv,o}^{<}-S_{vc,o}^{<}\bar{\Delta}_{hh,cv}%
^{\ r}\right]  ,\ \label{atom4}%
\end{equation}
where we made use of the relation:\cite{note5}
\begin{equation}
S_{vv,o}^{>T}=-S_{vv,o}^{<}, \label{greaterLessRel}%
\end{equation}
in our two-level atomic or flat-band limiting case.\cite{note6}

The pseudo-spin correlation functions are derived by first expressing the
$2\times2$ matrix in the band indices into a spin-canonical form as,%
\begin{align}
\left(
\begin{array}
[c]{cc}%
S_{cc,o}^{<} & S_{cv,o}^{<}\\
S_{vc,o}^{<} & S_{vv,o}^{<}%
\end{array}
\right)   &  =\frac{1}{2}\left(  S_{o,o}I+\vec{S}_{o}\cdot\vec{\sigma}\right)
\nonumber\\
&  =\frac{1}{2}\left(
\begin{array}
[c]{cc}%
S_{o,o}+S_{z,o} & S_{x,o}-iS_{y,o}\\
S_{x,o}+iS_{y,o} & S_{o,o}-S_{z,o}%
\end{array}
\right)  , \label{scalarCanon}%
\end{align}
where,
\begin{align}
S_{x,o}  &  =\left(  S_{vc,o}^{<}+S_{cv,o}^{<}\right)  ,\nonumber\\
iS_{y,o}  &  =\left(  S_{vc,o}^{<}-S_{cv,o}^{<}\right)  ,\nonumber\\
S_{z,o}  &  =\left(  S_{cc,o}^{<}-S_{vv,o}^{<}\right)  ,\nonumber\\
S_{o,o}  &  =\left(  S_{cc,o}^{<}+S_{vv,o}^{<}\right)  , \label{scalarVec}%
\end{align}
where we drop the '$^{<}$' superscript in the pseudo-spin correlation
functions, $S_{j,o}$. We note that $S_{o,o}$ represent the trace of the
original $4\times4$ spin matrix for the two bands. Thus, $S_{o,o}$ represent
the total charge of the system which may vary in space and time, with the
caveat that the background positive charge have to be subtracted from the
charge represented by the correlation density $S_{o,o}$ to obtain the net charge.

Similarly, we write the 'energy'-matrix in terms of the Pauli matrix as%
\begin{align}
\left(
\begin{array}
[c]{cc}%
\bar{H}_{c} & \bar{\Delta}_{hh,cv}^{\ r}\\
\ \bar{\Delta}_{ee,vc}^{r} & \bar{H}_{v}%
\end{array}
\right)   &  =\frac{1}{2}\left(  \breve{H}I+\mathcal{\vec{B}}\cdot\vec{\sigma
}\right)  ,\nonumber\\
&  =\frac{1}{2}\left(
\begin{array}
[c]{cc}%
\breve{H}+\mathcal{B}_{z} & \mathcal{B}_{x}-i\mathcal{B}_{y}\\
\mathcal{B}_{x}+i\mathcal{B}_{y} & \breve{H}-\mathcal{B}_{z}%
\end{array}
\right)  , \label{SenergyCanon}%
\end{align}
where we defined $\mathcal{\vec{B}}$ and $\breve{H}$ as,%
\begin{align}
\mathcal{B}_{x}  &  =\left(  \bar{\Delta}_{ee,vc}^{r}+\bar{\Delta}%
_{hh,cv}^{\ r}\right)  ,\nonumber\\
i\mathcal{B}_{y}  &  =\left(  \bar{\Delta}_{ee,vc}^{r}-\bar{\Delta}%
_{hh,cv}^{\ r}\right)  ,\nonumber\\
\mathcal{B}_{z}  &  =\left(  \bar{H}_{c}-\bar{H}_{v}\right)  =\hbar\omega
_{o},\nonumber\\
\breve{H}  &  =\left(  \bar{H}_{c}+\bar{H}_{v}\right)  =0. \label{SenergyVec}%
\end{align}
Therefore, we have the transport equations for the pseudo-spin correlation
functions,
\begin{align*}
i\hbar\frac{\partial}{\partial t}S_{x,o}  &  =\frac{1}{2}\left[  -\hbar
\omega_{o}S_{vc,o}^{<}+\bar{\Delta}_{ee,vc}^{r}\ \left(  S_{cc,o}%
^{<}\ \ -S_{vv,o}^{<}\right)  \right] \\
&  +\frac{1}{2}\left[  \hbar\omega_{o}S_{cv,o}^{<}+\bar{\Delta}_{hh,cv}%
^{r}\ \left(  S_{vv,o}^{<}\ \ \ -S_{cc,o}^{<}\right)  \right] \\
&  =\frac{i}{2}\left(  \mathcal{B}_{y}S_{o,z}-\mathcal{B}_{z}S_{o,y}\right)  ,
\end{align*}%
\begin{align*}
i\hbar\frac{\partial}{\partial t}iS_{y,o}  &  =\frac{1}{2}\left[  -\hbar
\omega_{o}S_{vc,o}^{<}+\bar{\Delta}_{ee,vc}^{r}\ \left(  S_{cc,o}%
^{<}\ \ -S_{vv,o}^{<}\right)  \right] \\
&  +\frac{1}{2}\left[  -\hbar\omega_{o}S_{cv,o}^{<}-\bar{\Delta}_{hh,cv}%
^{r}\ \left(  S_{vv,o}^{<}\ \ \ -S_{cc,o}^{<}\right)  \right] \\
&  =\frac{1}{2}\left[  \mathcal{B}_{x}S_{o,z}-\mathcal{B}_{z}S_{o,x}\right]  ,
\end{align*}%
\begin{align*}
i\hbar\frac{\partial}{\partial t}S_{z,o}  &  =\frac{1}{2}\left[  \bar{\Delta
}_{hh,cv}^{r}S_{vc,o}^{<}-\ S_{cv,o}^{<}\bar{\Delta}_{ee,vc}^{r}\right] \\
&  +\frac{1}{2}\left[  -\ \bar{\Delta}_{ee,vc}^{r}S_{cv,o}^{<}+S_{vc,o}%
^{<}\bar{\Delta}_{hh,cv}^{\ r}\right] \\
&  =\frac{i}{2}\left(  \mathcal{B}_{x}S_{o,y}-\mathcal{B}_{y}S_{o,x}\right)  .
\end{align*}
In vector equation form, the pseudo-spin magnetization equation is,
\begin{equation}
\frac{\partial}{\partial t}\vec{S}_{o}=\frac{1}{\hbar}\mathcal{\vec{B}\times
}\frac{1}{2}\vec{S}_{o}, \label{pseudoVecEq}%
\end{equation}
The above equation can also be written as%
\begin{align}
\frac{\partial}{\partial t}\left(  \frac{\hbar}{2}\vec{S}_{o}\right)   &
=\frac{mc}{2e\hbar}\mathcal{\vec{B}\times}\frac{e}{mc}\left(  \frac{\hbar}%
{2}\vec{S}_{o}\right)  ,\nonumber\\
&  =\vec{B}\times\vec{\mu}_{B} \label{pseudoVecEq2}%
\end{align}
where%
\[
\frac{e}{mc}\left(  \frac{\hbar}{2}\vec{S}_{o}\right)  =\vec{\mu}_{B}%
\]
corresponds to the magnetic moment of the pseudo-spin (pseudo-Bohr magneton),
and
\[
\frac{mc}{2e\hbar}\mathcal{\vec{B}=}\vec{B}%
\]
has the units of the magnetic field, $\vec{B}$. Thus we have realized a
pseudo-spin angular momentum $\frac{\hbar}{2}\vec{S}_{o}$, with effective
magnetic field, $\vec{B}$, determined by the interband terms in $\Delta$'s and
energy gap, Eq. (\ref{SenergyVec}). Equations (\ref{pseudoVecEq}) and
(\ref{pseudoVecEq2}) are also known as the Bloch equations. Note that $\vec
{S}_{o}$ rotates about the $z$-axis with frequency $\frac{\mathcal{B}_{z}%
}{\hbar}=\omega_{o}$\cite{mandelWolf} in the counterclockwise sense.

Indeed, the pseudo-spin angular momentum has eigenvalues $\pm\frac{1}{2}\hbar$
by virtue of the two discrete quantum-energy labels, since from Eq.
(\ref{scalarVec}) we have,
\begin{align*}
S_{z,o}  &  =1\text{ for }S_{vv,o}=0\text{ for the excited state,}\\
S_{z,o}  &  =-1\text{ for }S_{cc,o}=0\text{ for the unexcited or ground
state.}%
\end{align*}

\subsection{Spin-Independent Contribution}

We now consider the full spin-independent terms of Eqs. (\ref{pseudoA}) -
(\ref{pseudo4}), The pseudo-spin correlation functions are given by Eq.
(\ref{scalarVec}) We expressed in spin-canonical form the following matrix in
band indices,%

\begin{equation}
\left(
\begin{array}
[c]{cc}%
\bar{H}_{c}+\operatorname{Re}\bar{\Sigma}_{cc}^{r} & \operatorname{Re}%
\bar{\Delta}_{hh,cv}^{\ r}\\
\ \operatorname{Re}\bar{\Delta}_{ee,vc}^{r} & \bar{H}_{v}+\operatorname{Re}%
\bar{\Sigma}_{vv}^{r}%
\end{array}
\right)  =\frac{1}{2}\left(
\begin{array}
[c]{cc}%
\bar{H}+\mathcal{\beta}_{z} & \mathcal{\beta}_{x}-i\mathcal{\beta}_{y}\\
\mathcal{\beta}_{x}+i\mathcal{\beta}_{y} & \bar{H}-\mathcal{\beta}_{z}%
\end{array}
\right)  , \label{greekb}%
\end{equation}%
\begin{equation}
\left(
\begin{array}
[c]{cc}%
\operatorname{Im}\bar{\Sigma}_{cc}^{r} & \operatorname{Im}\bar{\Delta}%
_{hh,cv}^{\ r}\\
\ \operatorname{Im}\bar{\Delta}_{ee,vc}^{r} & \operatorname{Im}\bar{\Sigma
}_{vv}^{r}%
\end{array}
\right)  =\frac{1}{2}\left(
\begin{array}
[c]{cc}%
\bar{\zeta}+\zeta_{z} & \zeta_{x}-i\zeta_{y}\\
\zeta_{x}+i\zeta_{y} & \bar{\zeta}-\zeta_{z}%
\end{array}
\right)  , \label{impsir}%
\end{equation}%
\begin{equation}
\left(
\begin{array}
[c]{cc}%
\bar{\Sigma}_{cc}^{<} & \bar{\Delta}_{hh,cv}^{\ <}\\
\ \bar{\Delta}_{ee,vc}^{<} & \bar{\Sigma}_{vv}^{<}%
\end{array}
\right)  =\frac{1}{2}\left(
\begin{array}
[c]{cc}%
\bar{\zeta}^{<}+\zeta_{z}^{<} & \zeta_{x}^{<}-i\zeta_{y}^{<}\\
\zeta_{x}^{<}+i\zeta_{y}^{<} & \bar{\zeta}^{<}-\zeta_{z}^{<}%
\end{array}
\right)  . \label{psiless}%
\end{equation}
Note the use of caligraphic $\mathcal{B}$ in Eq. (\ref{SenergyCanon}) in
contrast to the use of the Greek $\beta$ in Eq. (\ref{greekb}). The
calculation is tedious but the results can be expressed in vector form as%

\begin{align}
&  i\hbar\left(  \frac{\partial}{\partial t_{1}}+\frac{\partial}{\partial
t_{2}}\right)  \vec{S}_{o}\nonumber\\
&  =\frac{1}{4}\left[  \bar{H},\vec{S}_{o}\right]  +\frac{1}{4}\left[
\mathcal{\vec{\beta}},S_{o,o}\right]  +\frac{i}{4}\left[  \mathcal{\vec{\beta
}}\times\vec{S}_{o}-\vec{S}_{o}\times\mathcal{\vec{\beta}}\right]  \text{
\ }\nonumber\\
&  +\frac{i}{4}\left[  \left\{  \bar{\zeta},\vec{S}_{o}\right\}  +\left\{
\vec{\zeta},S_{o,o}\right\}  \right]  -\frac{1}{4}\left\{  \vec{\zeta}%
\times\vec{S}_{o}+\vec{S}_{o}\times\vec{\zeta}\right\} \nonumber\\
&  +\frac{1}{4}\left[  \vec{\zeta}^{<},\operatorname{Re}S_{o,o}^{r}\right]
+\frac{1}{4}\left[  \bar{\zeta}^{<},\operatorname{Re}\vec{S}_{o}^{r}\right]
\nonumber\\
&  +\frac{i}{4}\left[  \vec{\zeta}^{<}\times\operatorname{Re}\vec{S}_{o}%
^{r}-\operatorname{Re}\vec{S}_{o}^{r}\times\vec{\zeta}^{<}\right] \nonumber\\
&  -\frac{i}{4}\left[  \left\{  \bar{\zeta}^{<},\operatorname{Im}\vec{S}%
_{o}^{r}\right\}  +\left\{  \vec{\zeta}^{<},\operatorname{Im}S_{o,o}%
^{r}\right\}  \right] \nonumber\\
&  +\frac{1}{4}\left\{  \vec{\zeta}^{<}\times\operatorname{Im}\vec{S}_{o}%
^{r}+\operatorname{Im}\vec{S}_{o}^{r}\times\vec{\zeta}^{<}\right\}  ,
\label{scalarContrib}%
\end{align}
where we use the following defined relation,%
\begin{align*}
iS_{y,o}^{r}  &  =S_{vc,o}^{r}-S_{cv,o}^{r}\\
&  =i\left(  \operatorname{Re}S_{y,o}^{r}+i\operatorname{Im}S_{y,o}%
^{r}\right)  .
\end{align*}

\subsubsection{The Equation for the Scalar $S_{o,o}$}

We have the equation for $S_{o,o}$ representing the total charge is given by,
\begin{align}
&  i\hbar\left(  \frac{\partial}{\partial t_{1}}+\frac{\partial}{\partial
t_{2}}\right)  S_{o,o}=\frac{1}{4}\left[  \bar{H},S_{o,o}\right]  +\frac{1}%
{4}\left[  \bar{\zeta}^{<},\operatorname{Re}S_{o,o}^{r}\right] \nonumber\\
&  +\frac{i}{4}\left[  \left\{  \bar{\zeta},S_{o,o}\right\}  \right]
-\frac{i}{4}\left\{  \bar{\zeta}^{<},\operatorname{Im}S_{o,o}^{r}\right\}
\nonumber\\
&  +\frac{1}{4}\left[  \mathcal{\vec{\beta}}^{\cdot psp},\vec{S}_{o}\right]
+\frac{i}{4}\left\{  \vec{\zeta}^{\cdot psp},\vec{S}_{o}\right\} \nonumber\\
&  +\frac{1}{4}\left[  \vec{\zeta}^{<\cdot psp},\operatorname{Re}\vec{S}%
_{o}^{r}\right]  -\frac{i}{4}\left\{  \vec{\zeta}^{<\cdot psp}%
,\operatorname{Im}\vec{S}_{o}^{r}\right\}  , \label{SooEq1}%
\end{align}
where the superscript $^{\cdot psp}$ means taking the dot product with respect
to the pseudo-spin vector components. One observes that the first four terms
in the the right hand side of Eq. (\ref{SooEq1}) exactly correspond to the
terms in the equation of the particle or charge correlation function in system
without the spin degree of freedom.

\subsection{Spin-Dependent Contributions}

The spin-dependent terms ignored in Sec. \ref{limcase} represent the direct
coupling of pseudo-spin to real (Pauli-Dirac) spins. We extract from Eqs.
(\ref{scalarEqCC}) - (\ref{scalarEqVV}) the terms involving the dot products
of real-spin vectors for their contributions to the pseudo-spin magnetization
transport equations. Since the real-spin degree of freedom is essentially
integrated out, what is left are the discrete band indices. First, we
expressed all $2\times2$ quantities in the band indices into their pseudo-spin
canonical matrix form. We make use of the expressions already given by Eqs.
(\ref{scalarCanon}), (\ref{scalarVec}), (\ref{SenergyCanon}), and
(\ref{SenergyVec}) for the pseudo-spin canonical form of the scalar
quantities. The vector quantities must also be expressed into their
pseudo-spin canonical matrix form in the band indices producing dyadic tensors.

We have for the real-spin vector dot product portion of the pseudo-spin
equations, where the dot product can be written for convenience, using the
Einstein summation convention for the Pauli-Dirac spin vector dot product,%

\begin{align}
i\hbar &  \left(  \frac{\partial}{\partial t_{1}}+\frac{\partial}{\partial
t_{2}}\right)  S_{cc,o}^{<}\nonumber\\
&  =\frac{1}{2}\left[  \mathcal{B}_{cc,i}\ S_{cc,i}^{<}-S_{cc,i}%
^{<}\ \mathcal{B}_{cc,i}\right] \nonumber\\
&  +\frac{1}{2}\left[  \Xi_{cc,i}^{r}\ S_{cc,i}^{<}-S_{cc,i}^{<}\ \Xi
_{cc,i}^{a}\right]  \ \ +\frac{1}{2}\left[  \Xi_{cc,i}^{<}\ S_{cc,i}%
^{a}-S_{cc,i}^{r}\ \Xi_{cc,i}^{<}\right]  \ \nonumber\\
&  +\frac{1}{2}\left[  \delta_{hh,cv,i}^{r}\ S_{vc,i}^{<}-S_{cv,i}^{<}%
\ \delta_{ee,vc,i}^{a}\right]  +\frac{1}{2}\left[  \delta_{hh,cv,i}%
^{<}\ S_{vc,i}^{a}-S_{cv,i}^{r}\ \delta_{ee,vc,i}^{<}\right]  \ \ ,
\label{vproduct1}%
\end{align}

\begin{align}
&  i\hbar\left(  \frac{\partial}{\partial t_{1}}+\frac{\partial}{\partial
t_{2}}\right)  S_{cv,o}^{<}\nonumber\\
&  =\frac{1}{2}\left[  \left(  \mathcal{B}_{cc,i}+\Xi_{cc,i}^{r}\right)
\ \ S_{cv,i}^{<}-S_{cv,i}^{<}\ \ \left(  \mathcal{B}_{vv,i}-\Xi_{\ vv,i}%
^{rT}\right)  \right] \nonumber\\
&  -\frac{1}{2}\left[  S_{cc,i}^{<}\ \ \delta_{hh,cv,i}^{\ a}+\delta
_{hh,cv,i}^{r}\ \ \ S_{vv,i}^{>T}\right]  \ \ \ +\frac{1}{2}\left[  \Xi
_{cc,i}^{<}\ \ S_{cv,i}^{a}+S_{cv,i}^{r}\ \ \Xi_{\ vv,i}^{>T}\right]
\ \ \ \ \ \nonumber\\
&  -\frac{1}{2}\left[  \delta_{hh,cv,i}^{<}\ \ S_{vv,i}^{rT}+S_{cc,i}%
^{r}\ \ \delta_{hh,cv,i}^{<}\right]  , \label{vproduct2}%
\end{align}%
\begin{align}
i\hbar &  \left(  \frac{\partial}{\partial t_{1}}+\frac{\partial}{\partial
t_{2}}\right)  S_{vc,o}^{<}\nonumber\\
&  =\frac{1}{2}\left[  \left(  \mathcal{B}_{vv,i}-\Xi_{\ vv,i}^{aT}\right)
\ \ S_{vc,i}^{<}-S_{vc,i}^{<}\ \ \left(  \mathcal{B}_{cc,i}+\Xi_{cc,i}%
^{a}\right)  \right] \nonumber\\
&  +\frac{1}{2}\left[  \delta_{ee,vc,i}^{r}\ \ S_{cc,i}^{<}+S_{vv,i}%
^{>T}\ \ \delta_{ee,vc,i}^{a}\right]  \ -\frac{1}{2}\left[  \Xi_{\ vv,i}%
^{>T}\ \ S_{vc,i}^{a}+S_{z,vc,i}^{r}\ \ \Xi_{cc,i}^{<}\right]
\ \ \ \nonumber\\
&  +\frac{1}{2}\left[  \delta_{ee,vc,i}^{<}\ \ S_{cc,i}^{a}+S_{vv,i}%
^{aT}\ \ \delta_{ee,vc,i}^{<}\right]  ,\ \ \ \label{vproduct3}%
\end{align}%
\begin{align}
&  i\hbar\left(  \frac{\partial}{\partial t_{1}}+\frac{\partial}{\partial
t_{2}}\right)  S_{vv,o}^{<}\nonumber\\
&  =\frac{1}{2}\left[  \mathcal{B}_{vv,i}\ \ S_{vv,i}^{<}-S_{vv,i}%
^{<}\ \ \mathcal{B}_{vv,i}\right]  \ \nonumber\\
&  +\frac{1}{2}\left[  \Xi_{\ vv,i}^{r}\ \ S_{vv,i}^{<}-S_{vv,i}^{<}%
\ \ \Xi_{\ vv,i}^{a}\right]  \ +\frac{1}{2}\left[  \Xi_{\ vv,i}^{<}%
\ \ S_{vv,i}^{a}-S_{vv,i}^{r}\ \ \Xi_{\ vv,i}^{<}\right]  \ \nonumber\\
&  +\frac{1}{2}\left[  \delta_{ee,vc,i}^{r}\ \ S_{cv,i}^{<}-S_{vc,i}%
^{<}\ \ \delta_{hh,cv,i}^{\ a}\right]  \ \ +\frac{1}{2}\left[  \delta
_{ee,vc,i}^{<}\ \ S_{cv,i}^{a}-S_{vc,i}^{r}\ \ \delta_{hh,cv,i}^{<}\right]
.\ \ \label{vproduct4}%
\end{align}

We have the following pseudo-spin canonical matrix form for the real-spin
vectors labeled by the discrete band indices,%

\begin{equation}
\left(
\begin{array}
[c]{cc}%
\mathcal{B}_{cc,i}+\operatorname{Re}\Sigma_{cc,i}^{r} & \operatorname{Re}%
\delta_{cv,i}^{r}\\
\ \operatorname{Re}\delta_{vc,i}^{r} & \mathcal{B}_{vv,i}+\operatorname{Re}%
\Sigma_{vv,i}^{r}%
\end{array}
\right)  =\frac{1}{2}\left(
\begin{array}
[c]{cc}%
\mathcal{\bar{B}}_{i}+\mathcal{B}_{z,i} & \mathcal{B}_{x,i}-i\mathcal{B}%
_{y,i}\\
\mathcal{B}_{x,i}+i\mathcal{B}_{y,i} & \mathcal{\bar{B}}_{i}-\mathcal{B}_{z,i}%
\end{array}
\right)  , \label{pscanonical1}%
\end{equation}%
\begin{equation}
\left(
\begin{array}
[c]{cc}%
\operatorname{Im}\Sigma_{cc,i}^{r} & \operatorname{Im}\delta_{hh,cv,i}^{\ r}\\
\ \operatorname{Im}\delta_{ee,vc,i}^{r} & \operatorname{Im}\Sigma_{vv,i}^{r}%
\end{array}
\right)  =\frac{1}{2}\left(
\begin{array}
[c]{cc}%
\bar{\zeta}_{i}+\zeta_{z,i} & \zeta_{x,i}-i\zeta_{y,i}\\
\zeta_{x,i}+i\zeta_{y,i} & \bar{\zeta}_{i}-\zeta_{z,i}%
\end{array}
\right)  , \label{pscanonical2}%
\end{equation}%
\begin{equation}
\left(
\begin{array}
[c]{cc}%
\Sigma_{cc,i}^{<} & \delta_{hh,cv,i}^{\ <}\\
\ \delta_{ee,vc,i}^{<} & \Sigma_{vv,i}^{<}%
\end{array}
\right)  =\frac{1}{2}\left(
\begin{array}
[c]{cc}%
\bar{\zeta}_{i}^{<}+\zeta_{z,i}^{<} & \zeta_{x,i}^{<}-i\zeta_{y,i}^{<}\\
\zeta_{x,i}^{<}+i\zeta_{y,i}^{<} & \bar{\zeta}_{i}^{<}-\zeta_{z,i}^{<}%
\end{array}
\right)  , \label{pscanonical3}%
\end{equation}

\begin{align}
S_{\alpha\beta,i}  &  =\left(
\begin{array}
[c]{cc}%
S_{cc,i}^{<} & S_{cv,i}^{<}\\
S_{vc,i}^{<} & S_{vv,i}^{<}%
\end{array}
\right) \nonumber\\
&  =\frac{1}{2}\left(
\begin{array}
[c]{cc}%
\bar{S}_{i}+S_{z,i} & S_{x,i}-iS_{y,i}\\
S_{x,i}+iS_{y,i} & \bar{S}_{i}-S_{z,i}%
\end{array}
\right)  . \label{pscanonical4}%
\end{align}
We will also make use of the definition%
\begin{align}
iS_{y,i}^{r}  &  =i\left(  \operatorname{Re}S_{y,i}^{r}+i\operatorname{Im}%
S_{y,i}^{r}\right) \nonumber\\
&  =S_{vc,i}^{r}-S_{cv,i}^{r} \label{defrealimag}%
\end{align}

The results for the pseudo-spin magnetization equation from the contributions
of real-spin vector dot product terms in Eqs. (\ref{vproduct1}) -
(\ref{vproduct4}) can also be written as a pseudo-spin vector equation similar
to the scalar-contribution given by Eq. (\ref{scalarContrib}) as,%
\begin{align}
&  i\hbar\left(  \frac{\partial}{\partial t_{1}}+\frac{\partial}{\partial
t_{2}}\right)  \vec{S}_{o}\nonumber\\
&  =\frac{1}{4}\left[  \mathcal{\bar{B}}_{i},\vec{S}_{i}\right]  +\frac{1}%
{4}\left[  \mathcal{\vec{B}}_{i},\bar{S}_{i}\right]  +\frac{i}{4}\left[
\mathcal{\vec{B}}_{i}\times\vec{S}_{i}-\vec{S}_{i}\times\mathcal{\vec{B}}%
_{i}\right]  \text{ \ }\nonumber\\
&  +\frac{i}{4}\left[  \left\{  \bar{\zeta}_{i},\vec{S}_{i}\right\}  +\left\{
\vec{\zeta}_{i},\bar{S}_{i}\right\}  \right]  -\frac{1}{4}\left\{  \vec{\zeta
}_{i}\times\vec{S}_{i}+\vec{S}_{i}\times\vec{\zeta}_{i}\right\} \nonumber\\
&  +\frac{1}{4}\left[  \vec{\zeta}_{i}^{<},\operatorname{Re}\bar{S}_{i}%
^{r}\right]  +\frac{1}{4}\left[  \bar{\zeta}_{i}^{<},\operatorname{Re}\vec
{S}_{i}^{r}\right] \nonumber\\
&  +\frac{i}{4}\left[  \vec{\zeta}_{i}^{<}\times\operatorname{Re}\vec{S}%
_{i}^{r}-\operatorname{Re}\vec{S}_{i}^{r}\times\vec{\zeta}_{i}^{<}\right]
\nonumber\\
&  -\frac{i}{4}\left[  \left\{  \bar{\zeta}_{i}^{<},\operatorname{Im}\vec
{S}_{i}^{r}\right\}  +\left\{  \vec{\zeta}_{i}^{<},\operatorname{Im}\bar
{S}_{i}^{r}\right\}  \right] \nonumber\\
&  +\frac{1}{4}\left\{  \vec{\zeta}_{i}^{<}\times\operatorname{Im}\vec{S}%
_{i}^{r}+\operatorname{Im}\vec{S}_{i}^{r}\times\vec{\zeta}_{i}^{<}\right\}  .
\label{v_prodContrib}%
\end{align}

\subsubsection{The Equation for the Scalar $S_{o,o}$}

Similarly the contribution to the equation for $S_{o,o}$ goes as,
\begin{align}
&  i\hbar\left(  \frac{\partial}{\partial t_{1}}+\frac{\partial}{\partial
t_{2}}\right)  S_{o,o}\nonumber\\
&  =\frac{1}{4}\left[  \mathcal{\bar{B}}_{i},\bar{S}_{i}\right]  +\frac{1}%
{4}\left[  \mathcal{\vec{B}}_{i}^{\cdot psp},\vec{S}_{i}\right] \nonumber\\
&  +\frac{1}{4}i\left[  \left\{  \bar{\zeta}_{i},\bar{S}_{i}\right\}  \right]
+\frac{1}{4}i\left\{  \vec{\zeta}_{i}^{\cdot psp},\vec{S}_{i}\right\}
\nonumber\\
&  +\frac{1}{4}\left[  \bar{\zeta}_{i}^{<},\operatorname{Re}\bar{S}_{i}%
^{r}\right]  -\frac{1}{4}i\left\{  \bar{\zeta}_{i}^{<},\operatorname{Im}%
\bar{S}_{i}^{r}\right\} \nonumber\\
&  +\frac{1}{4}\left[  \vec{\zeta}_{i}^{<\cdot psp},\operatorname{Re}\vec
{S}_{i}^{r}\right]  -\frac{1}{4}i\left\{  \vec{\zeta}_{i}^{<\cdot
psp},\operatorname{Im}\vec{S}_{i}^{r}\right\}  , \label{v_prodContrib2}%
\end{align}
where the superscripts $^{\cdot psp}$ indicates the process of taking the
pseudo-spin vector dot products or the process of 'integrating out' the
pseudo-spin degree of freedoms, aside from the dot product of the Pauli-Dirac
spin vector indicated by the Einstein summation convention for the repeated
'$i$' index.

\section{Nonequilibrium Pseudo-Spin Equations}

The nonequilibrium pseudo-spin vector transport equations result by adding
Eqs. (\ref{scalarContrib}) and (\ref{v_prodContrib})%
\begin{align}
&  i\hbar\left(  \frac{\partial}{\partial t_{1}}+\frac{\partial}{\partial
t_{2}}\right)  \vec{S}_{o}\nonumber\\
&  =\frac{1}{4}\left[  \bar{H},\vec{S}_{o}\right]  +\frac{1}{4}\left[
\bar{\zeta}^{<},\operatorname{Re}\vec{S}_{o}^{r}\right]  +\frac{i}{4}\left\{
\bar{\zeta},\vec{S}_{o}\right\}  -\frac{i}{4}\left\{  \bar{\zeta}%
^{<},\operatorname{Im}\vec{S}_{o}^{r}\right\}  \text{ \ }\nonumber\\
&  +\frac{1}{2}\left[  \mathcal{\vec{\beta}},S_{o,o}\right]  +\frac{1}%
{4}\left[  \vec{\zeta}^{<},\operatorname{Re}S_{o,o}^{r}\right]  +\frac{i}%
{4}\left\{  \vec{\zeta},S_{o,o}\right\}  -\frac{i}{4}\left\{  \vec{\zeta}%
^{<},\operatorname{Im}S_{o,o}^{r}\right\} \nonumber\\
&  +\frac{i}{4}\left[  \mathcal{\vec{\beta}}\times\vec{S}_{o}-\vec{S}%
_{o}\times\mathcal{\vec{\beta}}\right]  +\frac{i}{4}\left[  \vec{\zeta}%
^{<}\times\operatorname{Re}\vec{S}_{o}^{r}-\operatorname{Re}\vec{S}_{o}%
^{r}\times\vec{\zeta}^{<}\right] \nonumber\\
&  -\frac{1}{4}\left\{  \vec{\zeta}\times\vec{S}_{o}+\vec{S}_{o}\times
\vec{\zeta}\right\}  +\frac{1}{4}\left\{  \vec{\zeta}^{<}\times
\operatorname{Im}\vec{S}_{o}^{r}+\operatorname{Im}\vec{S}_{o}^{r}\times
\vec{\zeta}^{<}\right\} \nonumber\\
&  +\frac{1}{4}\left[  \mathcal{\bar{B}}_{i},\vec{S}_{i}\right]  +\frac{1}%
{4}\left[  \bar{\zeta}_{i}^{<},\operatorname{Re}\vec{S}_{i}^{r}\right]
+\frac{i}{4}\left\{  \bar{\zeta}_{i},\vec{S}_{i}\right\}  -\frac{i}{4}\left\{
\bar{\zeta}_{i}^{<},\operatorname{Im}\vec{S}_{i}^{r}\right\} \nonumber\\
&  +\frac{1}{4}\left[  \mathcal{\vec{B}}_{i},\bar{S}_{i}\right]  +\frac{1}%
{4}\left[  \vec{\zeta}_{i}^{<},\operatorname{Re}\bar{S}_{i}^{r}\right]
+\frac{i}{4}\left\{  \vec{\zeta}_{i},\bar{S}_{i}\right\}  -\frac{i}{4}\left\{
\vec{\zeta}_{i}^{<},\operatorname{Im}\bar{S}_{i}^{r}\right\} \nonumber\\
&  +\frac{i}{4}\left[  \mathcal{\vec{B}}_{i}\times\vec{S}_{i}-\vec{S}%
_{i}\times\mathcal{\vec{B}}_{i}\right]  +\frac{i}{4}\left[  \vec{\zeta}%
_{i}^{<}\times\operatorname{Re}\vec{S}_{i}^{r}-\operatorname{Re}\vec{S}%
_{i}^{r}\times\vec{\zeta}_{i}^{<}\right] \nonumber\\
&  -\frac{1}{4}\left\{  \vec{\zeta}_{i}\times\vec{S}_{i}+\vec{S}_{i}\times
\vec{\zeta}_{i}\right\}  +\frac{1}{4}\left\{  \vec{\zeta}_{i}^{<}%
\times\operatorname{Im}\vec{S}_{i}^{r}+\operatorname{Im}\vec{S}_{i}^{r}%
\times\vec{\zeta}_{i}^{<}\right\}  .\nonumber\\
&  \label{pseudoSpinEq}%
\end{align}

The corresponding absolute scalar\cite{note7} equation is obtained by adding
Eqs. (\ref{SooEq1}) and (\ref{v_prodContrib2}),%
\begin{align}
&  i\hbar\left(  \frac{\partial}{\partial t_{1}}+\frac{\partial}{\partial
t_{2}}\right)  S_{o,o}\nonumber\\
&  =\ \frac{1}{4}\ \left[  \bar{H},S_{o,o}\right]  +\ \frac{1}{4}\ \left[
\bar{\zeta}^{<},\operatorname{Re}S_{o,o}^{r}\right]  +\frac{i}{4}\left[
\left\{  \bar{\zeta},S_{o,o}\right\}  \right]  -\frac{i}{4}\left\{  \bar
{\zeta}^{<},\operatorname{Im}S_{o,o}^{r}\right\} \nonumber\\
&  +\ \frac{1}{4}\ \left[  \mathcal{\vec{\beta}}^{\cdot psp},\vec{S}%
_{o}\right]  +\ \frac{1}{4}\ \left[  \vec{\zeta}^{<\cdot psp}%
,\operatorname{Re}\vec{S}_{o}^{r}\right]  +\frac{i}{4}\left\{  \vec{\zeta
}^{\cdot psp},\vec{S}_{o}\right\}  -\frac{i}{4}\left\{  \vec{\zeta}^{<\cdot
psp},\operatorname{Im}\vec{S}_{o}^{r}\right\} \nonumber\\
&  +\ \frac{1}{4}\ \left[  \mathcal{\bar{B}}_{i},\bar{S}_{i}\right]
+\ \frac{1}{4}\ \left[  \bar{\zeta}_{i}^{<},\operatorname{Re}\bar{S}_{i}%
^{r}\right]  +\frac{i}{4}\left\{  \bar{\zeta}_{i},\bar{S}_{i}\right\}
-\frac{i}{4}\left\{  \bar{\zeta}_{i}^{<},\operatorname{Im}\bar{S}_{i}%
^{r}\right\} \nonumber\\
&  +\ \frac{1}{4}\ \left[  \mathcal{\vec{B}}_{i}^{\cdot psp},\vec{S}%
_{i}\right]  +\ \frac{1}{4}\ \left[  \vec{\zeta}_{i}^{<\cdot psp}%
,\operatorname{Re}\vec{S}_{i}^{r}\right]  +\frac{i}{4}\left\{  \vec{\zeta}%
_{i}^{\cdot psp},\vec{S}_{i}\right\}  -\frac{i}{4}\left\{  \vec{\zeta}%
_{i}^{<\cdot psp},\operatorname{Im}\vec{S}_{i}^{r}\right\}  .\nonumber\\
&  \label{abs_scalarEq}%
\end{align}

Equations (\ref{genBloch1}) - (\ref{genBloch4}) and (\ref{pseudoSpinEq}) -
(\ref{abs_scalarEq}) constitute the nonequilibrium multi-band magnetization
quantum transport equations. These sixteen equations for the components of the
spin vectors are the main results of this paper.

\section{\label{sumSMQTE}Multi-Band Spin Quantum Transport Equations}

We summarize the main results off this paper by recasting the pertinent
equations into more meaningful expressions. We start by rewriting the RHS of
Pauli-Dirac SMQTEs into three groups as,%

\begin{align}
&  i\hbar\left(  \frac{\partial}{\partial t_{1}}+\frac{\partial}{\partial
t_{2}}\right)  \vec{S}_{cc}^{<}\nonumber\\
&  =+\frac{1}{2}\left[
\begin{array}
[c]{c}%
\left[  \bar{H}_{c},\vec{S}_{cc}^{<}\right] \\
+\left[  \left(  \bar{\Sigma}_{cc}^{r}\right)  \vec{S}_{cc}^{<}-\ \vec{S}%
_{cc}^{<}\left(  \bar{\Sigma}_{cc}^{a}\right)  \right]  +\left[  \bar{\Sigma
}_{cc}^{<}\left(  \vec{S}_{cc}^{a}\right)  -\left(  \ \vec{S}_{cc}^{r}\right)
\bar{\Sigma}_{cc}^{<}\right] \\
+\left[  \left(  \bar{\Delta}_{hh,cv}^{r}\right)  \vec{S}_{vc}^{<}-\vec
{S}_{cv}^{<}\left(  \bar{\Delta}_{ee,vc}^{a}\right)  \right]  +\left[
\bar{\Delta}_{hh,cv}^{<}\ \vec{S}_{vc}^{a}-\vec{S}_{cv}^{r}\bar{\Delta
}_{ee,vc}^{<}\right]  \ \
\end{array}
\right] \nonumber\\
&  +\frac{1}{2}\left[
\begin{array}
[c]{c}%
+i\left[  \mathcal{\vec{B}}_{cc}\times\vec{S}_{cc}^{<}-\vec{S}_{cc}^{<}%
\times\mathcal{\vec{B}}_{cc}\right] \\
+i\left[  \left(  \vec{\Xi}_{cc}^{r}\right)  \ \times\vec{S}_{cc}^{<}%
-\ \vec{S}_{cc}^{<}\times\left(  \vec{\Xi}_{cc}^{a}\right)  \right]  +i\left[
\vec{\Xi}_{cc}^{<}\times\left(  \vec{S}_{cc}^{a}\right)  -\left(  \ \vec
{S}_{cc}^{r}\right)  \times\vec{\Xi}_{cc}^{<}\right] \\
+i\left[  \vec{\delta}_{hh,cv}^{r}\times\vec{S}_{vc}^{<}-\vec{S}_{cv}%
^{<}\times\vec{\delta}_{ee,vc}^{a}\right]  +i\left[  \vec{\delta}_{hh,cv}%
^{<}\ \times\vec{S}_{vc}^{a}-\vec{S}_{cv}^{r}\times\vec{\delta}_{ee,vc}%
^{<}\right]
\end{array}
\right] \nonumber\\
&  +\frac{1}{2}\left[
\begin{array}
[c]{c}%
+\left[  \mathcal{\vec{B}}_{cc},S_{cc,o}^{<}\right] \\
+\left[  \left(  \vec{\Xi}_{cc}^{r}\right)  \ S_{cc,o}^{<}-S_{cc,o}^{<}\left(
\vec{\Xi}_{cc}^{a}\right)  \right]  +\left[  \vec{\Xi}_{cc}^{<}\left(
S_{cc,o}^{a}\right)  -\left(  S_{cc,o}^{r}\right)  \vec{\Xi}_{cc}^{<}\right]
\\
+\left[  \left(  \vec{\delta}_{hh,cv}^{r}\right)  S_{vc,o}^{<}-S_{cv,o}%
^{<}\left(  \vec{\delta}_{ee,vc}^{a}\right)  \right]  +\left[  \vec{\delta
}_{hh,cv}^{<}\ S_{vc,o}^{a}-S_{cv,o}^{r}\vec{\delta}_{ee,vc}^{<}\right]
\end{array}
\right]  , \label{genBloch1-2}%
\end{align}

\begin{align}
&  i\hbar\left(  \frac{\partial}{\partial t_{1}}+\frac{\partial}{\partial
t_{2}}\right)  \vec{S}_{cv}^{<}\nonumber\\
&  =\frac{1}{2}\left[
\begin{array}
[c]{c}%
\left[  \bar{H}_{c}\vec{S}_{cv}^{<}-\vec{S}_{cv}^{<}\bar{H}_{v}\right] \\
+\left[  \bar{\Sigma}_{cc}^{r}\vec{S}_{cv}^{<}-\vec{S}_{cc}^{<}\bar{\Delta
}_{hh,cv}^{\ a}\right]  +\left[  \bar{\Sigma}_{cc}^{<}\vec{S}_{cv}^{a}-\vec
{S}_{cc}^{r}\bar{\Delta}_{hh,cv}^{<}\right] \\
-\left[  \bar{\Delta}_{hh,cv}^{r}\vec{S}_{vv}^{>T}-\vec{S}_{cv}^{<}\bar
{\Sigma}_{\ vv}^{rT}\right]  -\left[  \bar{\Delta}_{hh,cv}^{<}\vec{S}%
_{vv}^{rT}-\vec{S}_{cv}^{r}\bar{\Sigma}_{\ vv}^{>T}\right]
\end{array}
\right] \nonumber\\
&  +\frac{1}{2}\left[
\begin{array}
[c]{c}%
+i\left[  \mathcal{\vec{B}}_{cc}\times\vec{S}_{cv}^{<}-\vec{S}_{cv}^{<}%
\times\mathcal{\vec{B}}_{vv}\right] \\
+i\left[  \vec{\Xi}_{cc}^{r}\times\vec{S}_{cv}^{<}\ -\vec{S}_{cc}^{<}%
\times\vec{\delta}_{hh,cv}^{\ a}\right]  +i\left[  \vec{\Xi}_{cc}^{<}%
\times\vec{S}_{cv}^{a}\ \ -\vec{S}_{cc}^{r}\times\vec{\delta}_{hh,cv}%
^{<}\right] \\
-i\left[  \vec{\delta}_{hh,cv}^{r}\times\vec{S}_{vv}^{,>T}-\vec{S}_{cv}%
^{<}\times\vec{\Xi}_{\ vv}^{rT}\right]  -i\left[  \vec{\delta}_{hh,cv}%
^{<}\times\vec{S}_{vv}^{rT}-\vec{S}_{cv}^{r}\times\vec{\Xi}_{\ vv}%
^{>T}\right]
\end{array}
\right] \nonumber\\
&  +\frac{1}{2}\left[
\begin{array}
[c]{c}%
+\left[  \mathcal{\vec{B}}_{cc}S_{cv,o}^{<}-S_{cv,o}^{<}\mathcal{\vec{B}}%
_{vv}\right] \\
+\left[  \vec{\Xi}_{cc}^{r}S_{cv,o}^{<}-S_{cc,o}^{<}\vec{\delta}_{hh,cv}%
^{\ a}\right]  +\left[  \vec{\Xi}_{cc}^{<}S_{cv,o}^{a}-S_{cc,o}^{r}%
\ \vec{\delta}_{hh,cv}^{<}\right] \\
-\left[  \vec{\delta}_{hh,cv}^{r}S_{vv,o}^{>T}-S_{cv,o}^{<}\vec{\Xi}%
_{\ vv}^{rT}\right]  -\left[  \vec{\delta}_{hh,cv}^{<}S_{vv,o}^{rT}%
-S_{cv,o}^{r}\vec{\Xi}_{\ vv}^{>T}\right]
\end{array}
\right]  , \label{genBloch22}%
\end{align}%
\begin{align}
&  i\hbar\left(  \frac{\partial}{\partial t_{1}}+\frac{\partial}{\partial
t_{2}}\right)  \vec{S}_{vc}^{<}\nonumber\\
&  =\frac{1}{2}\left[
\begin{array}
[c]{c}%
\left[  \bar{H}_{v}\vec{S}_{vc}^{<}-\vec{S}_{vc}^{<}\bar{H}_{c}\right] \\
-\left[  \bar{\Sigma}_{\ vv}^{a,T}\vec{S}_{vc}^{<}-\vec{S}_{vv}^{>T}%
\bar{\Delta}_{ee,vc}^{a}\right]  -\left[  \bar{\Sigma}_{\ vv}^{>T}\ \vec
{S}_{vc}^{a}-\vec{S}_{vv}^{aT}\bar{\Delta}_{ee,vc}^{<}\right] \\
+\left[  \bar{\Delta}_{ee,vc}^{r}\vec{S}_{cc}^{<}-\vec{S}_{vc}^{<}\bar{\Sigma
}_{cc}^{a}\right]  +\left[  \bar{\Delta}_{ee,vc}^{<}\vec{S}_{cc}^{a}-\vec
{S}_{vc}^{r}\bar{\Sigma}_{cc}^{<}\right]
\end{array}
\right] \nonumber\\
&  +\frac{1}{2}\left[
\begin{array}
[c]{c}%
+i\left[  \mathcal{\vec{B}}_{vv}\times\vec{S}_{vc}^{<}-\vec{S}_{vc}^{<}%
\times\mathcal{\vec{B}}_{cc}\right] \\
+i\left[  \vec{S}_{vv}^{aT}\times\vec{\delta}_{ee,vc}^{<}-\vec{\Xi}%
_{\ vv}^{>T}\times\vec{S}_{vc}^{a}\right]  +i\left[  \vec{S}_{vv}^{>T}%
\times\vec{\delta}_{ee,vc}^{a}-\vec{\Xi}_{\ vv}^{a,T}\times\vec{S}_{vc}%
^{<}\right] \\
+i\left[  \vec{\delta}_{ee,vc}^{r}\times\vec{S}_{cc}^{<}-\vec{S}_{vc}%
^{<}\times\vec{\Xi}_{cc}^{a}\right]  +i\left[  \vec{\delta}_{ee,vc}%
^{<}\ \times\vec{S}_{cc}^{a}-\vec{S}_{vc}^{r}\times\vec{\Xi}_{cc}^{<}\right]
\ \
\end{array}
\right] \nonumber\\
&  +\frac{1}{2}\left[
\begin{array}
[c]{c}%
+\left[  \mathcal{\vec{B}}_{vv}S_{vc,o}^{<}-S_{vc,o}^{<}\mathcal{\vec{B}}%
_{cc}\right] \\
-\left[  \vec{\Xi}_{\ vv}^{a,T}S_{vc,o}^{<}-S_{vv,o}^{>T}\vec{\delta}%
_{ee,vc}^{a}\right]  -\left[  \vec{\Xi}_{\ vv}^{>T}S_{vc,o}^{a}-S_{vv,o}%
^{aT}\vec{\delta}_{ee,vc}^{<}\right] \\
+\left[  \vec{\delta}_{ee,vc}^{r}S_{cc,o}^{<}-S_{vc,o}^{<}\vec{\Xi}_{cc}%
^{a}\right]  +\left[  \vec{\delta}_{ee,vc}^{<}\ S_{cc,o}^{a}-S_{vc,o}^{r}%
\vec{\Xi}_{cc}^{<}\right]  \ \
\end{array}
\right]  , \label{genBloch3-2}%
\end{align}

\begin{align}
&  i\hbar\left(  \frac{\partial}{\partial t_{1}}+\frac{\partial}{\partial
t_{2}}\right)  \vec{S}_{vv}^{<}\nonumber\\
&  =\frac{1}{2}\left\{
\begin{array}
[c]{c}%
\left[  \bar{H}_{v},\vec{S}_{vv}^{<}\right] \\
+\left[  \bar{\Sigma}_{\ vv}^{r}\vec{S}_{vv}^{<}-\vec{S}_{vv}^{<}\bar{\Sigma
}_{\ vv}^{a}\right]  +\left[  \bar{\Sigma}_{\ vv}^{<}\vec{S}_{vv}^{a}-\vec
{S}_{vv}^{r}\bar{\Sigma}_{\ vv}^{<}\right] \\
+\left[  \bar{\Delta}_{ee,vc}^{r}\vec{S}_{cv}^{<}-\vec{S}_{vc}^{<}\bar{\Delta
}_{hh,cv}^{\ a}\right]  +\left[  \bar{\Delta}_{ee,vc}^{<}\vec{S}_{cv}^{a}%
-\vec{S}_{vc}^{r}\bar{\Delta}_{hh,cv}^{<}\right]
\end{array}
\right\} \nonumber\\
&  +\frac{1}{2}\left[
\begin{array}
[c]{c}%
+i\left[  \mathcal{\vec{B}}_{vv}\times\vec{S}_{vv}^{<}-\vec{S}_{vv}^{<}%
\times\mathcal{\vec{B}}_{vv}\right] \\
+i\left[  \vec{\Xi}_{\ vv}^{r}\times\vec{S}_{vv}^{<}\ -\vec{S}_{vv}^{<}%
\times\vec{\Xi}_{\ vv}^{a}\right]  \ +i\left[  \vec{\Xi}_{\ vv}^{<}\times
\vec{S}_{vv}^{a}-\vec{S}_{vv}^{r}\times\vec{\Xi}_{\ vv}^{<}\right] \\
+i\left[  \vec{\delta}_{ee,vc}^{<}\times\vec{S}_{cv}^{a}\ -\vec{S}_{vc}%
^{r}\times\vec{\delta}_{hh,cv}^{<}\right]  +i\left[  \vec{\delta}_{ee,vc}%
^{r}\times\vec{S}_{cv}^{<}\ -\vec{S}_{vc}^{<}\times\vec{\delta}_{hh,cv}%
^{\ a}\right]
\end{array}
\right] \nonumber\\
&  +\frac{1}{2}\left[
\begin{array}
[c]{c}%
+\left[  \mathcal{\vec{B}}_{vv},S_{vv,o}^{<}\right] \\
+\left[  \vec{\Xi}_{\ vv}^{r}S_{vv,o}^{<}-S_{vv,o}^{<}\vec{\Xi}_{\ vv}%
^{a}\right]  +\left[  \vec{\Xi}_{\ vv}^{<}S_{vv,o}^{a}-S_{vv,o}^{r}\vec{\Xi
}_{\ vv}^{<}\right] \\
+\left[  \vec{\delta}_{ee,vc}^{<}S_{cv,o}^{a}-S_{vc,o}^{r}\vec{\delta}%
_{hh,cv}^{<}\right]  +\left[  \vec{\delta}_{ee,vc}^{r}S_{cv,o}^{<}%
-S_{vc,o}^{<}\vec{\delta}_{hh,cv}^{\ a}\right]  \ \
\end{array}
\right]  . \label{genBloch4-2}%
\end{align}

We also write the RHS of the pseudo-spin transport equation into a similar
three group of terms not involving the real spin, plus term invoving real spin
as,%
\begin{align}
&  i\hbar\left(  \frac{\partial}{\partial t_{1}}+\frac{\partial}{\partial
t_{2}}\right)  \vec{S}_{o}\nonumber\\
&  =\left[
\begin{array}
[c]{c}%
\frac{1}{4}\left[  \bar{H},\vec{S}_{o}\right]  +\frac{1}{4}\left[  \bar{\zeta
}^{<},\operatorname{Re}\vec{S}_{o}^{r}\right] \\
+\frac{i}{4}\left\{  \bar{\zeta},\vec{S}_{o}\right\}  -\frac{i}{4}\left\{
\bar{\zeta}^{<},\operatorname{Im}\vec{S}_{o}^{r}\right\}
\end{array}
\right]  \text{ \ }\nonumber\\
&  +\left[
\begin{array}
[c]{c}%
\frac{i}{4}\left[  \mathcal{\vec{\beta}}\times\vec{S}_{o}-\vec{S}_{o}%
\times\mathcal{\vec{\beta}}\right]  +\frac{i}{4}\left[  \vec{\zeta}^{<}%
\times\operatorname{Re}\vec{S}_{o}^{r}-\operatorname{Re}\vec{S}_{o}^{r}%
\times\vec{\zeta}^{<}\right] \\
-\frac{1}{4}\left\{  \vec{\zeta}\times\vec{S}_{o}+\vec{S}_{o}\times\vec{\zeta
}\right\}  +\frac{1}{4}\left\{  \vec{\zeta}^{<}\times\operatorname{Im}\vec
{S}_{o}^{r}+\operatorname{Im}\vec{S}_{o}^{r}\times\vec{\zeta}^{<}\right\}
\end{array}
\right] \nonumber\\
&  +\left[
\begin{array}
[c]{c}%
\frac{i}{4}\left[  \mathcal{\vec{B}}_{i}\times\vec{S}_{o,i}-\vec{S}%
_{o,i}\times\mathcal{\vec{B}}_{i}\right] \\
+\frac{i}{4}\left[  \vec{\zeta}_{i}^{<}\times\operatorname{Re}\vec{S}%
_{o,i}^{r}-\operatorname{Re}\vec{S}_{o,i}^{r}\times\vec{\zeta}_{i}^{<}\right]
\\
-\frac{1}{4}\left\{  \vec{\zeta}_{i}\times\vec{S}_{o,i}+\vec{S}_{o,i}%
\times\vec{\zeta}_{i}\right\} \\
+\frac{1}{4}\left\{  \vec{\zeta}_{i}^{<}\times\operatorname{Im}\vec{S}%
_{o,i}^{r}+\operatorname{Im}\vec{S}_{o,i}^{r}\times\vec{\zeta}_{i}%
^{<}\right\}
\end{array}
\right] \nonumber\\
&  +\left[
\begin{array}
[c]{c}%
\frac{1}{4}\left[  \mathcal{\bar{B}}_{i},\vec{S}_{o,i}\right]  +\frac{1}%
{4}\left[  \mathcal{\vec{B}}_{i},\bar{S}_{o,i}\right] \\
+\frac{i}{4}\left\{  \bar{\zeta}_{i},\vec{S}_{o,i}\right\}  +\frac{i}%
{4}\left\{  \vec{\zeta}_{i},\bar{S}_{o,i}\right\} \\
+\frac{1}{4}\left[  \bar{\zeta}_{i}^{<},\operatorname{Re}\vec{S}_{o,i}%
^{r}\right]  +\frac{1}{4}\left[  \vec{\zeta}_{i}^{<},\operatorname{Re}\bar
{S}_{o,i}^{r}\right] \\
-\frac{i}{4}\left\{  \bar{\zeta}_{i}^{<},\operatorname{Im}\vec{S}_{o,i}%
^{r}\right\}  -\frac{i}{4}\left\{  \vec{\zeta}_{i}^{<},\operatorname{Im}%
\bar{S}_{o,i}^{r}\right\}
\end{array}
\right] \nonumber\\
&  +\left[
\begin{array}
[c]{c}%
+\frac{1}{4}\left[  \mathcal{\vec{\beta}},S_{o,o}\right]  +\frac{1}{4}\left[
\vec{\zeta}^{<},\operatorname{Re}S_{o,o}^{r}\right] \\
+\frac{i}{4}\left\{  \vec{\zeta},S_{o,o}\right\}  -\frac{i}{4}\left\{
\vec{\zeta}^{<},\operatorname{Im}S_{o,o}^{r}\right\}
\end{array}
\right]  . \label{genBloch5}%
\end{align}
Equations (\ref{genBloch1-2}) -(\ref{genBloch5}) and (\ref{abs_scalarEq})
constitute the nonequilibrium spin magnetization quantum transport equations
of this paper, written into at least three groups of more meaningful terms.
One note that the first group of terms of the magnetization vector equations
is similar to that of the nonequilibrium spinless particle-correlation
transport equations by virtue of the spin-independent transport coefficients,
second group contains all the torque terms, and the last group generally
contains coupling between spin, pseudo-spin, and charge. Note that that the
equation for the pseudo-spin magnetization vector involves coupling to the
total charge represented by $S_{o,o}$.

\section{Retarded Green's Function and Self-Consistency}

Clearly, some important observables, described by the spectral correlation
function, $\operatorname{Im}G^{r}$, and occupation number correlation
function, $i\hbar G^{<}$, are the governing dynamical variables in quantum
transport physics. Thus, strictly speaking one must also solve for the
transport equation of the nonequilibrium retarded Green's function and
self-energy in the presence of spin. The nonequilibrium quantum superfield
formalism yields the equation for $G^{r}\left(  1,2\right)  $ as\cite{LaRNC}%
\begin{equation}
i\hbar\left(  \frac{\partial}{\partial t_{1}}+\frac{\partial}{\partial t_{2}%
}\right)  G^{r}\left(  1,2\right)  =\left[  \left(  \bar{v}+\Sigma^{r}\right)
,G^{r}\right]  _{1,2}, \label{GrEq}%
\end{equation}
where $\bar{v}$ is the single-particle Hamiltonian, and all quantum labels are
absorbed in the two-point arguments denoted by the numeral $1,2$. Then similar
procedure can be followed in bringing the above equation into a multiband
matrix form. Then the submatrices can be expressed in the spin-canonical form
as in Eq. (\ref{eq15}), and one solves for the retarded spin correlation
functions. This process will yield another $16$ nonequilibrium transport equations.

However, usually one is focused mainly in solving the spin magnetization
transport equations, and treat Eq. (\ref{GrEq}) as separate
calculations.\cite{note8} In most cases, the solution to Eq. (\ref{GrEq}) are
obtained by some sort of approximation and the results plugged into the spin
magnetization equations as was done in the simulation of quantum transport
equations for charge carriers in nanoelectronic devices. This is coupled with
some further simplification by ignoring the effect of $\ \operatorname{Re}%
S^{r}$ in the spin magnetization transport equations.\cite{boltzmann} For
these reasons, we will no longer spend time in rigorously including Eq.
(\ref{GrEq}) in our spin magnetization transport equations.

\subsection{Self-Consistent Electric Field and Potential}

There is also an important ingredient in the nonlinearity of the SMQTEs
introduced through the need for the self-consistency of the potential
distribution.\cite{johnson2, singh,zhangWangSu} This in turn affects the
spin-orbit coupling terms in the single particle Hamiltonian and even affects
the corresponding many-body aspects of spin-orbit coupling.\cite{rajagopal}
Thus one also need to solve the Poisson equation for the self-consistent
potential, similar to what was done in the numerical simulation of resonant
tunneling devices.\cite{buotjensen, jbpaper,zhangWangSu} Clearly then the
total charge, represented by $S_{o,o}$ obeying Eq. (\ref{abs_scalarEq}) is
also coupled, after correcting for the background positive charge, to the
potential through the Poisson equation affecting the single particle
Hamiltonian and many-body effects in a highly nonlinear self-consistent loop
to be accounted for in real device numerical simulations, where clever
approximations are eventually needed to bring the problem to a manageable proportion.

\section{Phase-Space Spin Quantum Transport Equations}

We will first give formal prescriptions on how to transform all the
\textit{nonlocal} \textit{two space-time-points} correlation-function quantum
transport equations, given in Eqs. (\ref{genBloch1}) - (\ref{genBloch4}) and
(\ref{pseudoSpinEq}) - (\ref{abs_scalarEq}) into kinetic QDF transport
equations defined in \textit{local phase-space points}, i.e., in $\left(
p,q;E,t\right)  $ phase space. This is achieved through the use of Buot's
discrete phase-space transformation for condensed matter,\cite{lwww} or his
discrete formulation of quantum mechanics leading to the lattice version of
the Weyl transformation. Then we will examine what has so far been attempted
in the literature towards the treatment of kinetic QDF transport in spintronic devices.

We observe that all of the terms entering in Eqs. (\ref{genBloch1}) -
(\ref{genBloch4}) and (\ref{pseudoSpinEq}) - (\ref{abs_scalarEq}) involve
commutators and anticommutators of two scalar functions and of a scalar and a
vector functions, as well as sum and difference of cross products of vector
functions. It is more convenient to cast everything in terms of commutators
and anticommutators before making the necessary transformation to kinetic
equations in phase space. Thus, we need to change the expressions involving
cross product of vectors to commutator or anti-commutator, as the case maybe.
We make use of the following identities for two vectors $\vec{A}$ and $\vec
{B}$,%
\begin{align*}
\vec{A}\times\vec{B}-\vec{B}\times\vec{A}  &  =\hat{I}_{i}\hat{\epsilon}%
_{ijk}\left\{  A_{j},B_{k}\right\}  ,\\
\vec{A}\times\vec{B}+\vec{B}\times\vec{A}  &  =\hat{I}_{i}\hat{\epsilon}%
_{ijk}\left[  A_{j},B_{k}\right]  ,
\end{align*}
where $\hat{I}_{i}$ is the unit dyadic symmetric tensor or \textit{idemfactor}%
, and $\hat{\epsilon}_{ijk}$ is the anti-symmetric unit tensor. As used
before, the square bracket stands for the commutator and the curly bracket
stands for anti-commutator of the two vector components separated by a comma.
We also have terms that goes like, $\vec{A}\times\vec{B}-\vec{C}\times\vec{D}$
occurring in Eqs. (\ref{genBloch2}) and (\ref{genBloch3}), for example the
term $\vec{\Xi}_{cc}^{<}\times\vec{S}_{cv}^{a}\ \ -\vec{S}_{cc}^{r}\times
\vec{\delta}_{hh,cv}^{<}$ in Eqs. (\ref{genBloch2}). For the typical cross
products involving four spin vectors in Eq. (\ref{fourCorrCrossProd}), we will
simply expand this as
\begin{equation}
\vec{A}\times\vec{B}-\vec{C}\times\vec{D}=\hat{I}_{i}\hat{\epsilon}_{ijk}%
A_{j}B_{k}-\hat{I}_{i}\hat{\epsilon}_{ijk}C_{j}D_{k} \label{fourCorrCrossProd}%
\end{equation}

\subsection{Transformation to $\left(  p,q,E,t\right)  $ Phase-Space}

The QDF kinetic transport equations in $\left(  p,q,E,t\right)  $-space are
obtained by applying the \textquotedblright lattice\textquotedblright\ Weyl
transformation of the correlation function equations by using the following
set of identities \cite{buot2} (although continuum approximation is
interchangeably used, this is not essential and we adapt the word
\textquotedblright lattice\textquotedblright\ when referring to solid-state
problems). For convenience, we give these here the \ $\ $LWtransformation of
the following two space-time-point correlation functions:%
\begin{equation}
i\hbar\left(  \frac{\partial}{\partial t_{1}}+\frac{\partial}{\partial t_{2}%
}\right)  F\left(  12\right)  \Leftrightarrow i\hbar\frac{\partial}{\partial
t}F_{w}\left(  \vec{p},\vec{q},E,t\right)  , \label{rnc6.1}%
\end{equation}
where the Bloch states has the representation, $\left\langle q\right\vert
\left.  p\right\rangle =\left(  \frac{1}{N\hbar^{3}}\right)  ^{\frac{1}{2}%
}\exp\frac{i}{\hbar}\left\{  \vec{p}\cdot\vec{q}-Et\right\}  $ [ $N\hbar
^{3}\Rightarrow h^{3}$ in the continuum limit], representing a traveling wave
in lattice space, with group velocity in a band $\lambda$ given by
$\frac{dE_{\lambda}\left\{  \vec{p}\right\}  }{d\vec{p}}$. The above
identities readily follow from the definition of the \ $\ $LWtransform, which
in the continuum approximation can simply be written as%

\begin{align*}
i\hbar\left(  \frac{\partial}{\partial t_{1}}+\frac{\partial}{\partial t_{2}%
}\right)  F\left(  12\right)   &  =i\hbar\frac{\partial}{\partial t}%
{\displaystyle\int}
dv\ d\tau e^{\frac{i}{\hbar}\left(  p\cdot v-E\tau\right)  }F\left(
q-\frac{v}{2},t-\frac{\tau}{2};q+\frac{v}{2},t+\frac{\tau}{2}\right) \\
&  \Leftrightarrow i\hbar\frac{\partial}{\partial t}F_{w}\left(  \vec{p}%
,\vec{q},E,t\right)  .
\end{align*}

The second set is the \ $\ $LWtransform of \ a product of two-point
correlation functions [ $p\equiv\left(  \vec{p},-E\right)  $ and $q=\left(
\vec{q},t\right)  $] in terms of \textquotedblright Poisson bracket
operator\textquotedblright,%
\begin{equation}
AB\left(  p,q\right)  =\exp\left[  \frac{\hbar}{2i}\left(  \frac
{\partial^{\left(  a\right)  }}{\partial p}\cdot\frac{\partial^{\left(
b\right)  }}{\partial q}-\frac{\partial^{\left(  a\right)  }}{\partial q}%
\cdot\frac{\partial^{\left(  b\right)  }}{\partial p}\right)  \right]
a\left(  p,q\right)  \ b\left(  p,q\right)  , \label{rnc6.2}%
\end{equation}
or in terms of integral operator,%
\begin{align}
AB\left(  p,q\right)   &  =\frac{1}{\left(  2\pi\hbar\right)  ^{8}}\int
dp^{\prime}dq^{\prime}K_{A}^{+}\left(  p,q;p^{\prime},q^{\prime}\right)
\ b\left(  p^{\prime},q^{\prime}\right) \nonumber\\
&  =\frac{1}{\left(  2\pi\hbar\right)  ^{8}}\int dp^{\prime}dq^{\prime
}a\left(  p^{\prime},q^{\prime}\right)  K_{B}^{-}\left(  p,q;p^{\prime
},q^{\prime}\right)  \ , \label{rnc6.3}%
\end{align}
where the factor $\frac{1}{\left(  2\pi\hbar\right)  ^{8}}$ accounts for the
proper normalization of the integration (counting of states in terms of unit
action) in $\left(  p,q,E,t\right)  $-space, and the integral kernels are
defined by%
\begin{equation}
K_{Y}^{\pm}\left(  p,q;p^{\prime},q^{\prime}\right)  =\int du\ dv\exp\left\{
\frac{i}{\hbar}\left[  \left(  p-p^{\prime}\right)  \cdot v+\left(
q-q^{\prime}\right)  \cdot u\right]  \right\}  \ y\left(  p\pm\frac{u}{2}%
,q\mp\frac{v}{2}\right)  . \label{rnc6.4}%
\end{equation}
For numerical purposes using discrete lattice points, the following
expressions of $K_{Y}^{\pm}\left(  p,q;p^{\prime},q^{\prime}\right)  $ is more
preferable%
\begin{equation}
K_{Y}^{\pm}\left(  p,q;p^{\prime},q^{\prime}\right)  =\int du\ dv\exp\left\{
\frac{2i}{\hbar}\left[  \left(  p-p^{\prime}\right)  \cdot v+\left(
q-q^{\prime}\right)  \cdot u\right]  \right\}  \ y\left(  p\pm u,q\mp
v\right)  .
\end{equation}

Thus, we may write the \ $\ $LWtransform of a commutator $\left[  A,B\right]
$ and an anticommutator $\left\{  A,B\right\}  $ in terms of Poisson bracket
differential operator, $\Lambda$, as%
\begin{align}
\left[  A,B\right]  \left(  p,q\right)   &  =\cos\Lambda\left[  a\left(
p,q\right)  b\left(  p,q\right)  -b\left(  p,q\right)  a\left(  p,q\right)
\right] \nonumber\\
&  -i\sin\Lambda\left\{  a\left(  p,q\right)  b\left(  p,q\right)  +b\left(
p,q\right)  a\left(  p,q\right)  \right\}  , \label{rnc6.5}%
\end{align}%
\begin{align}
\left\{  A,B\right\}  \left(  p,q\right)   &  =\cos\Lambda\left\{  a\left(
p,q\right)  b\left(  p,q\right)  +b\left(  p,q\right)  a\left(  p,q\right)
\right\} \nonumber\\
&  -i\sin\Lambda\left[  a\left(  p,q\right)  b\left(  p,q\right)  -b\left(
p,q\right)  a\left(  p,q\right)  \right]  , \label{rnc6.6}%
\end{align}
where $\Lambda=\frac{\hbar}{2}\left(  \frac{\partial^{\left(  a\right)  }%
}{\partial p}\cdot\frac{\partial^{\left(  b\right)  }}{\partial q}%
-\frac{\partial^{\left(  a\right)  }}{\partial q}\cdot\frac{\partial^{\left(
b\right)  }}{\partial p}\right)  $. In terms of integral operators, we have,%
\begin{equation}
\left[  A,B\right]  \left(  p,q\right)  =\frac{1}{\left(  2\pi\hbar\right)
^{8}}\int dp^{\prime}dq^{\prime}K_{A}^{+}\left(  p,q;p^{\prime},q^{\prime
}\right)  \ b\left(  p^{\prime},q^{\prime}\right)  -b\left(  p^{\prime
},q^{\prime}\right)  K_{A}^{-}\left(  p,q;p^{\prime},q^{\prime}\right)  ,
\label{rnc6.7}%
\end{equation}%
\begin{equation}
\left\{  A,B\right\}  \left(  p,q\right)  =\frac{1}{\left(  2\pi\hbar\right)
^{8}}\int dp^{\prime}dq^{\prime}K_{A}^{+}\left(  p,q;p^{\prime},q^{\prime
}\right)  \ b\left(  p^{\prime},q^{\prime}\right)  +b\left(  p^{\prime
},q^{\prime}\right)  K_{A}^{-}\left(  p,q;p^{\prime},q^{\prime}\right)  .
\label{rnc6.8}%
\end{equation}
The above expressions simplify considerably when the \ $\ $LWtransforms are
scalar functions. For this case, we have the 'lattice' Weyl transform of a
commutator and anti-commutator of two operators,$\left[  A,B\right]  $ and
$\left\{  A,B\right\}  $ , respectively, given by the following expressions%
\begin{equation}
\left[  A,B\right]  \left(  p,q\right)  =\frac{1}{\left(  2\pi\hbar\right)
^{8}}\int dp^{\prime}dq^{\prime}K_{A}^{s}\left(  p,q;p^{\prime},q^{\prime
}\right)  \ b\left(  p^{\prime},q^{\prime}\right)  , \label{rnc6.9}%
\end{equation}%
\begin{equation}
\left\{  A,B\right\}  \left(  p,q\right)  =\frac{1}{\left(  2\pi\hbar\right)
^{8}}\int dp^{\prime}dq^{\prime}K_{A}^{c}\left(  p,q;p^{\prime},q^{\prime
}\right)  \ b\left(  p^{\prime},q^{\prime}\right)  , \label{rnc6.10}%
\end{equation}
where,%
\begin{align}
K_{Y}^{s}\left(  p,q;p^{\prime},q^{\prime}\right)   &  =\int du\ dv\exp
\left\{  \frac{2i}{\hbar}\left[  \left(  p-p^{\prime}\right)  \cdot v+\left(
q-q^{\prime}\right)  \cdot u\right]  \right\} \nonumber\\
&  \times\left[  \ y\left(  p+u,q-v\right)  -y\left(  p-u,q+v\right)  \right]
. \label{rnc6.11}%
\end{align}
and%
\begin{align}
K_{Y}^{c}\left(  p,q;p^{\prime},q^{\prime}\right)   &  =\int du\ dv\exp
\left\{  \frac{2i}{\hbar}\left[  \left(  p-p^{\prime}\right)  \cdot v+\left(
q-q^{\prime}\right)  \cdot u\right]  \right\} \nonumber\\
&  \times\left[  \ y\left(  p+u,q-v\right)  +y\left(  p-u,q+v\right)  \right]
. \label{rnc6.11c}%
\end{align}

\section{Concluding Remarks}

What we have accomplished in this paper is the demonstration of the formal
mathematical structure of nonequilibrium multiband spin-correlation transport
equations. It has been demonstrated that spin-dependent self-energies due to
many-body effects give rise to torques in the system. For example, the
many-body effects in spin-orbit coupling has lead to the separation of
self-energy into spin-independent part and the corresponding spin
vector.\cite{rajagopal, zhukov,mower} Thus, it is expected that the results
here will serve as a fundamental basis for constructing realistic transport
equations, embodying various approximation schemes, for engineering expediency.

The range of their validity of all simpler and manageable approximate
equations can be assessed in the light of the present mathematical structure
given in this paper This must be accurate enough for treating the space and
time dependent spin relaxation and dephasing scattering mechanisms between
conduction electrons, between valence holes, between electrons and holes,
their coupling to pseudo-spin, and to the total charges. The total electron
charges represented by $S_{o,o}$ minus the positive background charge will be
fed to the Poisson equation for self-consistency in the potential. It is
expected that the accompanying physical interpretations will acquire deeper
insights when applied to highly nonequilibrium situations for spintronic
device-performance applications, based on the fundamental structure of SMQTEs
presented in this paper, .

In particular, we have shown that the pseudo-spin is highly coupled to
Pauli-Dirac spin transport equations and to the particle charge density. It is
not clear how this pseudo-spin aspect of spin transport is described by
conventional methods of Pauli-Dirac spin scattering physics so far employed in
the literature on the theory of spin relaxation and dephasing mechanisms in solids.

\begin{acknowledgement}
The author is grateful to Prof. R. E. S. Otadoy, R. A. Loberternos, and D. L.
Villarin of the Theoretical and Computational Sciences and Engineering Group,
Department of Physics, University of San Carlos, for their interest in spin
quantum transport problems, which helps motivate the present extension of
their joint previous work.
\end{acknowledgement}

\bigskip\medskip%

\appendix

\section{Matrix Equations for Bloch Electrons with Spin}

In the presence of spin degree of freedom, Eqs. (\ref{eq8}), (\ref{eq6}),
(\ref{eq10}), and (\ref{eq9}) become matrix equations. For the
conduction-electron band, Eq. (\ref{eq8}), we have the matrix equation with
spin indices given by,%
\begin{align}
&  i\hbar\left(  \frac{\partial}{\partial t_{1}}+\frac{\partial}{\partial
t_{2}}\right)  \left(
\begin{array}
[c]{cc}%
G_{cc,\uparrow\uparrow}^{e-h,<}\left(  12\right)  & G_{cc,\uparrow\downarrow
}^{e-h,<}\left(  12\right) \\
G_{cc,\downarrow\uparrow}^{e-h,<}\left(  12\right)  & G_{cc,\downarrow
\downarrow}^{e-h,<}\left(  12\right)
\end{array}
\right) \nonumber\\
&  =\left(
\begin{array}
[c]{cc}%
\left[
\begin{array}
[c]{c}%
v_{c,\uparrow\sigma}\left(  1\xi\right)  G_{cc,\sigma\uparrow}^{<}\left(
\xi2\right) \\
-G_{cc,\uparrow\sigma}^{<}\left(  1\xi\right)  \ v_{c,\sigma\uparrow}%
^{T}\left(  \xi2\right)
\end{array}
\right]  & \left[
\begin{array}
[c]{c}%
v_{c,\uparrow\sigma}\left(  1\xi\right)  G_{cc,\sigma\downarrow}^{<}\left(
\xi2\right) \\
-G_{cc,\uparrow\sigma}^{<}\left(  1\xi\right)  \ v_{c,\sigma\downarrow}%
^{T}\left(  \xi2\right)
\end{array}
\right] \\
\left[
\begin{array}
[c]{c}%
v_{c,\downarrow\sigma}\left(  1\xi\right)  G_{cc,\sigma\uparrow}^{<}\left(
\xi2\right) \\
-G_{cc,\downarrow\sigma}^{<}\left(  1\xi\right)  \ v_{c,\sigma\uparrow}%
^{T}\left(  \xi2\right)
\end{array}
\right]  & \left[
\begin{array}
[c]{c}%
v_{c,\downarrow\sigma}\left(  1\xi\right)  G_{cc,\sigma\downarrow}^{<}\left(
\xi2\right) \\
-G_{cc,\downarrow\sigma}^{<}\left(  1\xi\right)  \ v_{c,\sigma\downarrow}%
^{T}\left(  \xi2\right)
\end{array}
\right]
\end{array}
\right) \nonumber\\
&  +\left(
\begin{array}
[c]{cc}%
\left[
\begin{array}
[c]{c}%
\Sigma_{cc,\uparrow\sigma}^{r}\ \left(  1\xi\right)  G_{cc,\sigma\uparrow}%
^{<}\left(  \xi2\right) \\
-G_{cc,\uparrow\sigma}^{<}\left(  1\xi\right)  \ \Sigma_{cc,\sigma\uparrow
}^{a}\left(  \xi2\right)
\end{array}
\right]  & \left[
\begin{array}
[c]{c}%
\Sigma_{cc,\uparrow\sigma}^{r}\ \left(  1\xi\right)  G_{cc,\sigma\downarrow
}^{<}\left(  \xi2\right) \\
-G_{cc,\uparrow\sigma}^{<}\left(  1\xi\right)  \ \Sigma_{cc,\sigma\downarrow
}^{a}\left(  \xi2\right)
\end{array}
\right] \\
\left[
\begin{array}
[c]{c}%
\Sigma_{cc,\downarrow\sigma}^{r}\ \left(  1\xi\right)  G_{cc,\sigma\uparrow
}^{<}\left(  \xi2\right) \\
-G_{cc,\downarrow\sigma}^{<}\left(  1\xi\right)  \ \Sigma_{cc,\sigma\uparrow
}^{a}\left(  \xi2\right)
\end{array}
\right]  & \left[
\begin{array}
[c]{c}%
\Sigma_{cc,\downarrow\sigma}^{r}\ \left(  1\xi\right)  G_{cc,\sigma\downarrow
}^{<}\left(  \xi2\right) \\
-G_{cc,\downarrow\sigma}^{<}\left(  1\xi\right)  \ \Sigma_{cc,\sigma
\downarrow}^{a}\left(  \xi2\right)
\end{array}
\right]
\end{array}
\right) \nonumber\\
&  +\left(
\begin{array}
[c]{cc}%
\left[
\begin{array}
[c]{c}%
\Sigma_{cc,\uparrow\sigma}^{<}\left(  1\xi\right)  \ G_{cc,\sigma\uparrow}%
^{a}\left(  \xi2\right) \\
-G_{cc,\uparrow\sigma}^{r}\left(  1\xi\right)  \ \Sigma_{cc,\sigma\uparrow
}^{<}\left(  \xi2\right)
\end{array}
\right]  & \left[
\begin{array}
[c]{c}%
\Sigma_{cc,\uparrow\sigma}^{<}\left(  1\xi\right)  \ G_{cc,\sigma\downarrow
}^{a}\left(  \xi2\right) \\
-G_{cc,\uparrow\sigma}^{r}\left(  1\xi\right)  \ \Sigma_{cc,\sigma\downarrow
}^{<}\left(  \xi2\right)
\end{array}
\right] \\
\left[
\begin{array}
[c]{c}%
\Sigma_{cc,\downarrow\sigma}^{<}\left(  1\xi\right)  \ G_{cc}^{a}\left(
\xi2\right) \\
-G_{cc,\downarrow\sigma}^{r}\left(  1\xi\right)  \ \Sigma_{cc,\sigma\uparrow
}^{<}\left(  \xi2\right)
\end{array}
\right]  & \left[
\begin{array}
[c]{c}%
\Sigma_{cc,\downarrow\sigma}^{<}\left(  1\xi\right)  \ G_{cc,\sigma\downarrow
}^{a}\left(  \xi2\right) \\
-G_{cc,\downarrow\sigma}^{r}\left(  1\xi\right)  \ \Sigma_{cc,\sigma
\downarrow}^{<}\left(  \xi2\right)
\end{array}
\right]
\end{array}
\right) \nonumber\\
&  +\left(
\begin{array}
[c]{cc}%
\left[
\begin{array}
[c]{c}%
\Delta_{hh,cv,\uparrow\sigma}^{e-h,r}\left(  1\xi\right)  g_{ee,vc,\sigma
\uparrow}^{e-h,<}\left(  \xi2\right) \\
-g_{hh,cv,\uparrow\sigma}^{e-h,<}\left(  1\xi\right)  \ \Delta_{ee,vc,\sigma
\uparrow}^{e-h,a}\left(  \xi2\right)
\end{array}
\right]  & \left[
\begin{array}
[c]{c}%
\Delta_{hh,cv,\uparrow\sigma}^{e-h,r}\left(  1\xi\right)  g_{ee,vc,\sigma
\downarrow}^{e-h,<}\left(  \xi2\right) \\
-g_{hh,cv,\uparrow\sigma}^{e-h,<}\left(  1\xi\right)  \ \Delta_{ee,vc,\sigma
\downarrow}^{e-h,a}\left(  \xi2\right)
\end{array}
\right] \\
\left[
\begin{array}
[c]{c}%
\Delta_{hh,cv,\downarrow\sigma}^{e-h,r}\left(  1\xi\right)  g_{ee,vc,\sigma
\uparrow}^{e-h,<}\left(  \xi2\right) \\
-g_{hh,cv,\downarrow\sigma}^{e-h,<}\left(  1\xi\right)  \ \Delta
_{ee,vc,\sigma\uparrow}^{e-h,a}\left(  \xi2\right)
\end{array}
\right]  & \left[
\begin{array}
[c]{c}%
\Delta_{hh,cv,\downarrow\sigma}^{e-h,r}\left(  1\xi\right)  g_{ee,vc,\sigma
\downarrow}^{e-h,<}\left(  \xi2\right) \\
-g_{hh,cv,\downarrow\sigma}^{e-h,<}\left(  1\xi\right)  \ \Delta
_{ee,vc,\sigma\downarrow}^{e-h,a}\left(  \xi2\right)
\end{array}
\right]
\end{array}
\right) \nonumber\\
&  +\left(
\begin{array}
[c]{cc}%
\left[
\begin{array}
[c]{c}%
\Delta_{hh,cv,\uparrow\sigma}^{e-h,<}\left(  1\xi\right)  \ g_{ee,vc,\sigma
\uparrow}^{e-h,a}\left(  \xi2\right) \\
-g_{hh,cv,\uparrow\sigma}^{e-h,r}\left(  1\xi\right)  \ \Delta_{ee,vc,\sigma
\uparrow}^{e-h,<}\left(  \xi2\right)
\end{array}
\right]  & \left[
\begin{array}
[c]{c}%
\Delta_{hh,cv,\uparrow\sigma}^{e-h,<}\left(  1\xi\right)  \ g_{ee,vc,\sigma
\downarrow}^{e-h,a}\left(  \xi2\right) \\
-g_{hh,cv,\uparrow\sigma}^{e-h,r}\left(  1\xi\right)  \ \Delta_{ee,vc,\sigma
\downarrow}^{e-h,<}\left(  \xi2\right)
\end{array}
\right] \\
\left[
\begin{array}
[c]{c}%
\Delta_{hh,cv,\downarrow\sigma}^{e-h,<}\left(  1\xi\right)  \ g_{ee,vc,\sigma
\uparrow}^{e-h,a}\left(  \xi2\right) \\
-g_{hh,cv,\downarrow\sigma}^{e-h,r}\left(  1\xi\right)  \ \Delta
_{ee,vc,\sigma\uparrow}^{e-h,<}\left(  \xi2\right)
\end{array}
\right]  & \left[
\begin{array}
[c]{c}%
\Delta_{hh,cv,\downarrow\sigma}^{e-h,<}\left(  1\xi\right)  \ g_{ee,vc,\sigma
\downarrow}^{e-h,a}\left(  \xi2\right) \\
-g_{hh,cv,\downarrow\sigma}^{e-h,r}\left(  1\xi\right)  \ \Delta
_{ee,vc,\sigma\downarrow}^{e-h,<}\left(  \xi2\right)
\end{array}
\right]
\end{array}
\right)  , \label{eq12}%
\end{align}
where the repeated spin subscript, $\sigma$, indicates the use of Einstein
summation convention over the spin degrees of freedom.

For the hole band, we have the corresponding matrix equation determined from
Eq. (\ref{eq6}) as,%

\begin{align}
&  i\hbar\left(  \frac{\partial}{\partial t_{1}}+\frac{\partial}{\partial
t_{2}}\right)  \left(
\begin{array}
[c]{cc}%
G_{vv,\uparrow\uparrow}^{e-h,<}\left(  12\right)  & G_{vv,\uparrow\downarrow
}^{e-h,<}\left(  12\right) \\
G_{vv,\downarrow\uparrow}^{e-h,<}\left(  12\right)  & G_{vv,\downarrow
\downarrow}^{e-h,<}\left(  12\right)
\end{array}
\right) \nonumber\\
&  =-\left(
\begin{array}
[c]{cc}%
\left[
\begin{array}
[c]{c}%
v_{vv,\uparrow\sigma}\left(  1\xi\right)  G_{vv,\sigma\uparrow}^{e-h,<}\left(
\xi2\right) \\
-G_{vv,\uparrow\sigma}^{e-h,<}\left(  1\xi\right)  v_{vv,\sigma\uparrow}%
^{T}\left(  \xi2\right)
\end{array}
\right]  & \left[
\begin{array}
[c]{c}%
v_{vv,\uparrow\sigma}\left(  1\xi\right)  G_{vv,\sigma\downarrow}%
^{e-h,<}\left(  \xi2\right) \\
-G_{vv,\uparrow\sigma}^{e-h,<}\left(  1\xi\right)  v_{vv,\sigma\downarrow}%
^{T}\left(  \xi2\right)
\end{array}
\right] \\
\left[
\begin{array}
[c]{c}%
v_{vv,\downarrow\sigma}\left(  1\xi\right)  G_{vv,\sigma\uparrow}%
^{e-h,<}\left(  \xi2\right) \\
-G_{vv,\downarrow\sigma}^{e-h,<}\left(  1\xi\right)  v_{vv,\sigma\uparrow}%
^{T}\left(  \xi2\right)
\end{array}
\right]  & \left[
\begin{array}
[c]{c}%
v_{vv,\downarrow\sigma}\left(  1\xi\right)  G_{vv,\sigma\downarrow}%
^{e-h,<}\left(  \xi2\right) \\
-G_{vv,\downarrow\sigma}^{e-h,<}\left(  1\xi\right)  v_{vv,\sigma\downarrow
}^{T}\left(  \xi2\right)
\end{array}
\right]
\end{array}
\right) \nonumber\\
&  -\left(
\begin{array}
[c]{cc}%
\left[
\begin{array}
[c]{c}%
\Sigma_{vv,\uparrow\sigma}^{e-h,r}\left(  1\xi\right)  G_{vv,\sigma\uparrow
}^{e-h,<}\left(  \xi2\right) \\
-G_{vv,\uparrow\sigma}^{e-h,<}\left(  1\xi\right)  \Sigma_{vv,\sigma\uparrow
}^{e-h,a}\left(  \xi2\right)
\end{array}
\right]  & \left[
\begin{array}
[c]{c}%
\Sigma_{vv,\uparrow\sigma}^{e-h,r}\left(  1\xi\right)  G_{vv,\sigma\downarrow
}^{e-h,<}\left(  \xi2\right) \\
-G_{vv,\uparrow\sigma}^{e-h,<}\left(  1\xi\right)  \Sigma_{vv,\sigma
\downarrow}^{e-h,a}\left(  \xi2\right)
\end{array}
\right] \\
\left[
\begin{array}
[c]{c}%
\Sigma_{vv,\downarrow\sigma}^{e-h,r}\left(  1\xi\right)  G_{vv,\sigma\uparrow
}^{e-h,<}\left(  \xi2\right) \\
-G_{vv,\downarrow\sigma}^{e-h,<}\left(  1\xi\right)  \Sigma_{vv,\sigma
\uparrow}^{e-h,a}\left(  \xi2\right)
\end{array}
\right]  & \left[
\begin{array}
[c]{c}%
\Sigma_{vv,\downarrow\sigma}^{e-h,r}\left(  1\xi\right)  G_{vv,\sigma
\downarrow}^{e-h,<}\left(  \xi2\right) \\
-G_{vv,\downarrow\sigma}^{e-h,<}\left(  1\xi\right)  \Sigma_{vv,\sigma
\downarrow}^{e-h,a}\left(  \xi2\right)
\end{array}
\right]
\end{array}
\right) \nonumber\\
&  -\left(
\begin{array}
[c]{cc}%
\left[
\begin{array}
[c]{c}%
\Sigma_{vv,\uparrow\sigma}^{e-h,<}\left(  1\xi\right)  G_{vv,\sigma\uparrow
}^{e-h,a}\left(  \xi2\right) \\
-G_{vv,\uparrow\sigma}^{e-h,r}\left(  1\xi\right)  \Sigma_{vv,\sigma\uparrow
}^{e-h,<}\left(  \xi2\right)
\end{array}
\right]  & \left[
\begin{array}
[c]{c}%
\Sigma_{vv,\uparrow\sigma}^{e-h,<}\left(  1\xi\right)  G_{vv,\sigma\downarrow
}^{e-h,a}\left(  \xi2\right) \\
-G_{vv,\uparrow\sigma}^{e-h,r}\left(  1\xi\right)  \Sigma_{vv,\sigma
\downarrow}^{e-h,<}\left(  \xi2\right)
\end{array}
\right] \\
\left[
\begin{array}
[c]{c}%
\Sigma_{vv,\downarrow\sigma}^{e-h,<}\left(  1\xi\right)  G_{vv,\sigma\uparrow
}^{e-h,a}\left(  \xi2\right) \\
-G_{vv,\downarrow\sigma}^{e-h,r}\left(  1\xi\right)  \Sigma_{vv,\sigma
\uparrow}^{e-h,<}\left(  \xi2\right)
\end{array}
\right]  & \left[
\begin{array}
[c]{c}%
\Sigma_{vv,\downarrow\sigma}^{e-h,<}\left(  1\xi\right)  G_{vv,\sigma
\downarrow}^{e-h,a}\left(  \xi2\right) \\
-G_{vv,\downarrow\sigma}^{e-h,r}\left(  1\xi\right)  \Sigma_{vv,\sigma
\downarrow}^{e-h,<}\left(  \xi2\right)
\end{array}
\right]
\end{array}
\right) \nonumber\\
&  -\left(
\begin{array}
[c]{cc}%
\left[
\begin{array}
[c]{c}%
\Delta_{ee,vc,\uparrow\sigma}^{e-h,r}\left(  1\xi\right)  g_{hh,cv,\sigma
\uparrow}^{e-h,<}\left(  \xi2\right) \\
-g_{ee,vc,\uparrow\sigma}^{e-h,<}\left(  1\xi\right)  \Delta_{hh,cv,\sigma
\uparrow}^{e-h,a}\left(  \xi2\right)
\end{array}
\right]  & \left[
\begin{array}
[c]{c}%
\Delta_{ee,vc,\uparrow\sigma}^{e-h,r}\left(  1\xi\right)  g_{hh,cv,\sigma
\downarrow}^{e-h,<}\left(  \xi2\right) \\
-g_{ee,vc,\uparrow\sigma}^{e-h,<}\left(  1\xi\right)  \Delta_{hh,cv,\sigma
\downarrow}^{e-h,a}\left(  \xi2\right)
\end{array}
\right] \\
\left[
\begin{array}
[c]{c}%
\Delta_{ee,vc,\downarrow\sigma}^{e-h,r}\left(  1\xi\right)  g_{hh,cv,\sigma
\uparrow}^{e-h,<}\left(  \xi2\right) \\
-g_{ee,vc,\downarrow\sigma}^{e-h,<}\left(  1\xi\right)  \Delta_{hh,cv,\sigma
\uparrow}^{e-h,a}\left(  \xi2\right)
\end{array}
\right]  & \left[
\begin{array}
[c]{c}%
\Delta_{ee,vc,\downarrow\sigma}^{e-h,r}\left(  1\xi\right)  g_{hh,cv,\sigma
\downarrow}^{e-h,<}\left(  \xi2\right) \\
-g_{ee,vc,\downarrow\sigma}^{e-h,<}\left(  1\xi\right)  \Delta_{hh,cv,\sigma
\downarrow}^{e-h,a}\left(  \xi2\right)
\end{array}
\right]
\end{array}
\right) \nonumber\\
&  -\left(
\begin{array}
[c]{cc}%
\left[
\begin{array}
[c]{c}%
\Delta_{ee,vc,\uparrow\sigma}^{e-h,<}\left(  1\xi\right)  g_{hh,cv,\sigma
\uparrow}^{e-h,a}\left(  \xi2\right) \\
-g_{ee,vc,\uparrow\sigma}^{e-h,r}\left(  1\xi\right)  \Delta_{hh,cv,\sigma
\uparrow}^{e-h,<}\left(  \xi2\right)
\end{array}
\right]  & \left[
\begin{array}
[c]{c}%
\Delta_{ee,vc,\uparrow\sigma}^{e-h,<}\left(  1\xi\right)  g_{hh,cv,\sigma
\downarrow}^{e-h,a}\left(  \xi2\right) \\
-g_{ee,vc,\uparrow\sigma}^{e-h,r}\left(  1\xi\right)  \Delta_{hh,cv,\sigma
\downarrow}^{e-h,<}\left(  \xi2\right)
\end{array}
\right] \\
\left[
\begin{array}
[c]{c}%
\Delta_{ee,vc,\downarrow\sigma}^{e-h,<}\left(  1\xi\right)  g_{hh,cv,\sigma
\uparrow}^{e-h,a}\left(  \xi2\right) \\
-g_{ee,vc,\downarrow\sigma}^{e-h,r}\left(  1\xi\right)  \Delta_{hh,cv,\sigma
\uparrow}^{e-h,<}\left(  \xi2\right)
\end{array}
\right]  & \left[
\begin{array}
[c]{c}%
\Delta_{ee,vc,\downarrow\sigma}^{e-h,<}\left(  1\xi\right)  g_{hh,cv,\sigma
\downarrow}^{e-h,a}\left(  \xi2\right) \\
-g_{ee,vc,\downarrow\sigma}^{e-h,r}\left(  1\xi\right)  \Delta_{hh,cv,\sigma
\downarrow}^{e-h,<}\left(  \xi2\right)
\end{array}
\right]
\end{array}
\right)  . \label{eq11}%
\end{align}

\section{Spin-Dependent Pairing Green's Functions}

\bigskip For the 'pairing' between electron and holes, we have from Eq.
(\ref{eq10})%

\begin{align}
&  i\hbar\left(  \frac{\partial}{\partial t_{1}}+\frac{\partial}{\partial
t_{2}}\right)  \left(
\begin{array}
[c]{cc}%
g_{hh,cv,\uparrow\uparrow}^{e-h,<}\left(  12\right)  & g_{hh,cv,\uparrow
\downarrow}^{e-h,<}\left(  12\right) \\
g_{hh,cv,\downarrow\uparrow}^{e-h,<}\left(  12\right)  & g_{hh,cv,\downarrow
\downarrow}^{e-h,<}\left(  12\right)
\end{array}
\right) \nonumber\\
&  =\left(
\begin{array}
[c]{cc}%
\left[
\begin{array}
[c]{c}%
v_{cc,\uparrow\sigma}\left(  \xi2\right)  g_{hh,cv,\sigma\uparrow}%
^{e-h,<}\left(  1\xi\right) \\
-g_{hh,cv,\uparrow\sigma}^{e-h,<}\left(  \xi2\right)  v_{vv,\sigma\uparrow
}^{T}\left(  1\xi\right)
\end{array}
\right]  & \left[
\begin{array}
[c]{c}%
v_{cc,\uparrow\sigma}\left(  \xi2\right)  g_{hh,cv,\sigma\downarrow}%
^{e-h,<}\left(  1\xi\right) \\
-g_{hh,cv,\uparrow\sigma}^{e-h,<}\left(  \xi2\right)  v_{vv,\sigma\downarrow
}^{T}\left(  1\xi\right)
\end{array}
\right] \\
\left[
\begin{array}
[c]{c}%
v_{cc,\downarrow\sigma}\left(  \xi2\right)  g_{hh,cv,\sigma\uparrow}%
^{e-h,<}\left(  1\xi\right) \\
-g_{hh,cv,\downarrow\sigma}^{e-h,<}\left(  \xi2\right)  v_{vv,\sigma\uparrow
}^{T}\left(  1\xi\right)
\end{array}
\right]  & \left[
\begin{array}
[c]{c}%
v_{cc,\downarrow\sigma}\left(  \xi2\right)  g_{hh,cv,\sigma\downarrow}%
^{e-h,<}\left(  1\xi\right) \\
-g_{hh,cv,\downarrow\sigma}^{e-h,<}\left(  \xi2\right)  v_{vv,\sigma
\downarrow}^{T}\left(  1\xi\right)
\end{array}
\right]
\end{array}
\right) \nonumber\\
&  +\left(
\begin{array}
[c]{cc}%
\left[
\begin{array}
[c]{c}%
\Sigma_{cc,\uparrow\sigma}^{e-h,r}\left(  1\xi\right)  g_{hh,cv,\sigma
\uparrow}^{e-h,<}\left(  \xi2\right) \\
-G_{cc,\uparrow\sigma}^{e-h,<}\left(  1\xi\right)  \Delta_{hh,cv,\sigma
\uparrow}^{e-h,a}\left(  \xi2\right)
\end{array}
\right]  & \left[
\begin{array}
[c]{c}%
\Sigma_{cc,\uparrow\sigma}^{e-h,r}\left(  1\xi\right)  g_{hh,cv,\sigma
\downarrow}^{e-h,<}\left(  \xi2\right) \\
-G_{cc,\uparrow\sigma}^{e-h,<}\left(  1\xi\right)  \Delta_{hh,cv,\sigma
\downarrow}^{e-h,a}\left(  \xi2\right)
\end{array}
\right] \\
\left[
\begin{array}
[c]{c}%
\Sigma_{cc,\downarrow\sigma}^{e-h,r}\left(  1\xi\right)  g_{hh,cv,\sigma
\uparrow}^{e-h,<}\left(  \xi2\right) \\
-G_{cc,\downarrow\sigma}^{e-h,<}\left(  1\xi\right)  \Delta_{hh,cv,\sigma
\uparrow}^{e-h,a}\left(  \xi2\right)
\end{array}
\right]  & \left[
\begin{array}
[c]{c}%
\Sigma_{cc,\downarrow\sigma}^{e-h,r}\left(  1\xi\right)  g_{hh,cv,\sigma
\downarrow}^{e-h,<}\left(  \xi2\right) \\
-G_{cc,\downarrow\sigma}^{e-h,<}\left(  1\xi\right)  \Delta_{hh,cv,\sigma
\downarrow}^{e-h,a}\left(  \xi2\right)
\end{array}
\right]
\end{array}
\right) \nonumber\\
&  +\left(
\begin{array}
[c]{cc}%
\left[
\begin{array}
[c]{c}%
\Sigma_{cc,\uparrow\sigma}^{e-h,<}\left(  1\xi\right)  g_{hh,cv,\sigma
\uparrow}^{e-h,a}\left(  \xi2\right) \\
-G_{cc,\uparrow\sigma}^{e-h,r}\left(  1\xi\right)  \Delta_{hh,cv,\sigma
\uparrow}^{e-h,<}\left(  \xi2\right)
\end{array}
\right]  & \left[
\begin{array}
[c]{c}%
\Sigma_{cc,\uparrow\sigma}^{e-h,<}\left(  1\xi\right)  g_{hh,cv,\sigma
\downarrow}^{e-h,a}\left(  \xi2\right) \\
-G_{cc,\uparrow\sigma}^{e-h,r}\left(  1\xi\right)  \Delta_{hh,cv,\sigma
\downarrow}^{e-h,<}\left(  \xi2\right)
\end{array}
\right] \\
\left[
\begin{array}
[c]{c}%
\Sigma_{cc,\downarrow\sigma}^{e-h,<}\left(  1\xi\right)  g_{hh,cv,\sigma
\uparrow}^{e-h,a}\left(  \xi2\right) \\
-G_{cc,\downarrow\sigma}^{e-h,r}\left(  1\xi\right)  \Delta_{hh,cv,\sigma
\uparrow}^{e-h,<}\left(  \xi2\right)
\end{array}
\right]  & \left[
\begin{array}
[c]{c}%
\Sigma_{cc,\downarrow\sigma}^{e-h,<}\left(  1\xi\right)  g_{hh,cv,\sigma
\downarrow}^{e-h,a}\left(  \xi2\right) \\
-G_{cc,\downarrow\sigma}^{e-h,r}\left(  1\xi\right)  \Delta_{hh,cv,\sigma
\downarrow}^{e-h,<}\left(  \xi2\right)
\end{array}
\right]
\end{array}
\right) \nonumber\\
&  +\left(
\begin{array}
[c]{cc}%
\left[
\begin{array}
[c]{c}%
-\Delta_{hh,cv,\uparrow\sigma}^{e-h,r}\left(  1\xi\right)  G_{vv,\sigma
\uparrow}^{e-h,>T}\left(  \xi2\right) \\
+g_{hh,cv,\uparrow\sigma}^{e-h,<}\left(  1\xi\right)  \Sigma_{vv,\sigma
\uparrow}^{e-h,rT}\left(  \xi2\right)
\end{array}
\right]  & \left[
\begin{array}
[c]{c}%
-\Delta_{hh,cv,\uparrow\sigma}^{e-h,r}\left(  1\xi\right)  G_{vv,\sigma
\downarrow}^{e-h,>T}\left(  \xi2\right) \\
+g_{hh,cv,\uparrow\sigma}^{e-h,<}\left(  1\xi\right)  \Sigma_{vv,\sigma
\downarrow}^{e-h,rT}\left(  \xi2\right)
\end{array}
\right] \\
\left[
\begin{array}
[c]{c}%
-\Delta_{hh,cv,\downarrow\sigma}^{e-h,r}\left(  1\xi\right)  G_{vv,\sigma
\uparrow}^{e-h,>T}\left(  \xi2\right) \\
+g_{hh,cv,\downarrow\sigma}^{e-h,<}\left(  1\xi\right)  \Sigma_{vv,\sigma
\uparrow}^{e-h,rT}\left(  \xi2\right)
\end{array}
\right]  & \left[
\begin{array}
[c]{c}%
-\Delta_{hh,cv,\downarrow\sigma}^{e-h,r}\left(  1\xi\right)  G_{vv,\sigma
\downarrow}^{e-h,>T}\left(  \xi2\right) \\
+g_{hh,cv,\downarrow\sigma}^{e-h,<}\left(  1\xi\right)  \Sigma_{vv,\sigma
\downarrow}^{e-h,rT}\left(  \xi2\right)
\end{array}
\right]
\end{array}
\right) \nonumber\\
&  +\left(
\begin{array}
[c]{cc}%
\left[
\begin{array}
[c]{c}%
-\Delta_{hh,cv,\uparrow\sigma}^{e-h,<}\left(  1\xi\right)  G_{vv,\sigma
\uparrow}^{e-h,rT}\left(  \xi2\right) \\
+g_{hh,cv,\uparrow\sigma}^{e-h,r}\left(  1\xi\right)  \Sigma_{vv,\sigma
\uparrow}^{e-h,>T}\left(  \xi2\right)
\end{array}
\right]  & \left[
\begin{array}
[c]{c}%
-\Delta_{hh,cv,\uparrow\sigma}^{e-h,<}\left(  1\xi\right)  G_{vv,\sigma
\downarrow}^{e-h,rT}\left(  \xi2\right) \\
+g_{hh,cv,\uparrow\sigma}^{e-h,r}\left(  1\xi\right)  \Sigma_{vv,\sigma
\downarrow}^{e-h,>T}\left(  \xi2\right)
\end{array}
\right] \\
\left[
\begin{array}
[c]{c}%
-\Delta_{hh,cv,\downarrow\sigma}^{e-h,<}\left(  1\xi\right)  G_{vv,\sigma
\uparrow}^{e-h,rT}\left(  \xi2\right) \\
+g_{hh,cv,\downarrow\sigma}^{e-h,r}\left(  1\xi\right)  \Sigma_{vv,\sigma
\uparrow}^{e-h,>T}\left(  \xi2\right)
\end{array}
\right]  & \left[
\begin{array}
[c]{c}%
-\Delta_{hh,cv,\downarrow\sigma}^{e-h,<}\left(  1\xi\right)  G_{vv,\sigma
\downarrow}^{e-h,rT}\left(  \xi2\right) \\
+g_{hh,cv,\downarrow\sigma}^{e-h,r}\left(  1\xi\right)  \Sigma_{vv,\sigma
\downarrow}^{e-h,>T}\left(  \xi2\right)
\end{array}
\right]
\end{array}
\right)  ,
\end{align}
and for its nonequilibrium (reverse) 'conjugate' process,\cite{note9} the
matrix equation derived from Eq. (\ref{eq9}), is,%
\begin{align}
&  i\hbar\left(  \frac{\partial}{\partial t_{1}}+\frac{\partial}{\partial
t_{2}}\right)  \left(
\begin{array}
[c]{cc}%
g_{ee,vc,\uparrow\uparrow}^{e-h,<}\left(  12\right)  & g_{ee,vc,\uparrow
\downarrow}^{e-h,<}\left(  12\right) \\
g_{ee,vc,\downarrow\uparrow}^{e-h,<}\left(  12\right)  & g_{ee,vc,\downarrow
\downarrow}^{e-h,<}\left(  12\right)
\end{array}
\right) \nonumber\\
&  =\left(
\begin{array}
[c]{cc}%
\left[
\begin{array}
[c]{c}%
v_{vv,\uparrow\sigma}\left(  1\xi\right)  g_{ee,vc,\sigma\uparrow}%
^{e-h,<}\left(  \xi2\right) \\
-g_{ee,vc,\uparrow\sigma}^{e-h,<}\left(  1\xi\right)  v_{c,\sigma\uparrow}%
^{T}\left(  \xi2\right)
\end{array}
\right]  & \left[
\begin{array}
[c]{c}%
v_{vv,\uparrow\sigma}\left(  1\xi\right)  g_{ee,vc,\sigma\downarrow}%
^{e-h,<}\left(  \xi2\right) \\
-g_{ee,vc,\uparrow\sigma}^{e-h,<}\left(  1\xi\right)  v_{c,\sigma\downarrow
}^{T}\left(  \xi2\right)
\end{array}
\right] \\
\left[
\begin{array}
[c]{c}%
v_{vv,\downarrow\sigma}\left(  1\xi\right)  g_{ee,vc,\sigma\uparrow}%
^{e-h,<}\left(  \xi2\right) \\
-g_{ee,vc,\downarrow\sigma}^{e-h,<}\left(  1\xi\right)  v_{c,\sigma\uparrow
}^{T}\left(  \xi2\right)
\end{array}
\right]  & \left[
\begin{array}
[c]{c}%
v_{vv,\downarrow\sigma}\left(  1\xi\right)  g_{ee,vc,\sigma\downarrow}%
^{e-h,<}\left(  \xi2\right) \\
-g_{ee,vc,\downarrow\sigma}^{e-h,<}\left(  1\xi\right)  v_{c,\sigma\downarrow
}^{T}\left(  \xi2\right)
\end{array}
\right]
\end{array}
\right) \nonumber\\
&  +\left(
\begin{array}
[c]{cc}%
\left[
\begin{array}
[c]{c}%
-\Sigma_{vv,\uparrow\sigma}^{e-h,aT}\left(  1\xi\right)  g_{ee,vc,\sigma
\uparrow}^{e-h,<}\left(  \xi2\right) \\
+G_{vv,\uparrow\sigma}^{e-h,>T}\left(  1\xi\right)  \Delta_{ee,vc,\sigma
\uparrow}^{e-h,a}\left(  \xi2\right)
\end{array}
\right]  & \left[
\begin{array}
[c]{c}%
-\Sigma_{vv,\uparrow\sigma}^{e-h,aT}\left(  1\xi\right)  g_{ee,vc,\sigma
\downarrow}^{e-h,<}\left(  \xi2\right) \\
+G_{vv,\uparrow\sigma}^{e-h,>T}\left(  1\xi\right)  \Delta_{ee,vc,\sigma
\downarrow}^{e-h,a}\left(  \xi2\right)
\end{array}
\right] \\
\left[
\begin{array}
[c]{c}%
-\Sigma_{vv,\downarrow\sigma}^{e-h,aT}\left(  1\xi\right)  g_{ee,vc,\sigma
\uparrow}^{e-h,<}\left(  \xi2\right) \\
+G_{vv,\downarrow\sigma}^{e-h,>T}\left(  1\xi\right)  \Delta_{ee,vc,\sigma
\uparrow}^{e-h,a}\left(  \xi2\right)
\end{array}
\right]  & \left[
\begin{array}
[c]{c}%
-\Sigma_{vv,\downarrow\sigma}^{e-h,aT}\left(  1\xi\right)  g_{ee,vc,\sigma
\downarrow}^{e-h,<}\left(  \xi2\right) \\
+G_{vv,\downarrow\sigma}^{e-h,>T}\left(  1\xi\right)  \Delta_{ee,vc,\sigma
\downarrow}^{e-h,a}\left(  \xi2\right)
\end{array}
\right]
\end{array}
\right) \nonumber\\
&  +\left(
\begin{array}
[c]{cc}%
\left[
\begin{array}
[c]{c}%
-\Sigma_{vv,\uparrow\sigma}^{e-h,>T}\left(  1\xi\right)  g_{ee,vc,\sigma
\uparrow}^{e-h,a}\left(  \xi2\right) \\
+G_{vv,\uparrow\sigma}^{e-h,aT}\left(  1\xi\right)  \Delta_{ee,vc,\sigma
\uparrow}^{e-h,<}\left(  \xi2\right)
\end{array}
\right]  & \left[
\begin{array}
[c]{c}%
-\Sigma_{vv,\uparrow\sigma}^{e-h,>T}\left(  1\xi\right)  g_{ee,vc,\sigma
\downarrow}^{e-h,a}\left(  \xi2\right) \\
+G_{vv,\uparrow\sigma}^{e-h,aT}\left(  1\xi\right)  \Delta_{ee,vc,\sigma
\downarrow}^{e-h,<}\left(  \xi2\right)
\end{array}
\right] \\
\left[
\begin{array}
[c]{c}%
-\Sigma_{vv,\downarrow\sigma}^{e-h,>T}\left(  1\xi\right)  g_{ee,vc,\sigma
\uparrow}^{e-h,a}\left(  \xi2\right) \\
+G_{vv,\downarrow\sigma}^{e-h,aT}\left(  1\xi\right)  \Delta_{ee,vc,\sigma
\uparrow}^{e-h,<}\left(  \xi2\right)
\end{array}
\right]  & \left[
\begin{array}
[c]{c}%
-\Sigma_{vv,\downarrow\sigma}^{e-h,>T}\left(  1\xi\right)  g_{ee,vc,\sigma
\downarrow}^{e-h,a}\left(  \xi2\right) \\
+G_{vv,\downarrow\sigma}^{e-h,aT}\left(  1\xi\right)  \Delta_{ee,vc,\sigma
\downarrow}^{e-h,<}\left(  \xi2\right)
\end{array}
\right]
\end{array}
\right) \nonumber\\
&  +\left(
\begin{array}
[c]{cc}%
\left[
\begin{array}
[c]{c}%
\Delta_{ee,vc,\uparrow\sigma}^{e-h,r}\left(  1\xi\right)  G_{cc,\sigma
\uparrow}^{e-h,<}\left(  \xi2\right) \\
-g_{ee,vc,\uparrow\sigma}^{e-h,<}\left(  1\xi\right)  \Sigma_{cc,\sigma
\uparrow}^{e-h,a}\left(  \xi2\right)
\end{array}
\right]  & \left[
\begin{array}
[c]{c}%
\Delta_{ee,vc,\uparrow\sigma}^{e-h,r}\left(  1\xi\right)  G_{cc,\sigma
\downarrow}^{e-h,<}\left(  \xi2\right) \\
-g_{ee,vc,\uparrow\sigma}^{e-h,<}\left(  1\xi\right)  \Sigma_{cc,\sigma
\downarrow}^{e-h,a}\left(  \xi2\right)
\end{array}
\right] \\
\left[
\begin{array}
[c]{c}%
\Delta_{ee,vc,\downarrow\sigma}^{e-h,r}\left(  1\xi\right)  G_{cc,\sigma
\uparrow}^{e-h,<}\left(  \xi2\right) \\
-g_{ee,vc,\downarrow\sigma}^{e-h,<}\left(  1\xi\right)  \Sigma_{cc,\sigma
\uparrow}^{e-h,a}\left(  \xi2\right)
\end{array}
\right]  & \left[
\begin{array}
[c]{c}%
\Delta_{ee,vc,\downarrow\sigma}^{e-h,r}\left(  1\xi\right)  G_{cc,\sigma
\downarrow}^{e-h,<}\left(  \xi2\right) \\
-g_{ee,vc,\downarrow\sigma}^{e-h,<}\left(  1\xi\right)  \Sigma_{cc,\sigma
\downarrow}^{e-h,a}\left(  \xi2\right)
\end{array}
\right]
\end{array}
\right) \nonumber\\
&  +\left(
\begin{array}
[c]{cc}%
\left[
\begin{array}
[c]{c}%
\Delta_{ee,vc,\uparrow\sigma}^{e-h,<}\left(  1\xi\right)  G_{cc,\sigma
\uparrow}^{e-h,a}\left(  \xi2\right) \\
-g_{ee,vc,\uparrow\sigma}^{e-h,r}\left(  1\xi\right)  \Sigma_{cc,\sigma
\uparrow}^{e-h,<}\left(  \xi2\right)
\end{array}
\right]  & \left[
\begin{array}
[c]{c}%
\Delta_{ee,vc,\uparrow\sigma}^{e-h,<}\left(  1\xi\right)  G_{cc,\sigma
\downarrow}^{e-h,a}\left(  \xi2\right) \\
-g_{ee,vc,\uparrow\sigma}^{e-h,r}\left(  1\xi\right)  \Sigma_{cc,\sigma
\downarrow}^{e-h,<}\left(  \xi2\right)
\end{array}
\right] \\
\left[
\begin{array}
[c]{c}%
\Delta_{ee,vc,\downarrow\sigma}^{e-h,<}\left(  1\xi\right)  G_{cc,\sigma
\uparrow}^{e-h,a}\left(  \xi2\right) \\
-g_{ee,vc,\downarrow\sigma}^{e-h,r}\left(  1\xi\right)  \Sigma_{cc,\sigma
\uparrow}^{e-h,<}\left(  \xi2\right)
\end{array}
\right]  & \left[
\begin{array}
[c]{c}%
\Delta_{ee,vc,\downarrow\sigma}^{e-h,<}\left(  1\xi\right)  G_{cc,\sigma
\downarrow}^{e-h,a}\left(  \xi2\right) \\
-g_{ee,vc,\downarrow\sigma}^{e-h,r}\left(  1\xi\right)  \Sigma_{cc,\sigma
\downarrow}^{e-h,<}\left(  \xi2\right)
\end{array}
\right]
\end{array}
\right)  .
\end{align}


\begin{thebibliography}{99}
\bibitem {note1}Gilbert\cite{gilbert} has shown that the damping terms can
contribute to the torque in the system. In quantum field theoretical language
this means that the spin-dependent self-energies will contribute to the torque
in the system.
\bibitem {note2}In this paper, we consider only pairing between fermions of
different species, namely, electrons and holes. The pairing potential is
derived from the interband self-energies, $\Sigma_{\alpha\beta}$, and hence
the electron-hole coupling is denoted by $\Delta_{\alpha\beta}^{e-h}$ in the
electron-hole picture or 'defect' representation. Remarkably, interband
pairing has been treated in the context of quantum chromodynamics QCD and cold
atoms, heavy fermions, cuprates, and BCS superconductivity.\cite{moreo,
gubankova, liu} In heavy fermions, the interband pairing occurs if two Fermi
surfaces arising from different bands are very close to one other. In QCD and
cold atoms the pairing is between particles with very different masses.
\bibitem {note3}$S_{cc,o}^{<}$ represents the trace of $2\times2$ spin
matrices, which can be equated to unity in pure quantum state. However, in
nonequilibrium quantum transport physics, $S_{cc,o}^{<}$ represent the
spin-independent charge of the conduction electrons, which could vary in space
and time. Therefore the transport equation for $S_{cc,o}^{<}$ must also be
solved for $S_{cc,o}^{<}$ to be used in calculating the charge to be fed to
the Poisson equation for self-consistency.\cite{arxiv}. For now, we postpone
this treatment as this will be incorporated later in the pseudo-spin
formulation, where it is the total charge in our multi-band system that is the
relevant variable to be fed to the Poisson equation for self-consistency in
the potential distribution.
\bibitem {note5}This relation is used to bring Eqs. (\ref{atom2}) and
(\ref{atom3}) to the electron picture. In fact this substitution is trivial by
virtue of the locality in space and time so that the correlation functions
involved are proportional to the absolute value of the field variables.
\bibitem {note6}In the electron picture, the right-hand sides of Eqs.
(\ref{atom1}){} and (\ref{atom4}) have opposite sign, whereas in the
electon-hole picture they have the same sign or have the same rate of change
as discussed before.
\bibitem {note7}$S_{o,o}$ represents the trace of the $4\times4$ matrix of Eq.
(\ref{eq15}) and hence independent of the Pauli-Dirac and pseudo-spin degrees
of freedom. The time vaiation does not in general vanish by virtue of the
nonlocality of the space and time arguments of the two-point correlations
involved in its equation.
\bibitem {note8}In the same sense perhaps where we usually treat the band
structure as a separate calculations already done before we construct our
two-band model of the semiconductor.
\bibitem {boltzmann}That the gradient expansion of SMQTEs, with the $\left[
\Sigma^{<},\operatorname{Re}S^{r}\right]  $ term neglected, yields the
classical Boltzmann transport equation for spinless system is often viewed as
justifying the neglect of $\left[  \Sigma^{<},\operatorname{Re}S^{r}\right]  $
in the quantum transport equations\cite{kadanoff}.
\bibitem {note9}For highly nonequilibrium situation, the reverse process is
not simply the Hermitian conjugate of the 'forward' process.
\end{thebibliography}
\end{document}